\documentclass[12pt]{article}
\usepackage{\jobname,epsfig,amsmath,amssymb,graphics,color,calc,cite,graphpap}
\usepackage{graphicx}

\newcommand{\sezione}[2]{
\refstepcounter{section}\label{#2}
\setcounter{equation}{0}
\setcounter{subsection}{0}
\addcontentsline{toc}{section}
      {\normalsize\textbf{\thesection.\ #1}}
\bigskip\bigskip\noindent
\normalsize\textbf{\thesection.\ #1}\nopagebreak\smallskip\nopagebreak}
\def\thesection{{\normalsize\arabic{section}}}
\newcommand{\subsec}[2]{
\refstepcounter{subsection}\label{#2}
\addcontentsline{toc}{subsection}
      {\normalsize\normalfont\textit{\thesubsection.\ #1}}
\medskip\medskip\noindent
\normalsize\normalfont
\textit{\thesubsection. \ #1}\nopagebreak\smallskip\nopagebreak}
\def\thesubsection{{\normalsize
{\arabic{section}.\arabic{subsection}}}}
\newcounter{appendice}
\def\theappendice{{\normalsize\Alph{appendice}}}
\newcommand{\appendice}[2]{
\refstepcounter{appendice}\label{#2}
\setcounter{equation}{0}
\setcounter{subsection}{0}
\setcounter{theorem}{0}
\def\theequation{\Alph{appendice}.\arabic{equation}}
\def\thelemma{\Alph{appendice}.\arabic{lemma}}
\def\thetheorem{\Alph{appendice}.\arabic{theorem}}
\def\thecorollary{\Alph{appendice}.\arabic{corollary}}
\def\theproposition{\Alph{appendice}.\arabic{proposition}}
\addcontentsline{toc}{section}
      {\normalsize\textbf{\theappendice.\ #1}}
\bigskip\bigskip\noindent
\normalsize\textbf{\theappendice.\ #1}\smallskip\nopagebreak}
\def\thelemma{\arabic{section}.\arabic{lemma}}
\def\thetheorem{\arabic{section}.\arabic{theorem}}
\def\theproposition{\arabic{section}.\arabic{proposition}}
\def\theequation{\arabic{section}.\arabic{equation}}

\newtheorem{theorem}{Theorem}[section]
\newtheorem{proposition}[theorem]{Proposition}

\newtheorem{lemma}[theorem]{Lemma}

\newcommand{\rr}[1]{{\normalfont\textrm{#1}}}
\newcommand{\cc}[1]{{\mathcal{#1}}}
\newcommand{\bb}[1]{{\normalfont\mathbb{#1}}}

\newcommand{\newatop}[2]{\genfrac{}{}{0pt}{}{#1}{#2}}
\newcommand{\qed}{$\phantom.$\hfill $\Box$\bigskip}
\newcommand{\dis}{\mathrm d}

\newcommand{\ul}[1]{\underline{#1}}
\newcommand{\ol}[1]{\overline{#1}}

\newcommand{\puno}{{+\underline{1}}}
\newcommand{\muno}{{-\underline{1}}}

\definecolor{light}{gray}{.9}

\begin{document}
\begin{titlepage}
\par\vskip 1cm\vskip 2em

\begin{center}
{\Large 
Metastability for reversible probabilistic cellular automata with
\vskip 0.2 cm
self--interaction}

\medskip\bigskip
{\large
\begin{tabular}[t]{c}
$\mbox{Emilio N.M.\ Cirillo}^{\tt 1}
\phantom{m} \mbox{Francesca R.\ Nardi}^{\tt 2,4}
\phantom{m} \mbox{Cristian Spitoni}^{\tt 3,4}$
\\
\end{tabular}
\par
}
\medskip
{\small
\begin{tabular}[t]{ll}
{\tt 1} & {\it Dipartimento Me.\ Mo.\ Mat., Universit\`a di Roma La Sapienza}\\
&  via A.\ Scarpa 16, 00161 Roma, Italy\\
&  E--mail: {\tt cirillo@dmmm.uniroma1.it}\\
\\
{\tt 2} & {\it Department of Mathematics and Computer Science}\\
& {\it Eindhoven University of Technology}\\
& P.O.\ Box 513, 5600 MB Eindhoven, The Netherlands \\
& E--mail: {\tt F.R.Nardi@tue.nl}\\
\\
{\tt 3} & {\it
Mathematical Institute, Leiden University}\\
& P.O.\ Box 9512, 2300 RA Leiden, The Netherlands\\
& E--mail: {\tt spitoni@eurandom.tue.nl}\\
\\
{\tt 4} & {\it
Eurandom}\\
& P.O.\ Box 513, 5600 MB, Eindhoven, The Netherlands\\
\end{tabular}
}
\end{center}

\smallskip\bigskip
\centerline{\bf Abstract}
\smallskip
\par\noindent
\small{The problem of metastability for a stochastic dynamics with a
parallel updating rule is addressed in the Freidlin--Wentzel regime,
namely, finite volume, small magnetic field, and small temperature.
The model is characterized by the existence of many fixed points and cyclic
pairs of the zero temperature dynamics, in which the system can be trapped in 
its way to the stable phase. 
Our strategy is based on recent powerful approaches, not needing 
a complete description of the fixed points of the dynamics, but relying on
few model dependent results. 
We compute the exit time, in the sense of logarithmic equivalence, and 
characterize the critical droplet that is necessarily visited by the 
system during its excursion from the metastable to the stable state.
We need to supply two model dependent inputs: (1) the communication 
energy, that is the minimal 
energy barrier that the system must overcome to reach the stable 
state starting from the metastable one;
(2) a recurrence property stating that 
for any configuration different from the metastable state there 
exists a path, starting from such a configuration 
and reaching a lower energy state,
such that its maximal energy is lower than the communication energy.
}



\medskip\bigskip
\footnoterule
\vskip 1.0em
{\small
\noindent
One of the authors (ENMC) acknowledges Eurandom for the kind hospitality.
CS would like to thank Anton Bovier for his kind hospitality at 
Weierstrass--Institut (Berlin). 
The work of FRN was supported by Dipartimento di Matematica, 
Universit\`a di Roma Tre. The work of CS was partially supported by 
Dipartimento Me.\ Mo.\ Mat., Universit\`a di Roma ``La Sapienza."

\vskip 1.0em
\noindent
}
\end{titlepage}
\vfill\eject

\sezione{Introduction}{s:int}
\par\noindent
Metastable states are very common in nature and are typical of systems
close to a first order phase transition. It is often observed that a system
can persist for a long period of time in a phase which is not the one 
favored by the thermodynamic parameters; classical examples are 
the super--saturated vapor and the magnetic hysteresis. The rigorous description
of this phenomenon in the framework of well defined mathematical models is 
relatively recent, dating back to the pioneering paper \cite{[CGOV]}, and 
has experienced substantial progress in the last decade. See \cite{[OV]} for a 
list of the most important papers on this subject. 

A natural setup in which the phenomenon of metastability can be studied 
is that of Markov chains, or Markov processes, describing 
the time evolution of a statistical mechanical system. Think for instance to 
a stochastic lattice spin system. In this context powerful theories
(see \cite{[BEGM],[MNOS],[OS]}) 
have been developed with the aim to find answers valid with maximal 
generality and to reduce
to a minimum the number of model dependent inputs necessary to describe 
the metastable behavior of the system. Whatever approach is chosen, the key 
model dependent question is the computation of the minimal energy barrier,
called \textit{communication energy}, to be overcome
by a path connecting the metastable to the stable state. Such a problem is in 
general quite complicated and becomes particularly difficult when the dynamics 
has a parallel character. Indeed, if simultaneous updates are allowed 
on the lattice, then no constraint on the structure of the trajectories 
in the configuration space is imposed. Therefore, to compute the communication 
energy, one must take into account all the possible transitions in the 
configuration space. 

The problem of the computation of the communication energy in a parallel 
dynamics setup has been addressed in \cite{[C],[CN]}. In particular, in 
\cite{[CN]} the typical questions of
metastability, that is  the determination of the \textit{exit time} and 
of the \textit{exit tube}, have been answered for a reversible Probabilistic 
Cellular Automaton (see~\cite{[GLD],[R],[St],[To],[D],[CNP]}), in which each 
spin is coupled only with its nearest neighbors. 
In that paper it has been shown that, during the transition from the 
metastable minus state to the stable plus state, the system visits an 
intermediate chessboard--like phase. 
In the present paper we study the reversible PCA in 
which each spin interacts both with itself and with its nearest neighbors;
the metastable behavior of such a model has been investigated on heuristic and 
numerical grounds in \cite{[BCLS]}.
The addition of the self--interaction changes completely the metastability 
scenario; in particular we show that the chessboard-like 
phase plays no role in the exit from the metastable phase.  

Another very interesting feature of this model is the presence of a large 
number of fixed points of the zero--temperature dynamics in which the 
system can be trapped. Following the powerful approach of \cite{[MNOS]},
we can 
compute the exit time avoiding a 
complete description of the trapping states. However, we cannot describe the 
exit tube, i.e., the tube of trajectories followed by the system during its 
exit from the metastable to the stable phase. 
The only information on the exit path that we prove in this paper
is the existence 
of a particular set of configurations which is necessarily visited by the 
system during its excursion from the metastable to the stable state.
This set plays the role of the saddle configuration set, which is 
usually introduced in the study of the metastable behavior of 
sequential dynamics.

According to the approach of \cite{[MNOS]}, 
the model dependent ingredients that must be 
provided are essentially two: 
(1) the solution of the global variational problem for all the paths connecting 
the metastable and the stable state, i.e., the computation of the 
communication energy;
(2) a sort of recurrence property stating
that, starting from each configuration different from the metastable 
and the stable state, it is possible to reach a configuration at lower 
energy following a path with an energy cost strictly smaller than the 
communication energy. 

To solve the global variational problem 
(see items~\ref{i:modeest} and~\ref{i:modeest2} in Theorem~\ref{t:stime}),
we obtain an upper bound on the communication energy by exhibiting a path
connecting the metastable state to the stable state whose maximal 
energy is equal to the communication energy. To find the lower
bound, we perform a partition of the configuration space, study the 
transitions between configurations in these partitions, and 
reduce the computation to the optimal one (see Figure~\ref{f:strategia}). 
To prove the recurrence property 
(see item~\ref{i:recurrence} in Theorem~\ref{t:stime}), we have to face 
the problem of the existence of a large number of fixed points of the 
dynamics. We solve this problem by showing that, for each configuration 
different from the metastable state, it is possible to find a path 
connecting it to the stable state, i.e., to the unique global minimum of the 
energy, such that the energy along this path is strictly smaller than 
the communication energy. 

We finally give a brief description of the content of the paper.
In Section~\ref{s:mod} we define the model and state
our main result in Theorem~\ref{t:meta}. 
The proof of Theorem~\ref{t:meta}, based on the model dependent 
results in Theorem~\ref{t:stime} and on \cite{[MNOS]}, is given in 
Section~\ref{s:scappo}. 
Section~\ref{s:moddep} is devoted to the proof
of the estimates on the energy landscape stated in Theorem~\ref{t:stime},
namely, the global variational problem 
(items~\ref{i:modeest} and \ref{i:modeest2}) and the recurrence property
(item~\ref{i:recurrence}).
The proof of items~\ref{i:modeest} and \ref{i:modeest2} relies on 
Proposition~\ref{t:insiemeG}, which is proven in 
Section~\ref{s:minmax}.
The appendix is devoted to a brief review of results in \cite{[MNOS]}.

\sezione{Model and results}{s:mod}
\par\noindent
In this section we introduce the basic notation, define the model, and state
our main result. 
In particular, Sections~\ref{s:lat}--\ref{s:stat} are devoted to the 
definition of the Probabilistic Cellular Automaton which will be studied in 
the sequel. 
In Section~\ref{s:meta} we state Theorem~\ref{t:meta} with the results 
on the metastable behavior of the system. 
In Section~\ref{s:transition} we introduce the transition rates and the 
zero temperature dynamics;
in Section~\ref{s:heu} we develop an heuristic argument on which the proof of 
the theorem is based. 
In Section~\ref{s:scappo}, finally, we prove Theorem~\ref{t:meta}.

\subsec{The lattice}{s:lat}
\par\noindent
The spatial structure is modeled by the two--dimensional finite square
$\Lambda:=\{0,\dots,L-1\}^2$, where $L$ is a positive integer,
with periodic boundary conditions; note that $\Lambda$ is a torus.
We shall use the metric induced by the Euclidean distance on the flat torus.
An element of $\Lambda$ is called a \textit{site}. 
We use $X^\rr{c}:=\Lambda\setminus X$ to denote the complement of 
$X\subset\Lambda$.

Let $x\in\Lambda$; we say that $y\in\Lambda$ is a
\textit{nearest neighbor} of $x$ if and only if
the distance on the torus of $x$ from $y$ is equal to $1$.
For $X\subset\Lambda$, we say that $y\in X^\rr{c}$ is an element of the 
\textit{external boundary} $\partial X$ of $X$ if and only if at least 
one of its nearest neighbors belongs to $X$;
we let also $\overline{X}:=X\cup\partial X$ be the {\it closure} of $X$.
Two sets $X,Y\subset\Lambda$ are said to be {\it not interacting} if and only
if for any $x\in X$ and $y\in Y$ their distance on the torus is 
larger or equal to $\sqrt{5}$. 

Let $x=(x_1,x_2)\in\Lambda$;
for $\ell_1,\ell_2$ positive integers we let
$Q_{\ell_1,\ell_2}(x)$ be the collection of the sites 
$\big((x_1+s_1)\!\!\!\mod L,(x_2+s_2)\!\!\!\mod L\big)$ 
for $s_i=0,\dots,x_i+\ell_i-1$ where $i=1,2$. 
Roughly speaking, $Q_{\ell_1,\ell_2}(x)$ is
the rectangle on the torus of side lengths $\ell_1$ and $\ell_2$
drawn starting from $x$ and moving in the positive direction along 
the two coordinate axes.
For $\ell$ a positive integer we let $Q_\ell(x):=Q_{\ell,\ell}(x)$.

\subsec{The configuration space}{s:conf}
\par\noindent
The single spin state space is given by the finite set $\{-1,+1\}$;
the configuration space in $X\subset\Lambda$ is defined as
$\cc{S}_X:=\{-1,+1\}^X$ and considered
equipped with the discrete topology and the corresponding Borel $\sigma$
algebra $\cc{F}_X$.
The model and the related quantities that will be
introduced later on will all depend on $\Lambda$, but since $\Lambda$ is
fixed it will be dropped from the notation; in this spirit
we let $\cc{S}_\Lambda=:\cc{S}$ and $\cc{F}_\Lambda=:\cc{F}$.

Given a configuration $\sigma\in\cc{S}$ and $X\subset\Lambda$,
we denote by $\sigma_X$ the {\it restriction} of $\sigma$
to $X$.
Let $m$ be a positive integer and
let $X_1,\dots,X_m\subset\Lambda$ be pairwise disjoint subsets of
$\Lambda$; for 
$\sigma_k\in\cc{S}_{X_k}$, with $k=1,\dots,m$, we denote by
$\sigma_1\sigma_2\cdots\sigma_m$
the configuration in $\cc{S}_{X_1\cup\cdots\cup X_m}$ such that
$(\sigma_1\sigma_2\cdots\sigma_m)_{X_k}=\sigma_k$
for all $k\in\{1,\dots,m\}$.
Moreover, given $\sigma\in\cc{S}$ and $x\in\Lambda$, we denote by
$\sigma^x$ the configuration such that $\sigma^x(x)=-\sigma(x)$ and
$\sigma^x(y)=\sigma(y)$ for $y\neq x$.
Let $x\in\Lambda$, we define the shift $\Theta_x$
acting on $\cc{S}$ by setting $(\Theta_x\sigma)_y:=\sigma_{y+x}$
for all $y\in\Lambda$ and $\sigma\in\cc{S}$.

Given a function $f:\cc{S}\to\bb{R}$,
if $f\in\cc{F}_X$ we shall
sometimes write $f(\sigma_X)$ for $f(\sigma)$.
Let $f,g:\cc{S}\to\cc{S}$ be two functions, we consider the
{\it product} or {\it composed} function $fg:\cc{S}\to\cc{S}$ such
that $fg(\sigma):=f(g(\sigma))$ for any $\sigma\in\cc{S}$.
We also let $f^2:=ff$ and, for $n$ a positive integer,
$f^n:=ff^{n-1}$.
We say that a configuration $\sigma\in\cc{S}$ is a {\it fixed point}
for the map $f:\cc{S}\to\cc{S}$ if and only if $f(\sigma)=\sigma$. Let
$\sigma\in\cc{S}$, consider the sequence $f^n(\sigma)$ with $n\ge1$,
if there exists $n'$ such that $f^n(\sigma)=f^{n'}(\sigma)$ for any
$n\ge n'$, we then let $\ol{f}\sigma:=f^{n'}\sigma$.

\subsec{The model}{s:dinamica}
\par\noindent
Let $\beta>0$ and $h\in\bb{R}$ such that $|h|<1$ and $2/h$ is not integer.
We consider the Markov chain on $\cc{S}$ with {\it transition matrix}
\begin{equation}
\label{markov}
p(\sigma,\eta)
:=\prod_{x\in\Lambda}p_{x,\sigma}\left(\eta(x)\right)\;\;\;
\forall\sigma,\eta\in\cc{S}
\end{equation}
where, for each $x\in\Lambda$ and $\sigma\in\cc{S}$,
$p_{x,\sigma}(\cdot)$ is the probability measure on $\cc{S}_{\{x\}}$
defined as follows
\begin{equation}
\label{rule}
p_{x,\sigma}(s)
:=
\frac{1}
{1+\exp\left\{-2\beta s(S_\sigma(x)+h)\right\}}
=
\frac{1}{2}
\left[1+s\tanh\beta \left(
S_\sigma(x)+h\right)\right]
\end{equation}
with $s\in\{-1,+1\}$ and
\begin{equation}
\label{rule2}
S_{\sigma}(x):= \sum_{y\in\overline{\{x\}}}\sigma(y)
\end{equation}
The normalization condition
$p_{x,\sigma}(s)+p_{x,\sigma}(-s)=1$ is trivially satisfied.
Note that $p_{x,\cdot}(s)\in\cc{F}_{\overline{\{x\}}}$ 
for any $x$ and $s$, that is  
the probability $p_{x,\sigma}(s)$ for the spin at site $x$ to be equal to $s$
depends only on the values of the five spins of $\sigma$ inside the
cross $\overline{\{x\}}$ centered at $x$.

Such a Markov chain on the finite space $\cc{S}$ is an example
of {\it reversible probabilistic cellular automata} (PCA), see 
\cite{[D],[CNP]}. Let $n\in\bb{N}$ be the discrete time
variable and let $\sigma_n\in\cc{S}$ denote the state of the chain at time
$n$; the configuration at time $n+1$ is chosen according to the law
$p(\sigma_n,\cdot)$, see (\ref{markov}), hence all the spins are updated
simultaneously and independently at any time.
Finally, given $\sigma\in\cc{S}$ we consider the chain with
initial configuration $\sigma_0=\sigma$, we denote with
$\bb{P}_\sigma$ the probability measure on the space
of trajectories, by $\bb{E}_\sigma$ the corresponding expectation value,
and by
\begin{equation}
\label{hitting}
\tau_A^\sigma:=\inf\{t>0:\,\sigma_t\in A\}
\end{equation}
the {\it first hitting time on} $A\subset\cc{S}$. We shall drop the initial
configuration from the notation (\ref{hitting}) whenever it is equal to
$\muno$, i.e., we shall write $\tau_A$ instead of $\tau_A^{\muno}$.

\subsec{The stationary measure and the phase diagram}{s:stat}
\par\noindent
The model (\ref{markov}) has been studied
numerically in \cite{[BCLS]}; we refer to that paper for
a detailed discussion about its stationary properties.
Here we simply recall the main features.
It is straightforward, see for instance \cite{[CNP],[D]},
that the PCA (\ref{markov})
is reversible with respect to the finite volume Gibbs measure
$\mu(\sigma):=\exp\{-H(\sigma)\}/Z$
with
$Z:=\sum_{\eta\in\cc{S}}\exp\{-H(\eta)\}$
and
\begin{equation}
\label{ham}
H(\sigma):=
H_{\beta,h}(\sigma):=
-\beta h\sum_{x\in\Lambda}\sigma(x)
-       \sum_{x\in\Lambda}   \log\cosh\left[\beta
\left(
S_{\sigma}(x)+h\right)\right]
\end{equation}
In other words the \textit{detailed balance} condition
\begin{equation}
\label{dett}
p(\sigma,\eta)\,e^{-H(\sigma)}=
p(\eta,\sigma)\,e^{-H(\eta)}
\end{equation}
is satisfied for any $\sigma,\eta\in\cc{S}$;
hence, the measure $\mu$ is stationary for the PCA (\ref{markov}).
In order to
understand its most important features, it is useful to study the related
Hamiltonian.
Since the Hamiltonian has the form (\ref{ham}), we shall often refer to
$1/\beta$ as to the {\it temperature} and to $h$ as to the
{\it magnetic field}.

The interaction is short range and it is possible to extract the
potentials; following \cite{[BCLS]} we rewrite the Hamiltonian
as
\begin{equation}
\label{rewham}
H_{\beta,h}(\sigma)=\sum_{x\in\Lambda} U_{x,\beta,h}(\sigma)
-\beta h\sum_{x\in\Lambda}\sigma(x)
\end{equation}
where $U_{x,\beta,h}(\sigma)=U_{0,\beta,h}(\Theta_x\sigma)$, recall that the
shift operator $\Theta_x$ has been defined in Section~\ref{s:conf} and
that periodic boundary are considered on $\Lambda$, and
\begin{equation}
\label{eq:U}
U_{0,\beta,h}(\sigma)
  =-\sum_{X\subset\overline{\{0\}}} J_{|X|,\beta,h}\prod_{x\in X}\sigma(x)
\end{equation}
The six coefficients $J_{0,\beta,h},\dots,J_{5,\beta,h}$ are determined by
using (\ref{ham}), (\ref{rewham}), and (\ref{eq:U}).
In the case $h=0$ only even values of $|X|$ occur and we find that the
pair interactions are ferromagnetic while the four--spin
interactions are not.
For a more detailed discussion see \cite{[BCLS]}.

The definition of ground state is not completely trivial in our model,
indeed the Hamiltonian $H$ depends on $\beta$.
The ground states are those configurations on which the Gibbs
measure $\mu$ is concentrated when the limit
$\beta\to\infty$ is considered, so that they can be defined as the
minima of the \textit{energy}
\begin{equation}
\label{hl}
E(\sigma):=
\lim_{\beta\to\infty}\frac{H(\sigma)}{\beta}
=
-h\sum_{x\in\Lambda}\sigma(x)
-\sum_{x\in\Lambda}|S_{\sigma}(x)+h|
\end{equation}
Let $\cc{X}\subset\cc{S}$, if the energy $E$ is constant on $\cc{X}$, we shall
misuse the notation by denoting by $E(\cc{X})$ the energy of the 
configurations in $\cc{X}$.

We first consider the case $h=0$. Since
$E(\sigma)=-\sum_{x\in\Lambda}|S_{\sigma}(x)|$,
it is obvious that there exist the two minima
$\puno,\muno\in\cc{S}$, with $\pm\ul{1}(x)=\pm1$ for each $x\in\Lambda$, such
that $E(\puno)=E(\muno)=-5|\Lambda|$.
For $h\neq0$ we have
$E(\puno)=-|\Lambda|(h+|5+h|)$ and
$E(\muno)=-|\Lambda|(-h+|-5+h|)$; it is immediate to verify that
$E(\puno)<E(\muno)$ for $h>0$ and
$E(\muno)<E(\puno)$ for $h<0$. We conclude that at $h=0$
there exist the two ground states $\muno$ and $\puno$.
At $h>0$ the unique ground state is given by $\puno$ and
at $h<0$ the unique ground state is given by $\muno$.
The phase diagram at finite large $\beta$ and $h=0$ has been studied 
rigorously in \cite{[DLR]}.

\subsec{Metastable behavior}{s:meta}
\par\noindent
We pose now the problem of metastability and state the related theorem
on the exit time. 
In this context, configurations with all the spins equal to minus one excepted
those in rectangular subsets of the lattice will play a key role.
We then let
\begin{equation}
\label{lambdo}
\Lambda^{\pm}(\sigma):=
 \{x\!\in\!\Lambda\!:\sigma(x)=\pm1\}
\end{equation}
for any $\sigma\in\cc{S}$; the set $\Lambda^+(\sigma)$ will be called
the \textit{support} of $\sigma$.
We say that $\sigma\in\cc{S}$ is a
\textit{rectangular droplet with side lengths $\ell$ and $m$}, with
$\ell,m$ integers such that $2\le\ell,m\le L-2$, if and only if
there exists $x\in\Lambda$ such that either
$\Lambda^+(\sigma)=Q_{\ell,m}(x)$ or
$\Lambda^+(\sigma)=Q_{m,\ell}(x)$.
We say that $\sigma\in\cc{S}$ is a
\textit{$n$--rectangular droplet with side lengths
$\ell_1,m_1,\dots,\ell_n,m_n$}, with $n\ge1$ an integer and
$\ell_i,m_i$ integers such that $2\le\ell_i,m_i\le L-2$ for $i=1,\dots,n$,
if and only if $\Lambda^+(\sigma)$ is the union of $n$ pairwise
not interacting rectangles (see Section~\ref{s:lat})
with side lengths $\ell_i$ and $m_i$ for $i=1,\dots,n$.
We finally say that $\sigma\in\cc{S}$ is a
\textit{multi--rectangular droplet} if and only if $\sigma$ is a
$n$--rectangular droplet for some integer $n\ge1$.
Note that a $1$--rectangular droplet
is indeed a rectangular droplet.
\textit{Square droplets} are defined similarly.

Consider, now, the model (\ref{markov}) with $0<h<1$ and
suppose that the system is prepared in the state $\sigma_0=\muno$;
in the infinite time limit the system tends to the phase with positive
magnetization.
We shall show that the minus one state is metastable in
the sense that the system spends a huge amount of time close to $\muno$
before visiting $\puno$; more precisely the first hitting
time $\tau_{\puno}$ to $\puno$ (recall (\ref{hitting}) and the remark below)
is an exponential random variable with mean
exponentially large in $\beta$.

Moreover, we give some information on the exit path that the system
follows during the escape from minus one to plus one. More precisely,
we show that there exists a class of configurations $\cc{C}\subset\cc{S}$,
called set of {\it critical droplets},
which is visited with high probability by the system during its escape from
$\muno$ to $\puno$. 
Let the {\it critical length} $\lambda$ be defined as
\begin{equation}
\label{lcritica}
\lambda:=\Big\lfloor\frac{2}{h}\Big\rfloor+1
\end{equation}
where, for any positive real $x$, we denote by $\lfloor x\rfloor$ the 
{\it integer part} of $x$,
i.e., the largest integer smaller than or equal to $x$.
Since $h$ has been chosen such that $2/h$ is not integer, see
Section~\ref{s:dinamica}, we have that $\lambda=2/h+\delta_h$ with
$\delta_h\in(0,1)$.
The set $\cc{C}$ is defined as the collection of
configurations with all the spins equal to $-1$ excepted those in a
rectangle of sides $\lambda-1$ and $\lambda$ and in a pair of
neighboring sites adjacent to one of the longer sides of the rectangle.

Given $\gamma\in\cc{C}$ we let
\begin{equation}
\label{prot-h}
\Gamma:=
E(\gamma)-
E(\muno)+2(1+h)=
-4h\lambda^2+16\lambda+4h(\lambda-2)+2(1+h)
\end{equation}
Note that by (\ref{prot-h}) and (\ref{lcritica}) it follows
\begin{equation}
\label{stimagamma}
\Gamma<8\lambda +10 -2h
\end{equation}
The simple bound above will be used in Section~\ref{s:variazionale}
to prove (\ref{stimaerre}) and in Section~\ref{s:proof}.

As has been explained in the introduction,
the energy of the configurations in $\cc{C}$ is strictly connected to
the typical exit time from the metastable state, indeed we have the 
following theorem.

\begin{theorem}
\label{t:meta}
For $h>0$ small enough and $L=L(h)$ large enough, we have that
\begin{enumerate}
\item\label{i:meta2}
for any $\varepsilon>0$
\begin{equation}
\label{meta2}
\lim_{\beta\to\infty}
\bb{P}_{\muno}(
e^{\beta\Gamma-\beta\varepsilon}
<\tau_{\puno}<
e^{\beta\Gamma+\beta\varepsilon}
)
=1
\end{equation}
\item\label{i:meta3}
\begin{equation}
\label{meta3}
\lim_{\beta\to\infty}\frac{1}{\beta}\log\bb{E}_\muno[\tau_\puno]=\Gamma
\end{equation}
\item\label{i:meta1}
\begin{equation}
\label{meta1}
\lim_{\beta\to\infty}
\bb{P}_{\muno}(\tau_{\cc{C}}<\tau_{\puno})=1
\end{equation}
\end{enumerate}
\end{theorem}
In other words, the above theorem states that the random variable
$(1/\beta)\log\tau_{\puno}$ converges in probability to $\Gamma$ as
$\beta\to\infty$ and that the logarithm of 
the mean value of $\tau_\puno$ divided 
times $\beta$ converges to $\Gamma$ in the same limit. 
Moreover, the last item ensures that, before reaching the stable state 
$\puno$, the system started at $\muno$
must necessarily visit the set of critical droplets 
$\cc{C}$.

The proof of Theorem~\ref{t:meta} will be given in
Section~\ref{s:scappo}. We note that, as usual in Probabilistic Cellular
Automata (see also \cite{[CN]}), the highest energy $\Gamma$ reached along 
the exit path is not achieved in a configuration, which is the typical 
situation in Glauber dynamics. Such a $\Gamma$ is the transition energy 
(see definition (\ref{comm-ene})) of 
the jump from the ``largest subcritical" configuration to the 
``smallest supercritical" 
one, see also the heuristic discussion in Section~\ref{s:heu}.

\subsec{Transition rate and zero temperature dynamics}{s:transition}
\par\noindent
In our problem (see also \cite{[CN]}) the energy difference
between two configurations $\sigma$ and $\eta$ is not sufficient to 
establish whether the
system prefers to jump from $\sigma$ to $\eta$ or vice versa. Indeed, for some
pairs of configurations a sort of 
barrier is seen in both directions; more precisely, it is possible to find 
$\sigma$ and $\eta$ such that both $p(\sigma,\eta)$ and $p(\eta,\sigma)$ 
tend to zero in the zero temperature limit $\beta\to\infty$.
As an example of such a behavior, consider the two 
following configurations: $\sigma$ is such that all the spins are 
equal to minus one excepted those associated with the sites belonging
to an $\ell\times\ell$ rectangle, with $3\le\ell\le L-2$, and to 
a two--site protuberance attached to one of the sides of the rectangle;
$\eta$ is a configuration obtained starting from $\sigma$ and flipping 
the spin associated with one of the sites neighboring both the 
rectangle and the protuberance.
By using (\ref{markov})--(\ref{rule2}), it is easy to show that
$p(\sigma,\eta)\sim\exp\{-2\beta(1-h)\}$ and 
$p(\eta,\sigma)\sim\exp\{-2\beta(1+h)\}$ for large $\beta$; 
see also Figure~\ref{f:tabella},
where we have reproduced
the table in \cite[FIG.~1]{[BCLS]} with the list of the single site event 
probabilities.
In that figure, the large $\beta$ behavior of the
probability, associated to the flip of the spin at the center, is
computed. 

\begin{figure}
\begin{center}
\begin{tabular}{cclccccccccl}
\begin{picture}(24,24)(0,+16)
\put(12,0){$+$}
\put(0,12){$+$}
\put(12,12){$+$}
\put(12,24){$+$}
\put(24,12){$+$}
\end{picture}
&& $e^{-2\beta(5+h)}$ &&&&&&
\begin{picture}(12,12)(0,+16)
\put(12,0){$-$}
\put(0,12){$-$}
\put(12,12){$-$}
\put(12,24){$-$}
\put(24,12){$-$}
\end{picture}
&&& $e^{-2\beta(5-h)}$
\\
&&&&&&\\
\begin{picture}(24,24)(0,+16)
\put(12,0){$-$}
\put(0,12){$+$}
\put(12,12){$+$}
\put(12,24){$+$}
\put(24,12){$+$}
\end{picture}
&& $e^{-2\beta(3+h)}$ &&&&&&
\begin{picture}(12,12)(0,+16)
\put(12,0){$+$}
\put(0,12){$-$}
\put(12,12){$-$}
\put(12,24){$-$}
\put(24,12){$-$}
\end{picture}
&&& $e^{-2\beta(3-h)}$
\\
&&&&&&\\
\begin{picture}(24,24)(0,+16)
\put(12,0){$-$}
\put(0,12){$+$}
\put(12,12){$+$}
\put(12,24){$+$}
\put(24,12){$-$}
\end{picture}
&& $e^{-2\beta(1+h)}$ &&&&&&
\begin{picture}(12,12)(0,+16)
\put(12,0){$+$}
\put(0,12){$-$}
\put(12,12){$-$}
\put(12,24){$-$}
\put(24,12){$+$}
\end{picture}
&&& $e^{-2\beta(1-h)}$
\\
&&&&&&\\
\begin{picture}(24,24)(0,+16)
\put(12,0){$-$}
\put(0,12){$+$}
\put(12,12){$+$}
\put(12,24){$-$}
\put(24,12){$-$}
\end{picture}
&& $1 - e^{-2\beta(1-h)}$ &&&&&&
\begin{picture}(12,12)(0,+16)
\put(12,0){$+$}
\put(0,12){$-$}
\put(12,12){$-$}
\put(12,24){$+$}
\put(24,12){$+$}
\end{picture}
&&& $1 - e^{-2\beta(1+h)}$
\\
&&&&&\\
\begin{picture}(24,24)(0,+16)
\put(12,0){$-$}
\put(0,12){$-$}
\put(12,12){$+$}
\put(12,24){$-$}
\put(24,12){$-$}
\end{picture}
&& $1 - e^{-2\beta(3-h)}$ &&&&&&
\begin{picture}(12,12)(0,+16)
\put(12,0){$+$}
\put(0,12){$+$}
\put(12,12){$-$}
\put(12,24){$+$}
\put(24,12){$+$}
\end{picture}
&&& $1 - e^{-2\beta(3+h)}$
\\
\end{tabular}
\end{center}
\vskip 0.5 cm
\caption{Large $\beta$ behavior of the probabilities for the flip of the
central spin for all possible configurations in the $5$--spin neighborhood.}
\label{f:tabella}
\end{figure}

To manage those barriers we associate the
{\it transition Hamiltonian} $H(\sigma,\eta)$ to each pair of configurations
$\sigma,\eta\in\cc{S}$. More precisely we extend the Hamiltonian (\ref{ham})
to the function $H:\cc{S}\cup\cc{S}\times\cc{S}\to\bb{R}$ so that
\begin{equation}
\label{comm}
H(\sigma,\eta):= H(\sigma)-\log p(\sigma,\eta)
\end{equation}
By the detailed balance principle (\ref{dett}), we have
$H(\sigma,\eta)=H(\eta,\sigma)$ for any $\sigma,\eta\in\cc{S}$.
Note that, by definition, $H(\sigma,\eta)\ge\max\{H(\sigma),H(\eta)\}$ and 
$p(\sigma,\eta)=\exp\{-[H(\sigma,\eta)-H(\sigma)]\}$; it is then reasonable
to think to $H(\sigma,\eta)$ as to the Hamiltonian level reached in the 
transition from $\sigma$ to $\eta$.
As already noted in Section~\ref{s:stat}, since the Hamiltonian 
depends on $\beta$, it is useful to compute its limiting behavior.
We then define the {\it transition energy}
\begin{equation}
\label{comm-ene}
E(\sigma,\eta):=\lim_{\beta\to\infty}\frac{1}{\beta}H(\sigma,\eta)
\end{equation}
Note that by using (\ref{hl}), (\ref{comm}), and
the symmetry of the transition hamiltonian, we get
\begin{equation}
\label{sim-ene}
E(\sigma,\eta)=E(\eta,\sigma)
\;\;\;\;\;\textrm{ and }\;\;\;\;
E(\sigma,\eta)=E(\sigma)+\Delta(\sigma,\eta)
\geq
\max\{E(\sigma),E(\eta)\}
\end{equation}
for any $\sigma,\eta\in\cc{S}$, with
\begin{equation}
\label{defdelta}
\Delta(\sigma,\eta):=
-\lim_{\beta\to\infty}
 \frac{1}{\beta}\log p(\sigma,\eta)=
\sum_{\newatop{x\in\Lambda:}{\eta(x)(S_\sigma(x)+h)<0}}2|S_\sigma(x)+h|\ge0
\end{equation}
the \emph{transition rate}; notice that 
in the second equality we have used the definition (\ref{hl}) of
$E(\sigma)$, (\ref{comm}), (\ref{markov}), and (\ref{rule}). 

The non--negative transition rate $\Delta$ will play a crucial role
in the study of the low temperature 
dynamics of the model (\ref{markov}); indeed it can 
be proven that the model satisfies the FW condition in \cite[Chapter~6]{[OV]},
that is for any $\sigma,\eta\in\cc{S}$ and 
$\beta>0$ large enough
\begin{equation}
\label{cri01}
e^{-\beta\Delta(\sigma,\eta)-\beta\gamma(\beta)}
\le
p(\sigma,\eta)
\le
e^{-\beta\Delta(\sigma,\eta)+\beta\gamma(\beta)}
\end{equation}
where $\gamma(\beta)$ does not depend
on $\sigma,\eta$ and tends to zero in the limit $\beta\to\infty$. 
From (\ref{cri01}) it follows that $p(\sigma,\eta)\to1$ for $\beta\to\infty$
if and only if $\Delta(\sigma,\eta)=0$. On the other hand, if 
$\Delta(\sigma,\eta)>0$, then $p(\sigma,\eta)\to0$ exponentially 
fast and with rate $\Delta(\sigma,\eta)$ in the limit $\beta\to\infty$,
so that $\Delta$ can be 
interpreted as the cost of the transition from $\sigma$ to $\eta$.

To get (\ref{cri01}) we first prove that 
for $\beta$ large enough 
\begin{equation}
\label{carne}
\Big|-\frac{1}{\beta}[H(\sigma,\eta)-H(\sigma)]+[E(\sigma,\eta)-E(\sigma)]\Big|
\le 
e^{-\beta(1-h)}
\end{equation}
The bound (\ref{cri01}) shall follow easily from (\ref{carne}),
(\ref{comm}), and the second equality in (\ref{sim-ene}) relating the 
transition energy to the transition rate.
To prove (\ref{carne}) we note that by using (\ref{comm}), (\ref{markov}),
(\ref{rule}), and (\ref{defdelta}) we get 
\begin{equation}
\label{accaede}
\frac{1}{\beta}[H(\sigma,\eta)-H(\sigma)]-[E(\sigma,\eta)-E(\sigma)]
=
 \frac{1}{\beta}
 \sum_{x\in\Lambda}
 \log(1+e^{-2\beta|S_\sigma(x)+h|})
\end{equation}
indeed,
\begin{displaymath}
\begin{array}{l}
{\displaystyle
\frac{1}{\beta}[H(\sigma,\eta)-H(\sigma)]-[E(\sigma,\eta)-E(\sigma)]
=
}
\\
{\displaystyle
\phantom{mer}
=\frac{1}{\beta}\sum_{x\in\Lambda}
 \log(1+e^{-2\beta\eta(x)[S_\sigma(x)+h]})
+
\!\!\!\!
\sum_{\newatop{x\in\Lambda:}{\eta(x)(S_\sigma(x)+h)<0}}
\!\!\!\!\!
     2\eta(x)[S_\sigma(x)+h]
} \\
{\displaystyle
\phantom{mer}
=
 \frac{1}{\beta}
 \!\!\!\!
 \sum_{\newatop{x\in\Lambda:}{\eta(x)(S_\sigma(x)+h)>0}}
 \!\!\!\!
 \log(1+e^{-2\beta\eta(x)[S_\sigma(x)+h]})
+\frac{1}{\beta}
 \!\!\!\!
 \sum_{\newatop{x\in\Lambda:}{\eta(x)(S_\sigma(x)+h)<0}}
 \log(1+e^{-2\beta\eta(x)[S_\sigma(x)+h]})
} \\
{\displaystyle
\phantom{mer=}
+
 \!\!\!\!
 \sum_{\newatop{x\in\Lambda:}{\eta(x)(S_\sigma(x)+h)<0}}
 \!\!\!\!
     2\eta(x)[S_\sigma(x)+h]
} \\
{\displaystyle
\phantom{mer}
=
 \frac{1}{\beta}
 \!\!\!\!
 \sum_{\newatop{x\in\Lambda:}{\eta(x)(S_\sigma(x)+h)>0}}
 \!\!\!\!
 \log(1+e^{-2\beta\eta(x)[S_\sigma(x)+h]})
+\frac{1}{\beta}
 \!\!\!\!
 \sum_{\newatop{x\in\Lambda:}{\eta(x)(S_\sigma(x)+h)<0}}
 \!\!\!\!
 \log(e^{+2\beta\eta(x)[S_\sigma(x)+h]}+1)
} \\
\end{array}
\end{displaymath}
yielding (\ref{accaede}).
The bound (\ref{carne}) follows once we note that 
$\log(1+\exp\{-2\beta|S_\sigma(x)+h|\})\ge0$ for any $x\in\Lambda$ and
$|S_\sigma(x)+h|\ge 1-h$ uniformly in $\sigma\in\cc{S}$ and $x\in\Lambda$, 
and choose $\beta\ge(\log|\Lambda|)/(1-h)$.

We finally introduce the zero temperature dynamics.
Consider a configuration $\sigma\in\cc{S}$ and $s\in\{-1,+1\}$;
since $|h|<1$, from (\ref{rule}) it follows that 
the probability $p_{x,\sigma}(s)$
tends either to $0$ or to $1$ in the limit $\beta\to\infty$.
Thus, due to the product structure of
(\ref{markov}), given $\sigma$ there exists a unique configuration
$\eta$ such that $p(\sigma,\eta)\to1$ in the limit $\beta\to\infty$. This
configuration is the one such that each spin $\eta(x)$ is chosen
so that $p_{x,\sigma}(\eta(x))\to1$ for $\beta\to\infty$.
We introduce the map $T:\cc{S}\to\cc{S}$, called the {\em zero temperature
dynamics}, which associates to each $\sigma\in\cc{S}$ the unique configuration
$T\sigma$ such that
$p(\sigma,T\sigma)\to1$ in the limit $\beta\to\infty$.

\begin{lemma}
\label{t:propdelta}
Given $\sigma,\eta\in\cc{S}$, we have that
$\Delta(\sigma,\eta)=0$ if and only if $\eta=T\sigma$.
\end{lemma}

\smallskip
\par\noindent
\textit{Proof of Lemma~\ref{t:propdelta}.\/}
The lemma follows immediately by using the definition of the zero temperature
dynamics $T$ and the remarks below (\ref{cri01}).
\qed

\begin{figure}[t]
\vskip 0.5 cm
\begin{center}
\includegraphics[height=6cm]{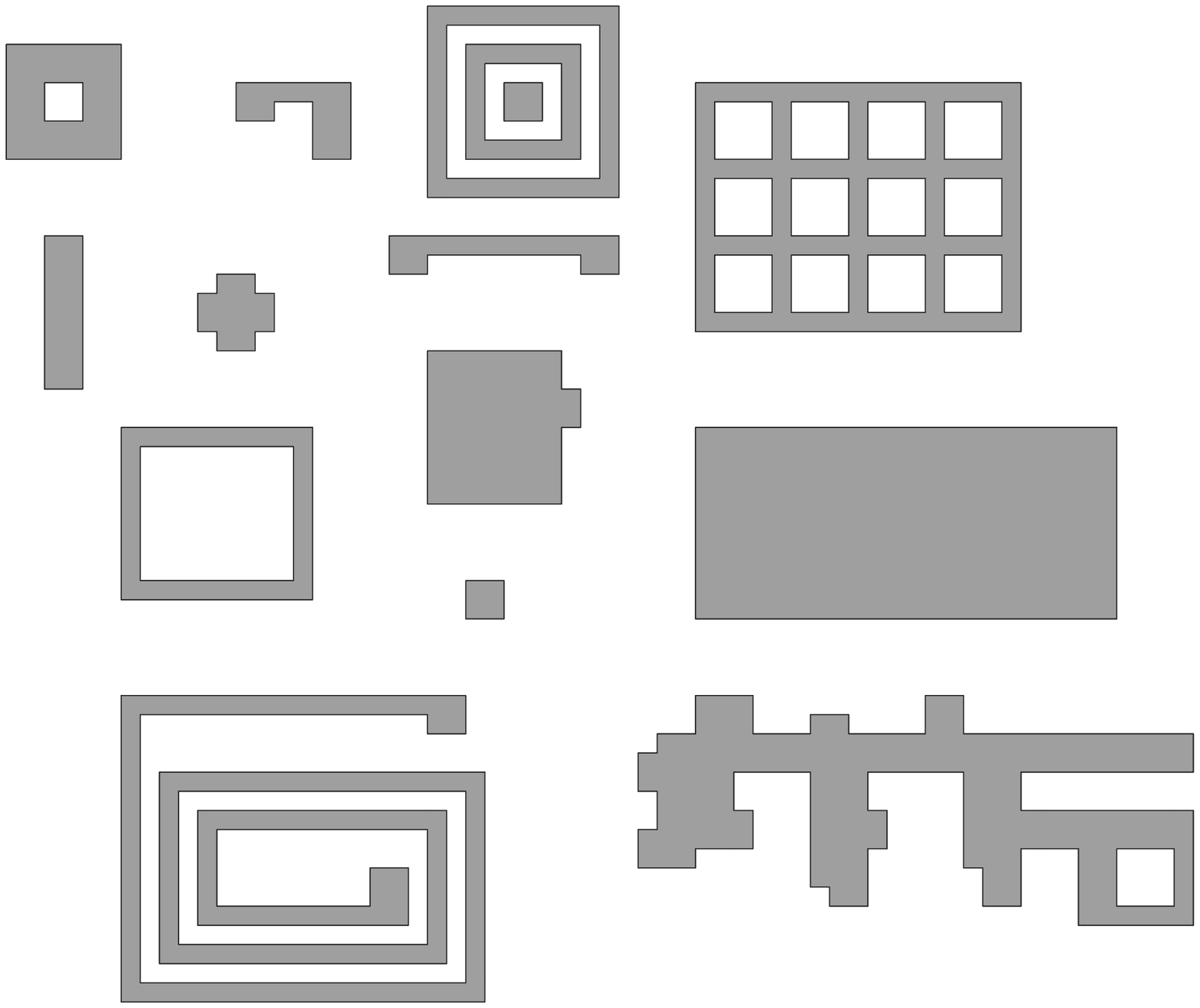}
\caption{Examples of stable states, pluses and minuses are represented 
respectively by grey and white regions.}
\label{f:stabili}
\end{center}
\end{figure}

\subsec{Stable states and stable pairs}{s:heu}
\par\noindent
The proof of Theorem~\ref{t:meta}, although mathematically complicated,
relies on a very straightforward physical argument based on a careful
description of the low temperature, i.e., large $\beta$, dynamics.
In this section we give an heuristic explanation of the
exponential estimate (\ref{meta2}) for the exit time $\tau_\puno$.

We introduce, first, the notion of stable configurations. 
If $T\sigma=\sigma$ the configuration
$\sigma$ is called {\it stable};
equivalently, we say that $\sigma\in\cc{S}$ is stable if and only if
for any $\eta\in\cc{S}\setminus\{\sigma\}$ one has
$p(\sigma,\eta)\to0$ in the limit $\beta\to\infty$.
If $\sigma$ is not stable and $T^2\sigma=\sigma$, we say that
$(\sigma,T\sigma)$ is the \textit{stable pair} associated to $\sigma$,
equivalently we say that $(\sigma,T\sigma)$ is a stable pair if and only if
$p(\sigma,T\sigma)\to1$ and $p(T\sigma,\sigma)\to1$
in the limit $\beta\to\infty$.
Recall $\Delta$ is non--negative, 
by (\ref{sim-ene}) and Lemma~\ref{t:propdelta}, it follows
\begin{equation}
\label{heavy}
E(\sigma,T\sigma)=E(\sigma)
\;\;\;\textrm{ and }\;\;\;
E(\sigma)\ge E(T\sigma)
\end{equation}
for any $\sigma\in\cc{S}$.

Note that a stable pair $(\sigma,\eta)$ is a $2$--cycle of the map $T$, 
indeed $T\sigma=\eta$ and $T\eta=\sigma$. It is easy to show that 
cycles longer than two do not exist for such a map. 
Suppose, by the way of contradiction, that 
$\sigma_1,\dots,\sigma_n\in\cc{S}$, with $n\ge3$ integer, are such that 
$\sigma_i\neq\sigma_j$ for $i\neq j$, 
$T\sigma_i=\sigma_{i+1}$ for $i=1,\dots,n-1$, and $T\sigma_n=\sigma_1$. 
By the inequality in (\ref{heavy}), it follows that
$E(\sigma_1)\ge\cdots\ge E(\sigma_n)\ge E(\sigma_1)$, which implies 
$E(\sigma_1)=\cdots=E(\sigma_n)$. This result, together with the equality
in (\ref{heavy}) and (\ref{sim-ene}), implies that 
$\Delta(\sigma_1,\sigma_2)=\Delta(\sigma_1,\sigma_n)=0$. Hence, by recalling
Lemma~\ref{t:propdelta}, we get 
$T\sigma_1=\sigma_2$ and $T\sigma_1=\sigma_n$. By definition of the map $T$,
we finally get $\sigma_n=\sigma_2$, which contradicts the hypothesis 
that $\sigma_i\neq\sigma_j$ for $i\neq j$.

\begin{figure}[t]
\vskip 0.5 cm
\begin{center}
\includegraphics[height=6cm]{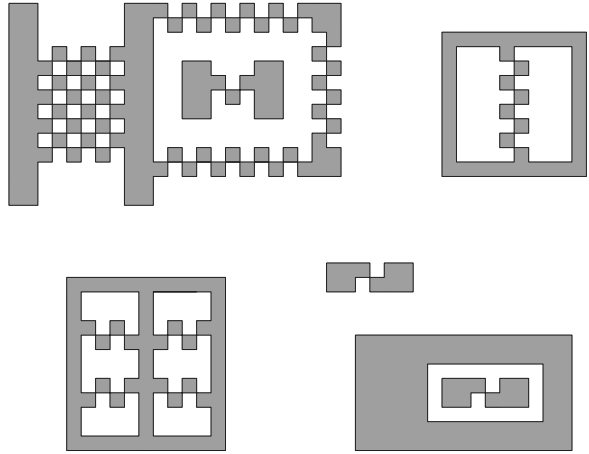}
\caption{Examples of stable pairs, pluses and minuses are represented 
respectively by grey and white regions.}
\label{f:pairs}
\end{center}
\end{figure}

As mentioned above, our model is characterized by the presence of a large 
number of stable configurations. 
Indeed, only those configurations in which there exists at least
one spin with a majority of opposite spins among its neighbors are not
stable, see Figure~\ref{f:tabella}.
All the configurations in which each spin is surrounded by at least 
two spins of the same sign are, instead, stable; some of the possible 
situations are shown in Figure~\ref{f:stabili}. 
In particular, notice that plus squared 
rings plunged into the sea of minuses are stable states. This scenario is
complicated by the presence of stable pairs; some of them are depicted 
in Figure~\ref{f:pairs}. Notice, in particular, the chessboards leaned to 
stable pluses regions. As we shall see in the sequel, the stable pairs do 
not play any important role in the study of metastability in 
model (\ref{markov}). We also recall that, in the case of a similar 
model studied in \cite{[CN]}, due to the presence of such pairs, the system 
was forced to visit an intermediate chessboard phase in its 
way from the minus metastable phase to the stable plus phase.

We describe, now, the typical low temperature behavior of the dynamics. 
Suppose that the initial condition is $\sigma_0=\sigma\in\cc{S}$; at low
temperature, with high probability, the system follows the unique 
{\it zero temperature trajectory}
\begin{displaymath}
\sigma_0=\sigma,\,
\sigma_1=T\sigma,\,
\sigma_2=T\sigma_1=T^2\sigma,\,
\dots,\,
\sigma_t=T(T^{t-1}\sigma)=T^t\sigma,
\dots
\end{displaymath}
Once the zero temperature trajectory ends up in a stable
configuration, it remains there forever.
Different trajectories are observed with probability exponentially
small in $\beta$.

We can now depict the typical behavior of the system at very low temperature.
Recall the definitions given in the last paragraph of Section~\ref{s:conf}.
Starting from $\sigma$, the system will reach in a time of order one either
the stable configuration $\ol{T}\sigma$ or the stable pair associated to
$\ol{T^2}\sigma$; note that $\ol{T}\sigma$ and $\ol{T^2}\sigma$ are
unique.
After a time exponentially large in $\beta$, the chain will depart from
the stable configuration, or from the stable pair,
and possibly reach a different stable configuration,
where it will remain for another exponentially long time. And so on.
It is then clear that, in the study of the low temperature dynamics, a key
role is played by stable configurations and stable pairs; indeed a
large amount of the time of each trajectory is spent there.

Among the large number of possible stable states, there are those
configurations in which the plus spins fill a rectangular region;
recall the definition of rectangular 
droplets given at the beginning of Section~\ref{s:meta}.
In \cite{[BCLS]} it has been conjectured that those
rectangular stable configurations are the relevant ones for
metastability. 
Moreover, there has been developed an heuristic argument to
show that $\lambda$, see (\ref{lcritica}), is the {\it critical length}
in the sense explained below. 
Rectangular droplets with smallest side length smaller or equal to
$\lambda-1$ are {\em subcritical}, namely, starting
from such a configuration the system visits $\muno$ before $\puno$ with
probability tending to one in the limit $\beta\to\infty$.
Rectangular droplets with smallest side length larger or equal to
$\lambda$ are {\em supercritical}, namely, starting
from such a configuration the system visits $\puno$ before $\muno$ 
with probability tending to one in the limit $\beta\to\infty$.

We reproduce shortly the heuristic argument in \cite[Section~IV]{[BCLS]} 
yielding the above conclusions. 
Consider a square droplet of side length $\ell$; 
we shall identify the best growth and shrinking mechanisms and, by
comparing the related typical times, get the critical length.
First note that the configuration obtained by attaching a single site 
protuberance to one of the sides of the droplet is not stable 
(see Figure~\ref{f:tabella}); 
it is needed at least a two--site protuberance to get a stable configuration.
The parallel dynamics allows the formation of a two--site protuberance
in one step; 
from Figure~\ref{f:tabella} and the product structure of (\ref{markov}),
it follows that the typical time for this process is 
$\tau_{\textrm{one}}\sim\exp\{4\beta(3-h)\}$.
On the other hand, the protuberance can be formed in two
consecutive steps: 
first a single site protuberance appears and, then, one of the two minuses 
adjacent both to the square droplet and to the protuberance is flipped. 
By using again the data in Figure~\ref{f:tabella},
we get that the typical time for the two--step process is
$\tau_{\textrm{two}}\sim\exp\{2\beta(3-h)+4\beta(1-h)\}$, where 
$2\beta(3-h)$ is the cost of the first step and 
$4\beta(1-h)$ is the sum of the costs paid in the second step 
to keep the single site protuberance and to flip the adjacent spin.
Clearly $\tau_{\textrm{two}}\ll\tau_{\textrm{one}}$ for $\beta$ large;  
hence, the most efficient mechanism to produce a two--site protuberance is the 
two--step one. 

The presence of a two--site protuberance is sufficient to ensure the 
growth of the droplet. 
Indeed, noted that $\exp\{2\beta(1-h)\}$ is the smallest typical time 
needed to leave a stable configuration (see Figure~\ref{f:tabella}), we 
have that the side with the two--site protuberance is filled by pluses via a 
sequence of $\ell-2$ flips of a minus spin with two neighboring pluses. 
Since each of those flips happens with typical time of order 
$\exp\{2\beta(1-h)\}$, we conclude that the growth time 
$\tau_{\textrm{growth}}$ is equal to $\tau_{\textrm{two}}$.

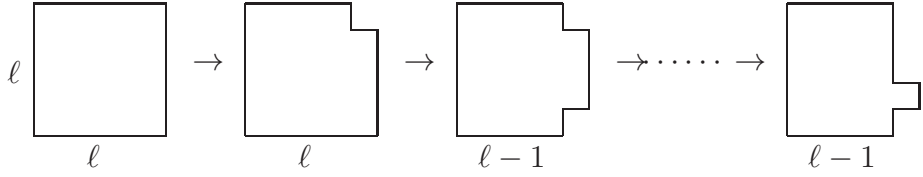
\begin{figure}[t]
\begin{center}
\begin{picture}(425,50)(-20,0)
\put(20,0){\line(1,0){50}}
\put(20,0){\line(0,1){50}}
\put(20,50){\line(1,0){50}}
\put(70,0){\line(0,1){50}}

\put(100,0){\line(1,0){50}}
\put(100,0){\line(0,1){50}}
\put(100,50){\line(1,0){40}}
\put(150,0){\line(0,1){40}}
\put(140,40){\line(1,0){10}}
\put(140,40){\line(0,1){10}}

\put(180,0){\line(1,0){40}}
\put(180,0){\line(0,1){50}}
\put(180,50){\line(1,0){40}}
\put(220,0){\line(0,1){10}}
\put(220,10){\line(1,0){10}}
\put(220,40){\line(1,0){10}}
\put(220,40){\line(0,1){10}}
\put(230,10){\line(0,1){30}}

\put(305,0){\line(1,0){40}}
\put(305,0){\line(0,1){50}}
\put(305,50){\line(1,0){40}}
\put(345,0){\line(0,1){10}}
\put(345,10){\line(1,0){10}}
\put(355,10){\line(0,1){10}}
\put(345,20){\line(1,0){10}}
\put(345,20){\line(0,1){30}}

\put(80,25){$\rightarrow$}
\put(160,25){$\rightarrow$}
\put(240,25){$\rightarrow$}
\put(250,25){$\cdots\cdots$}
\put(285,25){$\rightarrow$}
\put(10,20){$\ell$}
\put(40,-12){$\ell$}
\put(120,-12){$\ell$}
\put(188,-12){$\ell-1$}
\put(313,-12){$\ell-1$}
\end{picture}
\end{center}
\vskip 0.5 cm
\caption{Shrinking mechanism.}
\label{f:erosione}
\end{figure}

For what concerns the shrinking mechanism, it is easy to show that 
the most efficient one is the flipping of plus spins having two neighboring 
minuses (corner erosion).
The shrinking is then performed via a sequence of configurations as in 
Figure~\ref{f:erosione}, requiring the erosion of $\ell-1$ corner pluses and 
the final flipping of the unstable single site protuberance. 
Note that the intermediate configurations, joining the starting 
$\ell\times\ell$ square droplet to the ending single protuberance 
configuration, are stable; 
their lifetime, i.e., the typical time that 
must be waited for to see the system 
performing a transition, is $\exp\{2\beta(1-h)\}$ (see Figure~\ref{f:tabella}).
It follows that suitably long persistence in the $\ell-2$ intermediate stable 
configurations must be provided for in the most efficient shrinking path.  
The rate at which the entire process occurs is thus estimated as the rate for 
one erosion, $\exp\{-2\beta (1+h)\}$, times the probability that $\ell-2$ 
further erosions occur within the lifetime $\exp\{2\beta (1-h)\}$, which is of 
order $[\exp\{-2\beta (1+h)\}\exp\{2\beta (1-h)\}]^{\ell-2}$. 
We then conclude that the shrinking time is estimated as
$\tau_{\textrm{shrinking}}
  \sim\exp\{2\beta (1+h)+(\ell-2)[2\beta (1+h)-2\beta (1-h)]\}$.

By comparing, finally, 
$\tau_{\textrm{shrinking}}$ and $\tau_{\textrm{growth}}$, we get that 
growth is favored w.r.t.\ shrinking if and only if 
$\ell\ge\lfloor2/h\rfloor+1$. This remark strongly suggests that the length 
$\lambda$, defined in (\ref{lcritica}), plays the role of the critical 
length for what concerns the metastable behavior of the model.

We come, finally, to the heuristic argument suggesting the 
estimate (\ref{meta2}) for the exit time.
It is reasonable to suppose that the exit path visits an increasing  
sequence of subcritical rectangular droplets, whose side lengths 
differ at most by one. 
The highest energy along such a path wil be attained in the 
segment leading from the largest subcritical $\lambda\times(\lambda-1)$ 
droplet to the smallest supercritical $\lambda\times\lambda$ droplet. 
More precisely, denote by $\pi$ the configuration obtained by attaching a
single site protuberance to one of the two longer sides of the 
$\lambda\times(\lambda-1)$ droplet
and by $\gamma$ the configuration obtained by flipping in $\pi$ a minus 
spin adjacent to the rectangle and
neighboring the single site protuberance.
Recall the discussion above about the growth mechanism. It follows that 
the highest energy along the exit path must be attained in the transition 
from $\pi$ to $\gamma$, so that it is equal to 
$E(\pi,\gamma)$ (see (\ref{comm-ene})).
It is then reasonable to expect that the typical exit time is of order
$\exp\{\beta[E(\pi,\gamma)-E(\muno)]\}$.
Using the expression
\begin{equation}
\label{energiaret}
E(\psi)-E(\muno)=-4h\ell_1\ell_2+8(\ell_1+\ell_2)
\end{equation}
for a rectangular droplet $\psi\in\cc{S}$ of side lengths $\ell_1$ and 
$\ell_2$, recall that in such a configuration $2\le\ell_1,\ell_2\le L-2$, 
it is an easy exercise to show that
$E(\pi,\gamma)-E(\muno)=\Gamma$, see (\ref{prot-h}).

\subsec{Escape time}{s:scappo}
\par\noindent
In this section we prove Theorem~\ref{t:meta}. The main ingredients
will be the general results \cite[Theorem~4.1, 4.9, and 5.4]{[MNOS]},
the solution of the model dependent variational problem (\ref{modeest1}), 
i.e., the computation of the energy barrier between $\muno$ and $\puno$,
and the recurrence estimate (\ref{modeest2}).
In \cite{[MNOS]} the theory has been developed with quite strict
hypotheses on the dynamics, see \cite[equation~(1.3)]{[MNOS]},
nevertheless it can be shown that the same results hold in the present 
setup, see Appendix~\ref{s:mnos}.

To state the estimates on the energy landscape we need few more definitions.
A finite sequence of configurations
$\omega=\{\omega_1,\dots,\omega_n\}$ is called 
\textit{path} with starting configuration $\omega_1$ and ending
configuration $\omega_n$; we let $|\omega|:=n$.
We let $\Omega:=\cc{S}^{\bb{N}\setminus\{0\}}$ 
be the collection of all the possible paths.
Given two paths
$\omega$ and $\omega'$,
such that $\omega_{|\omega|}=\omega'_1$, we let
$\omega+\omega':=
 \{\omega_1,\dots,\omega_{|\omega|},\omega'_2,\dots,\omega'_{|\omega'|}\}$;
note that $|\omega+\omega'|=|\omega|+|\omega'|-1$.
Given a path $\omega$, we define the {\it height} along $\omega$ as
\begin{equation}
\label{pizza}
\Phi_{\omega}:=
\left\{
\begin{array}{ll}
E(\omega_1)
&\textrm{if } |\omega|=1\\
\max_{i=1,\dots,|\omega|-1} E(\omega_i,\omega_{i+1})
&\textrm{otherwise}\\
\end{array}
\right.
\end{equation}
Let $A,A'\subset\cc{S}$,
we denote by $\Theta(A,A')$ the set of all the paths $\omega\in\Omega$ 
such that $\omega_1\in A$ and $\omega_{|\omega|}\in A'$, that is the 
set of paths starting from a configuration in $A$ and 
ending in a configuration in $A'$.
The communication energy between $A,A'\subset\cc{S}$ is defined as
\begin{equation}
\label{pizza2}
\Phi(A,A'):=
\min_{\omega\in\Theta(A,A')} \Phi_\omega
\end{equation}
If $A=\{\sigma\}$, we shall 
misuse the notation by writing 
$\Theta(\sigma,A')$ instead of $\Theta(\{\sigma\},A')$ and
$\Phi(\sigma,A')$ instead of $\Phi(\{\sigma\},A')$. 

\begin{theorem}
\label{t:stime}
Recall the definition of $\Gamma$ in (\ref{prot-h}). Suppose that $h>0$
is chosen small enough. Then
\begin{enumerate}
\item\label{i:recurrence}
for any $\sigma\in\cc{S}\setminus\{\muno\}$ 
\begin{equation}
\label{modeest2}
\Phi(\sigma,\puno)-E(\sigma)<\Gamma
\end{equation}
\item\label{i:modeest}
\begin{equation}
\label{modeest1}
\Phi(\muno,\puno)-E(\muno)=\Gamma
\end{equation}
\item\label{i:modeest2}
for each path $\omega=\{\omega_1,\dots,\omega_n\}\in\Theta(\muno,\puno)$
such that $\Phi_\omega-E(\muno)=\Gamma$, there exists
$i\in\{2,\dots,n\}$ such that $\omega_i\in\cc{C}$ and
$E(\omega_{i-1},\omega_i)-E(\muno)=\Gamma$.
\end{enumerate}
\end{theorem}
Theorem~\ref{t:stime} will be proved in Section~\ref{s:moddep}.
Recall Theorems~\ref{t:mnos4.1}--\ref{t:mnos5.4} and the definitions 
given before them.

\smallskip
\par\noindent
\textit{Proof of Theorem~\ref{t:meta}.\/}
By using the results discussed at the end of Section~\ref{s:stat},
we have $\cc{S}^\rr{s}=\{\puno\}$.
We remark that, since $\cc{S}^\rr{s}=\{\puno\}$, then for any
$\sigma\in\cc{S}\setminus\cc{S}^\rr{s}$ we have $E(\puno)<E(\sigma)$;
this implies, together with (\ref{modeest1}) and (\ref{modeest2}),
that $\cc{S}^\rr{m}=\{\muno\}$ and $V_\muno=\Gamma$.
Finally, items~\ref{i:meta2} and \ref{i:meta3} follow
from Theorems~\ref{t:mnos4.1} and \ref{t:mnos4.9}, respectively.

Proof of item~\ref{i:meta1}. 
By using item~\ref{i:modeest2} in Theorem~\ref{t:stime}, we get that 
$\cc{C}$ is a gate for the transition from $\muno$ to $\puno$. 
The item~\ref{i:meta1} in Theorem~\ref{t:meta} 
follows from Theorem~\ref{t:mnos5.4}.
\qed

\sezione{The recurrence property and the variational problem}{s:moddep}
\par\noindent
In this section we prove the energy landscape estimates stated in
Theorem~\ref{t:stime}; in particular the
recurrence property (\ref{modeest2}) is proven in Section~\ref{s:recurrence}
and the variational problem (\ref{modeest1}) is solved in
Section~\ref{s:variazionale}. The proof of items~\ref{i:modeest} and 
\ref{i:modeest2} of Theorem~\ref{t:stime} 
relies on Proposition~\ref{t:insiemeG}, which is stated in 
Section~\ref{s:variazionale} and proven in Section~\ref{s:minmax}.

We give few more definitions.
Let $\sigma\in\cc{S}$ and $x\in\Lambda$, we say that the site $x$
is {\it stable} (resp.\ {\it unstable}) w.r.t.\ $\sigma$ if and only
if $\sigma(x)S_\sigma(x)>0$ (resp.\ $\sigma(x)S_\sigma(x)<0$).
Note that the stable sites are those that are not changed by the zero
temperature dynamics, more precisely $T\sigma(x)=\sigma(x)$ if and only
if $x$ is stable w.r.t.\ $\sigma$. Given $\sigma\in\cc{S}$ and
$k\in\{-5,-3,-1,+1,+3,+5\}$ we denote
by $\Lambda^{\pm}_k(\sigma)$
the collection of the sites $x\in\Lambda^{\pm}(\sigma)$ such that
$S_\sigma(x)=k$, i.e.,
\begin{equation}
\label{lambdi}
\Lambda^{\pm}_k(\sigma):=
 \{x\!\in\!\Lambda^{\pm}(\sigma)\!:S_\sigma(x)=k\}
\end{equation}
note that $\Lambda^+_{-5}(\sigma)=\emptyset$ and
$\Lambda^-_{+5}(\sigma)=\emptyset$; moreover, we set
\begin{equation}
\label{lambdi-mu}
\Lambda^{\pm}_{\le k}(\sigma):=
\Lambda^{\pm}_{-5}(\sigma)\cup\cdots\cup\Lambda^{\pm}_k(\sigma)
\;\;\;\textrm{ and }\;\;\;
\Lambda^{\pm}_{\ge k}(\sigma):=
\Lambda^{\pm}_k(\sigma)\cup\cdots\cup\Lambda^{\pm}_{+5}(\sigma)
\end{equation}
Finally, given $\sigma\in\cc{S}$, we denote
by $\Lambda^+_\rr{s}(\sigma)$ (resp.\ $\Lambda^+_\rr{u}(\sigma)$)
the collection of the sites $x\in\Lambda$ such that $\sigma(x)=+1$ and
$x$ is stable (resp.\ unstable) w..r.t.\ $\sigma$; similarly we define
$\Lambda^-_\rr{s}(\sigma)$ and $\Lambda^-_\rr{u}(\sigma)$.
By definition of stable and unstable sites we get that, for
any $\sigma\in\cc{S}$,
\begin{equation}
\label{lambdi-us}
\Lambda^+_\rr{u}(\sigma)=\Lambda^+_{\le-1}(\sigma),\;
\Lambda^-_\rr{u}(\sigma)=\Lambda^-_{\ge+1}(\sigma),\;
\Lambda^+_\rr{s}(\sigma)=\Lambda^+_{\ge+1}(\sigma),
\textrm{ and }
\Lambda^-_\rr{s}(\sigma)=\Lambda^-_{\le-1}(\sigma)
\end{equation}

\subsec{The recurrence property}{s:recurrence}
\par\noindent
Equation (\ref{modeest2}) in Theorem~\ref{t:stime}
states that, for any configuration $\sigma$
different from the metastable state $\muno$, it is possible to exhibit a path
$\omega$ joining $\sigma$ to the stable state $\puno$, i.e., to the absolute
minimum of the energy, such that $\Phi_\omega<E(\sigma)+\Gamma$.
On the heuristic ground, given $\sigma\in\cc{S}\setminus\{\muno\}$,
there exists
at least a plus spin; starting from such a plus it is possible to build a
supercritical $\lambda\times\lambda$ droplet of pluses paying an energy cost
strictly smaller than $E(\sigma)+\Gamma$. Indeed, by virtue of (\ref{modeest1}),
starting from $\muno$, the cost would be exactly $\Gamma$. On the other hand,
starting from $\sigma$, no energy must be paid to get the first plus spin and
the other pluses of $\sigma$, if any, help the production of the supercritical
droplet.

A rigorous proof needs the explicit construction of the path;
such a path will firstly realize the growth
of a supercritical $\lambda\times\lambda$ square with $\sigma$ as a background
and then its growth towards $\puno$. More precisely,
recall $\Lambda$ is a squared torus,
let $L$ be its side length and $0=(0,0)$ the origin;
recall the zero temperature dynamics mapping $T$ defined in
Section~\ref{s:transition} and let $\sigma\in\cc{S}$ be such that
$\sigma(x)=+1$ for any $x\in Q_{2,2}(0)$.
We define the path
\begin{equation}
\label{proricorrenza-1}
\Omega_\sigma:=
\Xi^2+\sum_{n=3}^L[\Psi^n+\Xi^n]
\end{equation}
where the paths $\Xi^n$, with $n=2,\dots,L$, and 
$\Psi^n$, with $n=3,\dots,L$,
are constructed algorithmically. 

We first describe informally the algorithms.
The path $\Xi^n$ starts 
from the configuration $\xi^n$ and ends in the configuration $\psi^{n+1}$.
The configuration 
$\xi^n$ is such that the square 
$Q_{n,n}(0)=\{0,\dots,n-1\}\times\{0,\dots,n-1\}$ is filled with pluses;
the path $\Xi^n$ fills with pluses 
the slice $Q_{1,n}(n,0)=\{(n,0),\dots,(n,n-1)\}$, adjacent to the square 
$Q_{n,n}(0)$, and produce 
$\psi^{n+1}$ in which the rectangle $Q_{n+1,n}(0)$ is filled with pluses.
Similarly, 
the path $\Psi^n$ starts 
from the configuration $\psi^n$ and ends in the configuration $\xi^n$.
The configuration 
$\psi^n$ is such that the rectangle 
$Q_{n,n-1}(0)=\{0,\dots,n-1\}\times\{0,\dots,n-2\}$ is filled with pluses;
the path $\Psi^n$ fills with pluses 
the slice $Q_{n,1}(0,n-1)=\{(0,n-1),\dots,(n-1,n-1)\}$, adjacent to the 
rectangle $Q_{n,n-1}(0)$, and produce 
$\xi^n$ in which the square $Q_{n,n}(0)$ is filled with pluses.

Definition of $\Xi^n$.
Let $\xi^2:=\sigma$, let $n\in\{2,\dots,L-1\}$, and suppose $\xi^n$ is
such that $\xi^n(x)=+1$ for $x\in Q_{n,n}(0)$, then
\par\noindent
\begin{enumerate}
\item\label{i:alg01}
 \texttt{set} $i=1$, $\xi^n_i=\xi^n$;
\item\label{i:alg02}
 \texttt{if} $T^2\xi^n_i=\xi^n_i$
 \texttt{then goto}~\ref{i:alg03}
 \texttt{else set} $i=i+1$ \texttt{and} $\xi^n_i=T\xi^n_{i-1}$
 \texttt{and goto}~\ref{i:alg02};
\item\label{i:alg03}
 \texttt{if} $\xi^n_i(x)=+1$ for all $x\in Q_{1,n}(n,0)$
 \texttt{then set} $\psi^{n+1}=\xi^n_i$ \texttt{and goto}~\ref{i:alg07};
 \item\label{i:alg04}
 \texttt{if} $Q_{1,n}(n,0)\cap\Lambda^+_\rr{s}(\xi^n_i)\neq\emptyset$,
 \texttt{then pick} $y,y'\in Q_{1,n}(n,0)$
         such that $\dis(y,y')=1$, $\xi^n_i(y)=-1$,
             and $y'\in\Lambda^+_\rr{s}(\xi^n_i)$,
 \texttt{set} $i=i+1$, $\xi^n_i(y)=+1$, $\xi^n_i(x)=T\xi^n_{i-1}(x)$
 $\forall x\in\Lambda\setminus\{y\}$
 \texttt{and goto}~\ref{i:alg03};
\item\label{i:alg05}
 \texttt{if} $Q_{1,n}(n,0)\cap\Lambda^+_\rr{u}(\xi^n_i)\neq\emptyset$,
 \texttt{then pick} $y,y'\in Q_{1,n}(n,0)$
         such that $\dis(y,y')=1$, $\xi^n_i(y)=-1$,
             and $y'\in\Lambda^+_\rr{u}(\xi^n_i)$,
 \texttt{set} $i=i+1$, $\xi^n_i(y)=+1$, $\xi^n_i(y')=+1$,
 $\xi^n_i(x)=T\xi^n_{i-1}(x)$ for any $x\in\Lambda\setminus\{y,y'\}$
 \texttt{and goto}~\ref{i:alg03};
\item\label{i:alg06}
 \texttt{set} $i=i+1$,
 $y=(n,0)$,
 $\xi^n_i(y)=+1$,
 $\xi^n_i(x)=T\xi^n_{i-1}(x)$ for any $x\in\Lambda\setminus\{y\}$
 \texttt{and goto}~\ref{i:alg03};
\item\label{i:alg07}
 \texttt{set} $h_n=i$, $\Xi^n=\{\xi^n_1,\dots,\xi^n_{h_n}\}$
 \texttt{and exit}.
\end{enumerate}

At step~\ref{i:alg02} the algorithm follows the zero temperature dynamics 
down to the stable pair or to the stable
state associated to $\xi^n$.
At step~\ref{i:alg03} the algorithm checks if the slice 
$Q_{1,n}(n,0)$ 
adjacent to the square $Q_{n,n}(0)$ is filled with pluses. 
In case of positive answer the algorithm jumps to step~\ref{i:alg07} and 
exits. If the answer is negative, then the slice is filled with pluses at 
steps~\ref{i:alg03}--\ref{i:alg06} as follows: 
a minus adjacent to a stable plus is flipped (step~\ref{i:alg04}); 
in absence of stable pluses, a minus adjacent to an unstable plus is flipped 
(step~\ref{i:alg05}). If the slice is filled with minuses, then the 
spin associated to the site $(n,0)$ is flipped (step~\ref{i:alg06}).

Definition of $\Psi^n$.
Let $n\in\{3,\dots,L\}$ and suppose $\psi^n$ is such that
$\psi^n(x)=+1$ for $x\in Q_{n,n-1}(0)$, then
\par\noindent
\begin{enumerate}
\item\label{i:palg01}
 \texttt{set} $i=1$, $\psi^n_i=\psi^n$;
\item\label{i:palg02}
 \texttt{if} $T^2\psi^n_i=\psi^n_i$
\texttt{then goto}~\ref{i:palg03}
 \texttt{else set} $i=i+1$ \texttt{and} $\psi^n_i=T\psi^n_{i-1}$
 \texttt{and goto}~\ref{i:palg02};
\item\label{i:palg03}
 \texttt{if} $\psi^n_i(x)=+1$ for all $x\in Q_{n,1}(0,n-1)$
 \texttt{then set} $\xi^n=\psi^n_i$ \texttt{and goto}~\ref{i:palg07};
\item\label{i:palg04}
 \texttt{if} $Q_{n,1}(0,n-1)\cap\Lambda^+_\rr{s}(\psi^n_i)\neq\emptyset$,
 \texttt{then pick} $y,y'\in Q_{n,1}(0,n-1)$
         such that $\dis(y,y')=1$, $\psi^n_i(y)=-1$,
             and $y'\in\Lambda^+_\rr{s}(\psi^n_i)$,
 \texttt{set} $i=i+1$, $\psi^n_i(y)=+1$, $\psi^n_i(x)=T\psi^n_{i-1}(x)$
 $\forall x\in\Lambda\setminus\{y\}$
 \texttt{and goto}~\ref{i:palg03};
\item\label{i:palg05}
 \texttt{if} $Q_{n,1}(0,n-1)\cap\Lambda^+_\rr{u}(\psi^n_i)\neq\emptyset$,
 \texttt{then pick} $y,y'\in Q_{n,1}(0,n-1)$
         such that $\dis(y,y')=1$, $\psi^n_i(y)=-1$,
             and $y'\in\Lambda^+_\rr{u}(\psi^n_i)$,
 \texttt{set} $i=i+1$, $\psi^n_i(y)=+1$, $\psi^n_i(y')=+1$,
 $\psi^n_i(x)=T\psi^n_{i-1}(x)$ for any $x\in\Lambda\setminus\{y,y'\}$
 \texttt{and goto}~\ref{i:palg03}
\item\label{i:palg06}
 \texttt{set} $i=i+1$,
 $y=(0,n)$,
 $\psi^n_i(y)=+1$,
 $\psi^n_i(x)=T\psi^n_{i-1}(x)$ for any $x\in\Lambda\setminus\{y\}$
 \texttt{and goto}~\ref{i:palg03};
\item\label{i:palg07}
 \texttt{set} $k_n=i$, $\Psi^n=\{\psi^n_1,\dots,\psi^n_{k_n}\}$
 \texttt{and exit}.
\end{enumerate}
\par\noindent

In the following lemma we summarize the main properties of the 
paths $\Xi^n$ and $\Psi^n$ defined above. In particular in 
items~\ref{i:cammino-2} and \ref{i:cammino-3} we state upper bounds 
on the energy levels reached by those paths. We show that the addition of a
slice of pluses to a rectangle of pluses can result in a net increment of the 
energy only if the length of the added slice does not exceed the 
critical length $\lambda$ (see (\ref{lcritica}), (\ref{cammino-2}), and 
(\ref{cammino-3})).

\begin{lemma}
\label{t:camminotilde}
Let $\sigma\in\cc{S}$ be such that $\sigma(x)=+1$ for any $x\in Q_{2,2}(0)$,
consider the path $\Omega_\sigma$ defined by (\ref{proricorrenza-1}). Then
\begin{enumerate}
\item\label{i:cammino-1}
for any $n=3,\dots,L$ the configuration $\psi^n$ is such that
$\psi^n(x)=+1$ for all $x\in Q_{n,n-1}(0)$,
for any $n=3,\dots,L$ the configuration $\xi^n$ is such that
$\xi^n(x)=+1$ for all $x\in Q_{n,n}(0),$
in particular $\xi^L=\puno$ and $\Xi^L=\{\xi^L\}$;
\item\label{i:cammino-2}
for any $n=2,\dots,L$ we have
\begin{equation}
\label{cammino-2}
E(\psi^{n+1})-E(\xi^n)\le(8-4hn)\vee0
\;\;\;\textrm{ and }\;\;\;
\Phi_{\Xi^n}\le E(\xi^n)+10-6h
\end{equation}
\item\label{i:cammino-3}
for any $n=3,\dots,L$ we have
\begin{equation}
\label{cammino-3}
E(\xi^n)-E(\psi^n)\le(8-4hn)\vee0
\;\;\;\textrm{ and }\;\;\;
\Phi_{\Psi^n}\le E(\psi^n)+10-6h
\end{equation}
\item\label{i:cammino-4}
we have
\begin{equation}
\label{cammino-4}
\Phi_{\Omega_\sigma}-E(\sigma)\le\Gamma-16(2-h)
\end{equation}
where we recall $\Gamma$ has been defined in (\ref{prot-h}).
\end{enumerate}
\end{lemma}

\smallskip
\par\noindent
\textit{Proof of Lemma~\ref{t:camminotilde}.\/}
Item~\ref{i:cammino-1} is an immediate consequence of the
algorithmic definition of $\Omega_\sigma$.
The proof of item~\ref{i:cammino-2} is similar to the proof of
item~\ref{i:cammino-3}.

Item~\ref{i:cammino-3}.\
Let $\Psi^n:=\{\psi^n_1,\dots,\psi^n_k,\dots,\psi^n_{k_n}\}$, with
$k_n\ge k\ge1$, such that $\psi^n_i=T\psi^n_{i-1}$ for $i=2,\dots,k$ and
$\psi^n_k=T^2\psi^n_k$; note that by construction
$\psi^n_1=\psi^n$, $\psi^n_{k_n}=\xi^n$, and $k_n-k\le n$.
By using (\ref{heavy}), we get
\begin{equation}
\label{cozza1}
\Phi_{\{\psi^n_1,\dots,\psi^n_k\}}=E(\psi^n_1)
\;\;\;\textrm{ and }\;\;\;
E(\psi^n_i)\ge E(\psi^n_{i+1})
\end{equation}
for $i=1\dots,k-1$. If $k_n=k$, then (\ref{cammino-3}) follows
immediately from (\ref{cozza1}). In the case $k_n\ge k+1$, we shall
prove that
\begin{equation}
\label{cozza2}
\Phi_{\{\psi^n_k,\psi^n_{k+1},\dots,\psi^n_{k_n}\}}
\le E(\psi^n_k)+10-6h
\;\;\;\textrm{ and }\;\;\;
E(\psi^n_{k_n})-E(\psi^n_k)\le(8-4hn)\vee0
\end{equation}
The bounds (\ref{cammino-3}) will then follow from (\ref{cozza1}) and
(\ref{cozza2}).

We are then left with the proof of (\ref{cozza2}), which can be achieved by
discussing the following three cases.

\par\noindent
Case 1.\ There exist $y,y'\in Q_{n,1}(0,n-1)$ such that
$\psi^n_k(y)=-1$, $y'\in\Lambda^+_\rr{s}(\psi^n_k)$.
The configuration $\psi^n_{k+1}$ is defined at the step~\ref{i:palg04}
of the algorithm; it is immediate to see that all the configurations
$\psi^n_{i}$, with $i=k+1,\dots,k_n$, are indeed defined at the
step~\ref{i:palg04}.
Then, by using (\ref{comm-ene}), see also
figure~\ref{f:tabella}, we get the following bounds on the transition
energies:
\begin{equation}
\label{cozza3}
E(\psi^n_i,\psi^n_{i+1})\le E(\psi^n_i)+2(1-h)
\;\textrm{ and }\;
E(\psi^n_{i+1},\psi^n_i)\ge E(\psi^n_{i+1})+2(1+h)
\end{equation}
for any $i=k,\dots,k_n-1$.
By using (\ref{cozza3}), (\ref{pizza}), and (\ref{sim-ene})
we get
\begin{equation}
\label{cozza4}
\Phi_{\{\psi^n_k,\psi^n_{k+1},\dots,\psi^n_{k_n}\}}
\le E(\psi^n_k)+2(1-h)
\;\textrm{ and }\;
E(\psi^n_{k_n})-E(\psi^n_k)\le-4h(k_n-k)
\end{equation}
which, recalling $k_n\ge k+1$, imply (\ref{cozza2}).

\par\noindent
Case 2.\ There exist $y,y'\in Q_{n,1}(0,n-1)$ such that
$\psi^n_k(y)=-1$, $\psi^n_k(y')=+1$, and
$\Lambda^+_\rr{s}(\psi^n_k)\cap Q_{n,1}(0,n-1)=\emptyset$.
The configuration $\psi^n_{k+1}$ is defined at the step~\ref{i:palg05}
of the algorithm; it is immediate to remark that all the configurations
$\psi^n_{i}$, with $i=k+1,\dots,k_n$, are instead defined at the
step~\ref{i:palg04}.

Let $i=k,\dots,k_n-1$, let $y,y'\in Q_{n,1}(0,n-1)$ be the two sites
which are picked up by the algorithm, let $\Delta_i$ be the collection of the
sites in $Q_{n,1}(0,n-1)$ different from $y,y'$ and such that they
become stable plus sites at this step of the path; more precisely, let
$\Delta_i:=\Lambda^+_\rr{s}(\psi^n_{i+1})
      \setminus(\Lambda^+_\rr{s}(\psi^n_i)\cup\{y,y'\})$.
Note that the update of the sites in $\Delta_i$ has no energy cost since
they follow the zero temperature dynamics $T$.

By using (\ref{comm-ene}), see also figure~\ref{f:tabella}, we get
the estimates
\begin{equation}
\label{cozza6}
\begin{array}{lcllcl}

E(\psi^n_k,\psi^n_{k+1})
&\!\!\!\le\!\!\!&
E(\psi^n_k)+4(1-h)
&
E(\psi^n_{k+1},\psi^n_k)
&\!\!\!\ge\!\!\!&
E(\psi^n_{k+1})+2(1+h)(1+|\Delta_k|)
\\
E(\psi^n_i,\psi^n_{i+1})
&\!\!\!\le\!\!\!&
E(\psi^n_i)+2(1-h)
&
E(\psi^n_{i+1},\psi^n_i)
&\!\!\!\ge\!\!\!&
E(\psi^n_{i+1})+2(1+h)(1+|\Delta_i|)
\\
\end{array}
\end{equation}
for any $i=k+1,\dots,k_n-1$.
If $|\Delta_i|=0$ for any $i=k,\dots,k_n-1$, then it must necessarily
be $k_n-k=n-1$. We get
\begin{equation}
\label{cozza9}
\Phi_{\{\psi^n_k,\psi^n_{k+1},\dots,\psi^n_{k_n}\}}
\le E(\psi^n_k)+4(1-h)
\;\;\;\textrm{ and }\;\;\;
E(\psi^n_{k_n})-E(\psi^n_k)\le2-2h-4h(n-1)
\end{equation}
Noted that $8-4hn=2-2h-4h(n-1)+(6-2h)$, 
the bound (\ref{cozza2}) follows since $h\le3$.
Suppose, finally, that there exists
$i\in\{k,\dots,k_n-1\}$ such that $|\Delta_i|\neq0$; hence
\begin{equation}
\label{cozza11}
\Phi_{\{\psi^n_k,\psi^n_{k+1},\dots,\psi^n_{k_n}\}}
\le E(\psi^n_k)+4(1-h)
\;\;\;\textrm{ and }\;\;\;
E(\psi^n_{k_n})-E(\psi^n_k)\le -4h(k_n-k+1)
\end{equation}
Recall $k_n\ge k+1$; the bounds (\ref{cozza11}) imply (\ref{cozza2}) trivially.

\par
\begin{figure}[t]
\begin{picture}(100,160)(10,0)
\qbezier(130,10)(140,45)(150,80)
\put(130,10){\circle*{3}}
\put(130,3){\scriptsize{$\psi^n_k$}}
\qbezier(150,80)(180,185)(190,100)
\put(150,80){\circle*{3}}
\put(150,73){\scriptsize{$\psi^n_{k+1}$}}
\qbezier(190,100)(210,160)(220,90)
\put(190,100){\circle*{3}}
\put(188,93){\scriptsize{$\psi^n_{k+2}$}}
\qbezier(220,90)(240,150)(250,80)
\put(220,90){\circle*{3}}
\put(218,83){\scriptsize{$\psi^n_{k+3}$}}
\put(250,80){\circle*{3}}
\put(248,73){\scriptsize{$\psi^n_{k+4}$}}
\qbezier[20](260,95)(282.5,87.5)(305,80)
\qbezier(310,60)(330,120)(340,50)
\put(310,60){\circle*{3}}
\put(308,53){\scriptsize{$\psi^n_{k_n-1}$}}
\put(340,50){\circle*{3}}
\put(338,43){\scriptsize{$\psi^n_{k_n}$}}
\put(105,50){\scriptsize{$2(3-h)$}}
\put(115,110){\scriptsize{$2\cdot2(1-h)$}}
\put(200,20){\scriptsize{$2(1-h)$}}
\qbezier[70](235,22.5)(380,50)(320,70)
\put(322,69){\vector(-3,1){5}}
\qbezier[50](225,30)(250,80)(230,100)
\put(233,97){\vector(-1,1){5}}
\qbezier[50](205,30)(220,90)(200,110)
\put(203,107){\vector(-1,1){5}}
\put(250,150){\scriptsize{$2(1+h)$}}
\qbezier[40](250,145)(220,137.5)(190,130)
\put(190,130){\vector(-4,-1){5}}
\qbezier[30](260,145)(240,130)(220,115)
\put(222,116){\vector(-4,-3){5}}
\qbezier[30](270,145)(260,125)(250,105)
\put(250,105){\vector(-1,-2){1}}
\qbezier[50](280,145)(380,110)(340,70)
\put(340,70){\vector(-1,-1){1}}
\end{picture}
\caption{Graphical representation of the estimates (\ref{cozza5}).}
\label{f:pathdimezzo}
\end{figure}
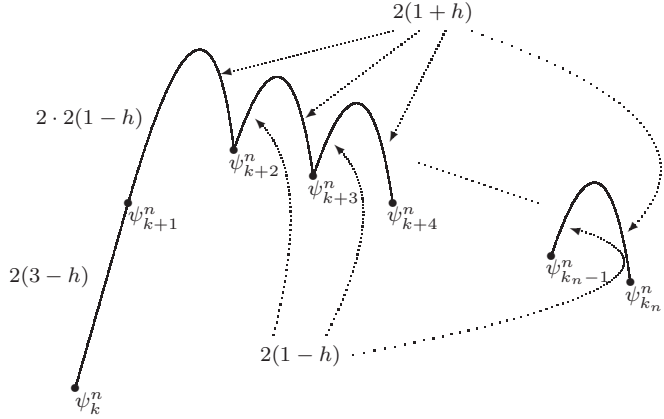

\noindent
Case 3.\
For each $y\in Q_{n,1}(0,n-1)$ we have $\psi^n_k(y)=-1$.
In this case $k_n-k=n$,
$\psi^n_{k+1}$ is defined at the step~\ref{i:palg06},
$\psi^n_{k+2}$ is defined either at the step~\ref{i:palg04} or at the
step~\ref{i:palg05}, and
$\psi^n_{k+i}$, with $i=3,\dots,k_n$, are defined at the step~\ref{i:palg04}
of the algorithm. By using (\ref{comm-ene}), see also
figure~\ref{f:tabella}, we get
\begin{equation}
\label{cozza5}
\begin{array}{lcllcl}
E(\psi^n_k,\psi^n_{k+1})
&\!\!\!\le\!\!\!&
E(\psi^n_k)+2(3-h)
&
E(\psi^n_{k+1},\psi^n_k)
&\!\!\!\ge\!\!\!&
E(\psi^n_{k+1})
\\
E(\psi^n_{k+1},\psi^n_{k+2})
&\!\!\!\le\!\!\!&
E(\psi^n_{k+1})+4(1-h)
&
E(\psi^n_{k+2},\psi^n_{k+1})
&\!\!\!\ge\!\!\!&
E(\psi^n_{k+2})+2(1+h)
\\
E(\psi^n_i,\psi^n_{i+1})
&\!\!\!\le\!\!\!&
E(\psi^n_i)+2(1-h)
&
E(\psi^n_{i+1},\psi^n_i)
&\!\!\!\ge\!\!\!&
E(\psi^n_{i+1})+2(1+h)
\end{array}
\end{equation}
for $i=k+2,\dots,k_n-1$; see the Figure~\ref{f:pathdimezzo} for a
graphical representation of the estimates (\ref{cozza5}).
Note that the equalities hold, for instance, in the case
$\psi^n_k(x)=-1$ for any $x\in\partial Q_{n,1}(0,n-1)\setminus Q_{n,n-1}(0)$.
By using (\ref{cozza5}), (\ref{pizza}), and (\ref{sim-ene})
we get
\begin{equation}
\label{cozza4-1}
\Phi_{\{\psi^n_k,\psi^n_{k+1},\dots,\psi^n_{k_n}\}}
\le E(\psi^n_k)+10-6h
\;\textrm{ and }\;
E(\psi^n_{k_n})-E(\psi^n_k)\le[8-4h(k_n-k)]=8-4hn
\end{equation}
which imply (\ref{cozza2}).

We remark that in this case~3 the path $\{\psi^n_k,\dots,\psi^n_{k_n}\}$
realizes the standard growth of the rectangular plus droplet $\psi^n$ up to the
square droplet $\psi^n_{k_n}$ via the formation of a unit plus protuberance in
the slice adjacent to one of the longer sides of the rectangle and the
bootstrap percolation plus filling of the same slice.

Item~\ref{i:cammino-4}.\  Let $\eta,\eta'$ two consecutive configurations of
the path $\Omega_\sigma$, we shall prove that
\begin{equation}
\label{pro-cammino-1}
E(\eta,\eta')-E(\sigma)\le\Gamma-16(2-h)
\end{equation}
The bound (\ref{cammino-4}) will then follow, see (\ref{pizza}).
Recall the critical length $\lambda$ has
been defined in (\ref{lcritica}) and consider the following four cases.

\par\noindent
Case 1.\
The configurations $\eta,\eta'$ belong to $\Xi^n$ for some $n\le\lambda-1$.
This case is similar to the case~2.

\par\noindent
Case 2.\
The configurations $\eta,\eta'$ belong to $\Psi^n$ for some $n\le\lambda$.
By using (\ref{pizza}), (\ref{cammino-2}), and (\ref{cammino-3}) we have
\begin{displaymath}
\begin{array}{rl}
E(\eta,\eta')\!\!\!
&
{\displaystyle
\le
\Phi_{\Psi^n}\le E(\psi^n)+10-6h
\phantom{\sum_{i-3}^{\lambda-1}}
}
\\
&
{\displaystyle
\le
E(\psi^n)-E(\xi^{n-1})+E(\xi^{n-1})-\cdots
  -E(\psi^3)+E(\psi^3)-E(\xi^2)+E(\xi^2)
}
\\
&
{\displaystyle
\phantom{\le}
  +10-6h
}
\\
&
{\displaystyle
\le
E(\sigma)+18-14h+8\sum_{i=3}^{n-1}[2-hi]
\le
E(\sigma)+18-14h+8\sum_{i=3}^{\lambda-1}[2-hi]
}
\end{array}
\end{displaymath}
where we have used that $2-hi>0$ for $i\le\lambda-1$ and $\xi^2=\sigma$.
The bound (\ref{pro-cammino-1}) follows easily.

\par\noindent
Case 3.\
The configurations $\eta,\eta'$ belong to $\Xi^n$ for some $n\ge\lambda$.
Note that for $n\ge\lambda$ the bounds (\ref{cammino-2}) and (\ref{cammino-3})
on the differences of energy become trivial since $8-4hn<0$, hence
$E(\xi^n)\le E(\psi^\lambda)$. Then
\begin{displaymath}
E(\eta,\eta')\le\Phi_{\Xi^n}
\le E(\xi^n)+10-6h\le E(\psi^\lambda)+10-6h
\end{displaymath}
where in the first inequality we used (\ref{pizza}),
in the second the bound (\ref{cammino-2}), and in the last the fact that
$E(\xi^n)\le E(\psi^\lambda)$.
To get (\ref{pro-cammino-1}) we then perform the same computation as in the
case~2.

\par\noindent
Case 4.\
The configurations $\eta,\eta'$ belong to $\Psi^n$ for some $n\ge\lambda+1$.
This case is similar to the case~3.
\qed

\smallskip
\par\noindent
\textit{Proof of item~\ref{i:recurrence} of Theorem~\ref{t:stime}.\/}
Let $\sigma\in\cc{S}\setminus\{\muno\}$. If $\sigma=\puno$ the
statement of the lemma is trivial; we then suppose $\sigma\neq\puno$.
Since by hypothesis $\sigma\neq\muno$, there exists
$x\in\Lambda$ such that $\sigma(x)=+1$; without loss of generality we suppose
$\sigma(0)=+1$.
Consider the path $\omega:=\{\sigma,\sigma^1,\sigma^2,\sigma^3\}$
with
\begin{itemize}
\item[--]
$\sigma^1$ is such that $\sigma^1(x)=+1$ for all $x\in Q_{2,1}(0)$ and
$\sigma^1(x)=T\sigma(x)$ for all $x\in\Lambda\setminus Q_{2,1}(0)$;
\item[--]
$\sigma^2$ is such that $\sigma^2(x)=+1$ for all
$x\in Q_{2,1}(0)\cup Q_1(0,1)$ and $\sigma^2(x)=T\sigma^1(x)$ for all
$x\in\Lambda\setminus[Q_{2,1}(0)\cup Q_1(0,1)]$;
\item[--]
$\sigma^3$ is such that $\sigma^3(x)=+1$ for all $x\in Q_{2,2}(0)$ and
$\sigma^3(x)=T\sigma^2(x)$ for all $x\in\Lambda\setminus Q_{2,2}(0)$.
\end{itemize}
By definition the path $\omega+\Omega_{\sigma^3}$ starts at $\sigma$ and ends
in $\puno$, i.e., $\omega+\Omega_{\sigma^3}\in\Theta(\sigma,\puno)$,
moreover we shall prove that
\begin{equation}
\label{proricorrenza-3}
\Phi_{\omega+\Omega_{\sigma^3}}<E(\sigma)+\Gamma
\end{equation}
The item~\ref{i:recurrence} of Theorem~\ref{t:stime} will then follow.

To prove (\ref{proricorrenza-3}) we first consider the path
$\omega$;
by using (\ref{comm-ene}), see also figure~\ref{f:tabella}, we get
the following bounds on the transition energies:
\begin{equation}
\label{proricorrenza-4}
\begin{array}{lcllcl}
E(\sigma,\sigma^1)&\!\!\!\le\!\!\!&E(\sigma)+2\cdot2(3-h)
&
E(\sigma^1,\sigma)&\!\!\!\ge\!\!\!&E(\sigma^1)
\\
E(\sigma^1,\sigma^2)&\!\!\!\le
                \!\!\!&E(\sigma^1)+2\cdot2(1-h)+2(3-h)
&
E(\sigma^2,\sigma^1)&\!\!\!\ge\!\!\!&E(\sigma^2)
\\
E(\sigma^2,\sigma^3)&\!\!\!\le\!\!\!&E(\sigma^2)+3\cdot2(1-h)
&
E(\sigma^3,\sigma^2)&\!\!\!\ge\!\!\!&E(\sigma^3)+2(1+h)
\\
\end{array}
\end{equation}
By using (\ref{proricorrenza-4}), (\ref{pizza}), (\ref{sim-ene}),
(\ref{prot-h}), and the definition (\ref{lcritica}) of the critical
length $\lambda$, it is easy to show that
\begin{equation}
\label{proricorrenza-6}
\Phi_\omega-E(\sigma)\le28-16h<\Gamma
\end{equation}
and
\begin{equation}
\label{proricorrenza-7}
E(\sigma^3)-E(\sigma)\le26-18h
\end{equation}
We consider, now, the path $\Omega_{\sigma^3}$;
by using (\ref{cammino-4}) and (\ref{proricorrenza-7}), we get
\begin{equation}
\label{proricorrenza-8}
\Phi_{\Omega_{\sigma^3}}-E(\sigma)=
\Phi_{\Omega_{\sigma^3}}-E(\sigma^3)+E(\sigma^3)-E(\sigma)\le
\Gamma-16(2-h)+26-18h=\Gamma-2(3+h)
\end{equation}
The inequality (\ref{proricorrenza-3}) follows from
(\ref{proricorrenza-6}) and (\ref{proricorrenza-8}).
\qed

\subsec{The variational problem}{s:variazionale}
\par\noindent
Item~\ref{i:modeest} of Theorem~\ref{t:stime}
deals with the determination of the
minimal energy barrier between the metastable state $\muno$ and the stable one
$\puno$, more precisely with the computation of $\Phi(\muno,\puno)$.
In the context of serial Glauber dynamics this problem has been faced with
different approaches each suited to the model under exam, see \cite{[OV]} and
\cite[Section~4.2]{[MNOS]}, where a quite general technique is described.
All these methods rely on the continuity of the dynamics, namely, on the
property that at each step only one spin is updated.

In the case of parallel dynamics, see \cite{[CN]}, the lacking of continuity
increases the difficulty of the computation of the communication energy 
between the
metastable and the stable state. We follow, here, the method proposed in
\cite{[CN]} which is based on the construction of a set $\cc{G}\subset\cc{S}$
containing $\muno$, but not $\puno$, and on the evaluation of the
transition energy for all the possible transitions from the interior to
the exterior of such a set $\cc{G}$.

To define the set $\cc{G}$ we need to introduce the two
mappings $A,B:\cc{S}\to\cc{S}$.
Let $\sigma\in\cc{S}$, we set $A\sigma:=\sigma$ if $E(\sigma^x)>E(\sigma)$
for any $x\in\Lambda^+_\rr{u}(\sigma)$, otherwise
$A\sigma:=\sigma^x$ where $x$ is the first element of
$\Lambda^+_\rr{u}(\sigma)$ in lexicographic order. The map $A$ flips the
first, in lexicographic order, unstable plus spin of $\sigma$ to which
corresponds a decrease of the energy. Under the effect of the map $A$ the
number of pluses decreases, but only unstable pluses are flipped.
Let $\sigma\in\cc{S}$, the configuration $B\sigma\in\cc{S}$ is such that
for each $x\in\Lambda$
\begin{equation}
\label{bootdef}
B\sigma(x):=
\Big\{
\begin{array}{ll}
-\sigma(x) & \textrm{ if } x\in\Lambda^-_{\ge-1}(\sigma)\\
\sigma(x)  & \textrm{ otherwise}
\end{array}
\end{equation}
Note that the operator $B$ performs a single step of {\it bootstrap
percolation}; relatively to $\sigma$, it flips all the minus unstable
spins and, among the stable minus spins, only those with two neighboring
minuses.

In the sequel a relevant role will be played by the configuration
$\ol{B}\,\ol{A}\sigma$, for any $\sigma\in\cc{S}$; recall 
the definition of fixed point of a map given at the end of 
Section~\ref{s:conf}.
The sole unstable positive spins in $\ol{A}\sigma$ are those
corresponding to energy increasing flips. Starting from $\ol{A}\sigma$, the
map $B$, which flips the minus spins with at least two plus spins among the
nearest neighbors, is applied iteratively until a fixed point is reached.
It is easy to show that the pluses in such a fixed point form well
separated rectangles or stripes winding around the torus; more precisely,
the pluses in $\ol{B}\,\ol{A}\sigma$ occupy the region
$\bigcup_{i=1}^nQ_{\ell_{i,1},\ell_{i,2}}(x_i)\subset\Lambda$, where
$n,\ell_{1,1},\ell_{1,2},\dots,\ell_{n,1},\ell_{n,2}$ are positive integers
and $x_i\in\Lambda$ for any $i=1,\dots,n$, with
$Q_{\ell_{i,1},\ell_{i,2}}(x_i)$ being pairwise not interacting 
(see Section~\ref{s:lat}). Note that, depending on the values of
$\ell_{i,1},\ell_{i,2}$, the set $Q_{\ell_{i,1},\ell_{i,2}}(x_i)$ can be
either a rectangle or a stripe winding around the torus.

We can now define the set $\cc{G}$. Let $\sigma\in\cc{S}$, consider
$\ol{B}\,\ol{A}\sigma$, and, provided $\ol{B}\,\ol{A}\sigma\neq\muno$,
denote by $Q_{\ell_{i,1},\ell_{i,2}}(x_i)$
the collection of pairwise not interacting rectangles (or stripes) obtained by
collecting all the sites $y\in\Lambda$ such that
$\ol{B}\,\ol{A}\sigma(y)=+1$.
We say that $\sigma\in\cc{G}$ if and only if $\ol{B}\,\ol{A}\sigma=\muno$ or
$\min\{\ell_{i,1},\ell_{i,2}\}\le\lambda-1$ and
$\max\{\ell_{i,1},\ell_{i,2}\}\le L-2$ for any $i=1,\dots,n$.
Note that configurations $\sigma$ such that $\ol{B}\,\ol{A}\sigma$
contains plus stripes winding around the torus $\Lambda$ do not
belong to $\cc{G}$.

In general $\ol{T}\sigma\neq\ol{B}\,\ol{A}\sigma$, this means that 
$\ol{B}\,\ol{A}\sigma$ is not necessarily the result of the zero temperature
dynamics started at $\sigma$. 
This is not a problem when looking for the minimal energy barrier
between $\muno$ and $\puno$, provided the energy of such
configurations is larger than $\Gamma$. The definition of $\cc{G}$
is indeed satisfactory because we can prove the following
Proposition~\ref{t:insiemeG} on which the proof of
items~\ref{i:modeest} and \ref{i:modeest2} of Theorem~\ref{t:stime}
is mostly based. To state the lemma we need one more definition:
recall the set $\cc{C}$ is defined as the collection of
configurations with all the spins equal to $-1$ excepted those in a
rectangle of sides $\lambda-1$ and $\lambda$ and in a pair of
neighboring sites adjacent to one of the longer sides of the
rectangle. Then, given $\gamma\in\cc{C}$, we let
$\pi(\gamma)\subset\cc{S}$ the set whose elements are the two
configurations that can be obtained from $\gamma$ by flipping one of
the two plus spins in the pair attached to one of the longer sides
of the plus spin $\lambda\times(\lambda-1)$ rectangle. We also let
$\cc{P}$ be the collection of all the configurations with all the
spins equal to $-1$ excepted those in a rectangle of sides
$\lambda-1$ and $\lambda$ and in a single site adjacent to one of
the longer sides of the rectangle. Finally, we let $\cc{R}$  be the
collection of rectangular droplets with sides $\lambda-1$ and
$\lambda$. By using (\ref{energiaret}), we have
\begin{equation}
\label{enerre}
E(\cc{R})-E(\muno)=-4h\lambda^2+4h\lambda+16\lambda-8=
\Gamma-10+6h 
\end{equation} 
where we have used in the last equality the definition (\ref{prot-h}) of 
$\Gamma$. By using (\ref{stimagamma}), we have the easy bound
\begin{equation}
\label{stimaerre}
E(\cc{R})-E(\muno)<8\lambda+4h
\end{equation} 

\begin{proposition}
\label{t:insiemeG} 
With the definitions above, for 
$h>0$ small enough and $L=L(h)$ large enough, we have
\begin{enumerate}
\item\label{i:insiemeG-1}
$\muno\in\cc{G}$, $\puno\in\cc{S}\setminus\cc{G}$, and
$\cc{C}\subset\cc{S}\setminus\cc{G}$;
\item\label{i:insiemeG-2}
for each $\eta\in\cc{G}$ and $\zeta\in\cc{S}\setminus\cc{G}$ we have
$E(\eta,\zeta)\ge E(\muno)+\Gamma$;
\item\label{i:insiemeG-3}
for each $\eta\in\cc{G}$ and $\zeta\in\cc{S}\setminus\cc{G}$ we have
$E(\eta,\zeta)=E(\muno)+\Gamma$ if and only if
$\zeta\in\cc{C}$ and $\eta\in\pi(\zeta)$.
\end{enumerate}
\end{proposition}

\smallskip
\par\noindent
\textit{Proof of item~\ref{i:modeest} of Theorem~\ref{t:stime}.\/}
Since $\muno\in\cc{G}$ and $\puno\in\cc{S}\setminus\cc{G}$, see
item~\ref{i:insiemeG-1} in Proposition~\ref{t:insiemeG}, we have that
any path $\omega=\{\omega_1,\dots,\omega_n\}$ such that $\omega_1=\muno$
and $\omega_n=\puno$ must necessarily contain a transition from $\cc{G}$
to $\cc{S}\setminus\cc{G}$, i.e., there must be $i\in\{2,\dots,n\}$ such
that $\omega_{i-1}\in\cc{G}$ and $\omega_i\in\cc{S}\setminus\cc{G}$.
Thus, item~\ref{i:insiemeG-2} in Proposition~\ref{t:insiemeG} implies that
$\Phi_\omega\ge E(\muno)+\Gamma$; since the
path $\omega$ is arbitrary, it follows that
\begin{equation}
\label{promodeest-1}
\Phi(\muno,\puno)\ge E(\muno)+\Gamma
\end{equation}

To complete the proof of (\ref{modeest1}) we need to exhibit a path
connecting $\muno$ to $\puno$ such that the height along such a path is less
than or equal to $E(\muno)+\Gamma$.
Consider the path $\omega:=\{\muno,\sigma^1,\sigma^2,\sigma^3,\sigma^4\}$
with $\sigma^1$ 
the configuration with all the spins equal to minus one excepted the one
at the origin,
$\sigma^2$
the configuration with all the spins equal to minus one excepted the ones
associated to the sites in the rectangle $Q_{2,1}(0)$,
$\sigma^3$
the configuration with all the spins equal to minus one excepted the ones
associated to the sites in $Q_{2,1}(0)\cup Q_1(0,1)$, and
$\sigma^4$
the configuration with all the spins equal to minus one excepted the ones
associated to the sites in the square $Q_2(0)$.

By definition, the path $\omega+\Omega_{\sigma^4}$ starts at $\muno$ and ends
in $\puno$, i.e., $\omega+\Omega_{\sigma^4}\in\Theta(\muno,\puno)$.
Moreover, we shall prove that
\begin{equation}
\label{promodeest-3}
\Phi_{\omega+\Omega_{\sigma^4}}-E(\muno)\le\Gamma
\end{equation}
The inequality (\ref{promodeest-3}), together with
(\ref{promodeest-1}), implies (\ref{modeest1}).

\begin{figure}[t]
\begin{picture}(100,180)(10,0)
\qbezier(130,10)(145,25)(160,40)
\put(130,10){\circle*{3}}
\put(130,3){\scriptsize{$\muno$}}
\qbezier(160,40)(180,160)(200,80)
\put(160,40){\circle*{3}}
\put(160,33){\scriptsize{$\sigma^1$}}
\qbezier(200,80)(220,180)(240,120)
\put(200,80){\circle*{3}}
\put(200,73){\scriptsize{$\sigma^2$}}
\qbezier(240,120)(260,200)(280,140)
\put(240,120){\circle*{3}}
\put(240,113){\scriptsize{$\sigma^3$}}
\put(280,140){\circle*{3}}
\put(280,133){\scriptsize{$\sigma^4$}}
\put(115,27){\scriptsize{$2(5-h)$}}
\put(124,73){\scriptsize{$2\cdot2(3-h)$}}
\put(129,125){\scriptsize{$2\cdot2(1-h)+2(3-h)$}}
\put(210,155){\scriptsize{$3\cdot2(1-h)$}}
\put(305,79){\scriptsize{$2(1-h)$}}
\qbezier[70](190,90)(130,30)(300,78)
\put(188,88){\vector(1,1){5}}
\qbezier[70](230,125)(200,105)(300,85)
\put(228,124){\vector(2,1){5}}
\put(278,152){\scriptsize{$2(1+h)$}}
\end{picture}
\caption{Energy landscape for the path $\omega$.}
\label{f:primopath}
\end{figure}
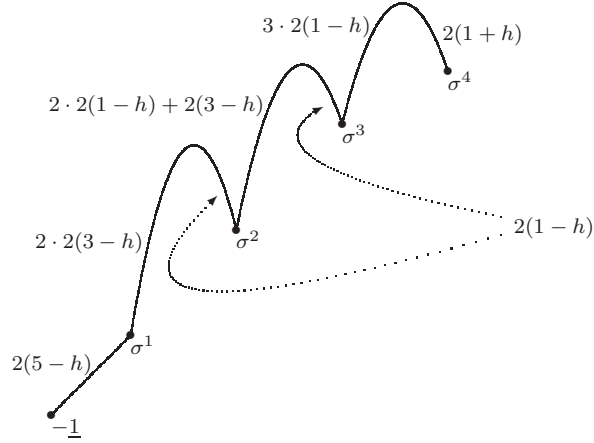

We are then left with the proof of (\ref{promodeest-3}).
We first consider the path $\omega$; by using (\ref{comm-ene}), see also
figure~\ref{f:tabella}, we get
\begin{equation}
\label{promodeest-6-1}
\begin{array}{lcllcl}
E(\muno,\sigma^1)&\!\!\!=\!\!\!&E(\muno)+2(5-h)
&
E(\sigma^1,\muno)&\!\!\!=\!\!\!&E(\sigma^1)
\\
E(\sigma^1,\sigma^2)&\!\!\!=\!\!\!&E(\sigma^1)+2\cdot2(3-h)
&
E(\sigma^2,\sigma^1)&\!\!\!=\!\!\!&E(\sigma^2)+2(1-h)
\\
E(\sigma^2,\sigma^3)&\!\!\!=\!\!\!&E(\sigma^2)+2\cdot2(1-h)+2(3-h)
&
E(\sigma^3,\sigma^2)&\!\!\!=\!\!\!&E(\sigma^3)+2(1-h)
\\
E(\sigma^3,\sigma^4)&\!\!\!=\!\!\!&E(\sigma^3)+3\cdot2(1-h)
&
E(\sigma^4,\sigma^3)&\!\!\!=\!\!\!&E(\sigma^4)+2(1+h)
\\
\end{array}
\end{equation}
see figure~\ref{f:primopath} for a graphical representation.

By using (\ref{promodeest-6-1}), (\ref{pizza}), (\ref{sim-ene}),
(\ref{prot-h}), and the definition (\ref{lcritica}) of the critical
length $\lambda$,
it is easy to show that, provided $h$ is chosen smaller than $3+\sqrt{5}$,
\begin{equation}
\label{promodeest-6-3}
\Phi_\omega-E(\muno)\le34-14h<\Gamma
\end{equation}
and
\begin{equation}
\label{promodeest-7-3}
E(\sigma^4)-E(\muno)\le32-16h
\end{equation}
We consider, now, the path $\Omega_{\sigma^4}$;
by using (\ref{cammino-4}) and (\ref{promodeest-7-3}), we get
\begin{equation}
\label{promodeest-8-3}
\Phi_{\Omega_{\sigma^4}}-E(\muno)=
\Phi_{\Omega_{\sigma^4}}-E(\sigma^4)+E(\sigma^4)-E(\muno)\le
\Gamma-16(2-h)+32-16h=\Gamma
\end{equation}
The inequality (\ref{promodeest-3}) follows from
(\ref{promodeest-6-3}) and (\ref{promodeest-8-3}).
This completes the proof of item~\ref{i:modeest} of Theorem~\ref{t:stime}.
\qed

\smallskip
\par\noindent
\textit{Proof of item~\ref{i:modeest2} of Theorem~\ref{t:stime}.\/}
The item follows from
item~\ref{i:modeest} of Theorem~\ref{t:stime} and
item~\ref{i:insiemeG-3} of Proposition~\ref{t:insiemeG}.
\qed

\sezione{Proof of Proposition~\ref{t:insiemeG}}{s:minmax}
\par\noindent
In Section~\ref{s:proof} we shall prove Proposition~\ref{t:insiemeG}
concerning the solution of the \emph{minmax} problem. 
Some preliminary lemmata are stated in advance. 
More precisely, 
in Section~\ref{s:auxiliary} we state the 
Lemmata~\ref{t:boots}--\ref{t:discesa}
concerning energy estimates for the maps $A$ and $B$ 
(see Section~\ref{s:variazionale}).
In Section~\ref{s:rectdrop} the Lemmata~\ref{lem1} and \ref{t:retto-ener},
concerning properties of rectangular droplets (see Section~\ref{s:meta}), 
are stated.
Section~\ref{s:rectdrop1} is devoted to the comparison of configurations 
in $\cc{G}$ (see Section~\ref{s:variazionale}) and in $\cc{G}^\rr{c}$.

\subsec{Energy estimates for the maps $A$ and $B$}{s:auxiliary}
\par\noindent
In Lemma~\ref{t:boots} we give estimates on the energy of the
configurations obtained by applying the maps $A$ and $B$. For any
$\sigma\in\cc{S}$ we let
\begin{equation}
\label{ciccio}
N_A(\sigma):=\sum_{x\in\Lambda}[1-\delta_{\sigma(x),\ol{A}\sigma(x)}]
\;\;\;\textrm{ and }\;\;\;
N_B(\sigma):=\sum_{x\in\Lambda}
             [1-\delta_{\ol{A}\sigma(x),\ol{B}\,\ol{A}\sigma(x)}]
\end{equation}
with $\delta$ the Kronecker $\delta$. Note that $N_A(\sigma)$ is the number of
plus spins which are flipped by the iterative application of the map $A$ to
$\sigma$, while $N_B(\sigma)$ is the number of minus spins
which are flipped by the iterative application of the bootstrap percolation
map $B$ to $\ol{A}\sigma$.

\begin{lemma}
\label{t:boots}
Let $\sigma\in\cc{S}$ and $h>0$ small enough. Then
\begin{enumerate}
\item\label{i:filam}
we have
\begin{equation}
\label{filam}
E(\sigma)\geq E(\ol{A}\sigma)+(2-10h)N_A(\sigma)
\end{equation}
\item\label{i:boots}
we have
\begin{equation}
\label{boots}
E(\ol{A}\sigma)\geq E(\ol{B}\,\ol{A}\sigma)+4h N_B(\sigma)
\end{equation}
\end{enumerate}
\end{lemma}

In order to prove Lemma~\ref{t:boots} we state
Lemma~\ref{t:p-boots} on some properties of unstable plus spins
and Lemma~\ref{t:p2-boots} concerning an energy estimate for a single
application of the bootstrap percolation map $B$.
Recall (\ref{lambdi}), (\ref{lambdi-mu}), and (\ref{lambdi-us}); recall
also that, given $\sigma\in\cc{S}$ and
$x\in\Lambda$, the configuration $\sigma^x$ has been defined in
Section~\ref{s:conf} as the one obtained by flipping the spin of $\sigma$
associated with the site $x$.

\begin{lemma}
\label{t:p-boots}
Let $\sigma\in\cc{S}$; for $h>0$ small enough, we have that the following
statements hold true:
\begin{enumerate}
\item
\label{i:p-boots-1}
if there exists $x\in\Lambda^+_\rr{u}(\sigma)$ such that
$E(\sigma^x)>E(\sigma)$, then 
$|\partial\{x\}\cap\Lambda^-_\rr{s}(\sigma)|\le1$, 
i.e., there exists at most one nearest neighbor of $x$ which is
stable w.r.t.\ $\sigma$ and such that the associated spin is minus one;
\item
\label{i:p-boots-2}
if there exists $x\in\Lambda^+_\rr{u}(\sigma)$ such that
$E(\sigma^x)\le E(\sigma)$, then $E(\sigma)\ge E(\sigma^x)+2-10h$;
\item
\label{i:p-boots-3}
if $E(\sigma^x)>E(\sigma)$ for any $x\in\Lambda^+_\rr{u}(\sigma)$, then
\begin{equation}
\label{moc-a-me}
2|\Lambda^+_{-1}(\sigma)|+
3|\Lambda^+_{-3}(\sigma)|\le
3|\Lambda^-_{+1}(\sigma)|+
4|\Lambda^-_{+3}(\sigma)|
\end{equation}
\end{enumerate}
\end{lemma}

\smallskip
\par\noindent
\textit{Proof of Lemma~\ref{t:p-boots}.\/}
Let $x\in\Lambda^+_\rr{u}(\sigma)$, then $\sigma(x)=+1$, $\sigma^x(x)=-1$, and
$S_\sigma(x)<0$; by using (\ref{hl}), we get
\begin{equation}
\label{cavolo}
E(\sigma^x)-E(\sigma)
=
2h-2
+\sum_{y\in\partial\{x\}}(|S_{\sigma}(y)+h|-|S_\sigma(y)-2+h|)
\end{equation}
Note that, since $\sigma(x)=+1$, we have that
$S_\sigma(y)$, with $y\in\partial\{x\}$, can assume the values
$-3,-1,+1,+3,+5$; by performing the direct computations
one shows that
\begin{equation}
\label{cavolo2}
|S_\sigma(y)+h|-|S_{\sigma}(y)-2+h|\in\{-2,2h,+2\}
\end{equation}
for $y\in\partial\{x\}$.

Item~\ref{i:p-boots-1}.\
Let $x\in\Lambda^+_\rr{u}(\sigma)$ such that $E(\sigma^x)>E(\sigma)$;
since $S_\sigma(y)<0$ for
$y\in\partial\{x\}\cap\Lambda^-_\rr{s}(\sigma)$, by using (\ref{cavolo}) we get
\begin{displaymath}
E(\sigma^x)-E(\sigma)
=
2h-2(1+|\partial\{x\}\cap\Lambda^-_\rr{s}(\sigma)|)
+\!\!\!\!\sum_{y\in\partial\{x\}\setminus\Lambda^-_\rr{s}(\sigma)}\!\!\!\!
(|S_{\sigma}(y)+h|-|S_\sigma(y)-2+h|)
\end{displaymath}
Suppose, by the way of contradiction, that 
$|\partial\{x\}\cap\Lambda^-_\rr{s}(\sigma)|\ge2$,
then we have
\begin{displaymath}
E(\sigma^x)-E(\sigma)
\le
2h-6
+\!\!\!\!\sum_{y\in\partial\{x\}\setminus\Lambda^-_\rr{s}(\sigma)}\!\!\!\!
(|S_\sigma(y)+h|-|S_{\sigma}(y)-2+h|)
\end{displaymath}
By (\ref{cavolo2}) we obtain
$|S_\sigma(y)+h|-|S_{\sigma}(y)-2+h|\le2$ for $y\in\partial\{x\}$,
and, noting that
$|\partial\{x\}\setminus\Lambda^-_\rr{s}(\sigma)|\le2$, we finally get
$E(\sigma^x)-E(\sigma)\le2h-6+4=2h-2<0$, which is in contradiction with
the hypothesis.

Item~\ref{i:p-boots-2}.\
Let $x\in\Lambda^+_\rr{u}(\sigma)$ such that
$E(\sigma^x)\le E(\sigma)$. 
Recalling (\ref{cavolo}) and (\ref{cavolo2}), we have that 
the number of sites $y\in\partial\{x\}$ such that 
$|S_\sigma(y)+h|-|S_{\sigma}(y)-2+h|=+2$ must be at most equal to 
the number of sites $y\in\partial\{x\}$ such that 
$|S_\sigma(y)+h|-|S_{\sigma}(y)-2+h|=-2$, otherwise it would be 
$E(\sigma^x)-E(\sigma)>0$.
Thus, un upper bound to the sum in (\ref{cavolo}) is found when
all the $y\in\partial\{x\}$ are such that 
$|S_\sigma(y)+h|-|S_{\sigma}(y)-2+h|=2h$. We then get
\begin{displaymath}
\sum_{y\in\partial\{x\}}(|S_{\sigma}(y)+h|-|S_\sigma(y)-2+h|)
\le2h|\partial\{x\}|=8h
\end{displaymath}
from which $E(\sigma^x)-E(\sigma)\le-2+10h$ follows.

Item~\ref{i:p-boots-3}.\ Consider $\sigma\in\cc{S}$ such that
$E(\sigma^x)>E(\sigma)$ for any $x\in\Lambda^+_\rr{u}(\sigma)$ and let
$r_\sigma(y)=1$ if $y\in\Lambda^-_\rr{u}$ and
$r_\sigma(y)=0$ otherwise. Recall that $\Lambda^+_{-5}(\sigma)=\emptyset$
and recall (\ref{lambdi-us}); by exploiting the first part of this lemma we get
\begin{displaymath}
\sum_{x\in\Lambda^+_\rr{u}(\sigma)}
\sum_{y\in\partial\{x\}} r_\sigma(y)=
\!\!\!\sum_{x\in\Lambda^+_{-1}(\sigma)}
\sum_{y\in\partial\{x\}} r_\sigma(y)+
\!\!\!\sum_{x\in\Lambda^+_{-3}(\sigma)}
\sum_{y\in\partial\{x\}} r_\sigma(y)\ge
2|\Lambda^+_{-1}(\sigma)|+
3|\Lambda^+_{-3}(\sigma)|
\end{displaymath}
On the other hand, a site in $\Lambda^-_{+1}(\sigma)$ is nearest neighbor
of at most three sites in $\Lambda^+_\rr{u}$, indeed the number of unstable
pluses neighboring such a site can be less than three since some of the
pluses can be stable ones,
and a site in $\Lambda^-_{+3}(\sigma)$ is nearest neighbor
of at most four sites in $\Lambda^+_\rr{u}$; then we have
\begin{displaymath}
\sum_{x\in\Lambda^+_\rr{u}(\sigma)}
\sum_{y\in\partial\{x\}} r_\sigma(y)\le
3|\Lambda^-_{+1}(\sigma)|+
4|\Lambda^-_{+3}(\sigma)|
\end{displaymath}
The inequality (\ref{moc-a-me}) follows trivially from the two bounds above.
\qed

\begin{lemma}
\label{t:p2-boots}
Suppose $h>0$ small enough.
Let $\sigma\in\cc{S}$, suppose $E(\sigma^x)>E(\sigma)$ for any
$x\in\Lambda^+_\rr{u}(\sigma)$. Then
\begin{equation}
\label{boots2}
E(\sigma)\geq E(B\sigma)+4h|\Lambda^-_{\ge-1}(\sigma)|
\end{equation}
\end{lemma}
Recall that $\Lambda^-_{\ge-1}(\sigma)$ is exactly the
set of sites whose associated spin flips under the action of the bootstrap
percolation map $B$ (see (\ref{bootdef})).

\smallskip
\par\noindent
\textit{Proof of  Lemma~\ref{t:p2-boots}.\/}
To compare $E(\sigma)$ and $E(B\sigma)$ we shall use (\ref{sim-ene}) and
suitable bounds on $E(\sigma,B\sigma)$ and $E(B\sigma,\sigma)$.
Recall (\ref{comm-ene}), see also figure~\ref{f:tabella}, and the definition
(\ref{bootdef}) of
the bootstrap percolation map $B$; we have that in the forward transition from
$\sigma$ to $B\sigma$ the energy costs are those associated to the flip of the
stable minuses with two neighboring pluses and those associated to the
permanence of the unstable pluses. More precisely, we have
\begin{equation}
\label{boot1}
E(\sigma,B\sigma)=E(\sigma)
                 +2(1-h)|\Lambda^-_{-1}(\sigma)|
                 +2(1-h)|\Lambda^+_{-1}(\sigma)|
                 +2(3-h)|\Lambda^+_{-3}(\sigma)|
\end{equation}
On the other hand in the backward transition 
from $B\sigma$ to $\sigma$ the energy
costs that must be surely paid are those associated to the reverse flipping of
the pluses that have been created in the forward transition; more precisely,
we have
\begin{equation}
\label{boot2}
E(B\sigma,\sigma)\ge E(B\sigma)
+2(1+h)|\Lambda^-_{-1}(\sigma)|
+2(3+h)|\Lambda^-_{+1}(\sigma)|
+2(5+h)|\Lambda^-_{+3}(\sigma)|
\end{equation}
Note that in (\ref{boot2}) it is not possible to take advantage from the
permanence of the possible unstable pluses in $B\sigma$, because, as we shall
see in the proof of item~\ref{i:boots} of the Lemma~\ref{t:boots}, we have
$\Lambda^+_\rr{u}(B\sigma)=\emptyset$.

To complete the proof we have to distinguish two cases. Suppose, first,
that
$\Lambda^+_{-1}(\sigma)=\Lambda^-_{+3}(\sigma)=\emptyset$; by using
(\ref{boot1}), (\ref{boot2}), and (\ref{sim-ene}), we get
\begin{displaymath}
E(\sigma)\ge E(B\sigma)+4h|\Lambda^-_{-1}(\sigma)|
  -2(3-h)|\Lambda^+_{-3}(\sigma)|+2(3+h)|\Lambda^-_{+1}(\sigma)|
\end{displaymath}
The bound (\ref{boots2}) follows noting that, in this case,
$\Lambda^-_{\ge-1}(\sigma)=\Lambda^-_{-1}(\sigma)\cup\Lambda^-_{+1}(\sigma)$
and (\ref{moc-a-me}) reduces to
$|\Lambda^+_{-3}(\sigma)|\le|\Lambda^-_{+1}(\sigma)|$.
Suppose, now, that either $\Lambda^+_{-1}(\sigma)\neq\emptyset$ or
$\Lambda^-_{+3}(\sigma)\neq\emptyset$.
By using (\ref{moc-a-me}) we have
\begin{displaymath}
\begin{array}{l}
|\Lambda^+_{-1}(\sigma)|+3|\Lambda^+_{-3}(\sigma)|\le
2|\Lambda^+_{-1}(\sigma)|+3|\Lambda^+_{-3}(\sigma)|\vphantom{_\big\{}\\
\phantom{merdonemerdonemerdone}
\le3|\Lambda^-_{+1}(\sigma)|+4|\Lambda^-_{+3}(\sigma)|\le
3|\Lambda^-_{+1}(\sigma)|+5|\Lambda^-_{+3}(\sigma)|
\end{array}
\end{displaymath}
Since either $|\Lambda^+_{-1}(\sigma)|\ge0$ or
$|\Lambda^-_{+3}(\sigma)|\ge0$, we have that 
\begin{equation}
\label{moc-a-me-intermedia}
|\Lambda^+_{-1}(\sigma)|+3|\Lambda^+_{-3}(\sigma)|<
 3|\Lambda^-_{+1}(\sigma)|+5|\Lambda^-_{+3}(\sigma)|
\end{equation}
We shall prove that, provided $h<1$, 
\begin{equation}
\label{moc-a-me-h}
(1-h)|\Lambda^+_{-1}(\sigma)|+
(3-h)|\Lambda^+_{-3}(\sigma)|<
(3-h)|\Lambda^-_{+1}(\sigma)|+
(5-h)|\Lambda^-_{+3}(\sigma)|
\end{equation}
First of all we note that the inequality (\ref{moc-a-me-h}) is equivalent to 
\begin{displaymath}
\begin{array}{l}
h[|\Lambda^-_{+3}(\sigma)|+|\Lambda^-_{+1}(\sigma)|
-|\Lambda^+_{-1}(\sigma)|-|\Lambda^+_{-3}(\sigma)|] 
\vphantom{_\big\{}\\
\phantom{me}
<
|\Lambda^-_{+3}(\sigma)|
+|\Lambda^-_{+1}(\sigma)|
-|\Lambda^+_{-1}(\sigma)|
-|\Lambda^+_{-3}(\sigma)|
+[2|\Lambda^-_{+1}(\sigma)|
+4|\Lambda^-_{+3}(\sigma)|
-2|\Lambda^+_{-3}(\sigma)|]
\end{array}
\end{displaymath}
which is trivially satisfied when the left hand side is negative or equal
to zero, since (\ref{moc-a-me-intermedia}) implies that the right hand side is 
strictly positive;
on the other hand, 
if the left hand side is strictly positive, recalling that $h<1$, the 
inequality will follow once we shall have proved that 
$2|\Lambda^-_{+1}(\sigma)|+4|\Lambda^-_{+3}(\sigma)|
  -2|\Lambda^+_{-3}(\sigma)|\ge0$.

To get this last bound we note that 
by using item~\ref{i:p-boots-1} in Lemma~\ref{t:p-boots}, it follows that
for each site belonging to $\Lambda^+_{-3}$, there are at least three unstable 
minus spins among the four nearest neighboring ones. Hence, we get 
$|\Lambda^-_\rr{u}(\sigma)|\ge(4/3)|\Lambda^+_{-3}(\sigma)|$. 
Moreover, noted that 
$|\Lambda^-_\rr{u}(\sigma)|=|\Lambda^-_{+1}(\sigma)|+|\Lambda^-_{+3}(\sigma)|$,
we also get 
\begin{displaymath}
\begin{array}{l}
2|\Lambda^-_{+1}(\sigma)|+4|\Lambda^-_{+3}(\sigma)|-2|\Lambda^+_{-3}(\sigma)|
\vphantom{x_\big\{}\\
\phantom{merdone}
=2|\Lambda^-_{+3}(\sigma)|+2\big[|\Lambda^-_{+1}(\sigma)|
          +|\Lambda^-_{+3}(\sigma)|-|\Lambda^+_{-3}(\sigma)|\big]
\vphantom{x_\big\{}\\
\phantom{merdone}
=2|\Lambda^-_{+3}(\sigma)|+2\big[|\Lambda^-_{u}(\sigma)|
       -|\Lambda^+_{-3}(\sigma)|\big]
\ge
       2\big[|\Lambda^-_{+3}(\sigma)|+1/3|\Lambda^+_{-3}(\sigma)|\big]
\ge0
\end{array}
\end{displaymath}

Finally,
the bound (\ref{boots2}) follows easily by using (\ref{sim-ene}),
(\ref{boot1}), (\ref{boot2}), and the inequality (\ref{moc-a-me-h}).
\qed

\smallskip
\par\noindent
\textit{Proof of Lemma~\ref{t:boots}.\/}
Item~\ref{i:filam}.\ The bound (\ref{filam}) is proven easily by applying
iteratively item~\ref{i:p-boots-2} of Lemma~\ref{t:p-boots}.

Item~\ref{i:boots}.\
Suppose $\ol{B}\,\ol{A}\sigma=B^n\ol{A}\sigma$ for some integer $n$.
We first note that by Lemma~\ref{t:p-boots} each site
$x\in\Lambda^+_\rr{u}(\ol{A}\sigma)$ has at least two neighboring minuses
which are unstable w.r.t.\ $\ol{A}\sigma$, more precisely
$|\partial\{x\}\cap\Lambda^-_\rr{u}(\ol{A}\sigma)|\ge2$. 
Recall the definition (\ref{bootdef}) of the bootstrap percolation map $B$;
since $\Lambda^-_\rr{u}(\ol{A}\sigma)\subset\Lambda^-_{\ge-1}(\ol{A}\sigma)$,
the minuses in $\partial\{x\}\cap\Lambda^-_\rr{u}(\ol{A}\sigma)$
flip under the action of $B$.
Hence, $|\partial\{x\}\cap\Lambda^+(B\ol{A}\sigma)|\ge2$. We then have
$\Lambda^+_\rr{u}(B\ol{A}\sigma)=\emptyset$; in other words all the unstable
pluses in $\ol{A}\sigma$ become stable after the application of a single
step of the bootstrap percolation.

By definition of the bootstrap percolation map we also have that
$\Lambda^+_\rr{u}(B^i\ol{A}\sigma)=\emptyset$ for
any $i=2,\dots,n$, i.e., no site in $\Lambda^+(B^i\ol{A}\sigma)$ is
unstable w.r.t.\ $B^i\ol{A}\sigma$.
Note, finally, that 
$E((\ol{A}\sigma)^x)>E(\ol{A}\sigma)$
for any $x\in\Lambda^+_\rr{u}(\ol{A}\sigma)$.
The theorem then follows
by applying iteratively Lemma~\ref{t:p2-boots}.
\qed

Let $\sigma\in\cc{S}$, we refine the estimate (\ref{filam}) by considering 
the plus spins that are flipped by the iterative application of the map $A$ and 
are associated with sites outside the support of the configuration 
$\ol{B}\,\ol{A}\sigma$. Let the \textit{branch} of $\sigma$ be 
\begin{equation}
\label{branch}
L(\sigma):=|\Lambda^+(\sigma)\setminus\Lambda^+(\ol{B}\,\ol{A}\sigma)|
\end{equation}
i.e., the number of pluses outside the
rectangles of $\ol{B}\,\ol{A}\sigma$ which are flipped by the map $A$; note 
that $L(\sigma)\le N_A(\sigma)$ (see (\ref{ciccio})).

\begin{lemma}
\label{t:discesa}
For any $\sigma\in\cc{S}$ such that $L(\sigma)\ge1$, we have that
\begin{equation}
\label{svolta}
E(\sigma)-E(\ol{A}\sigma)\ge
\left\{\begin{array}{ll}
6-2h & \textrm{ if } L(\sigma)=1\\
10-6h+(2-10h)(L(\sigma)-2)
& \textrm{ if } L(\sigma)\ge2\\
\end{array}
\right.
\end{equation}
\end{lemma}

\smallskip
\par\noindent
\textit{Proof of Lemma~\ref{t:discesa}.\/}
Let $\sigma\in\cc{S}$ such that $L(\sigma)=1$, the set
$\Lambda^+(\sigma)\setminus\Lambda^+(\ol{B}\,\ol{A}\sigma)$
has a unique element $x$. There exists a natural number $j$ such that
$A^{j-1}\sigma(x)=+1$ and $A^{j}\sigma(x)=-1$. 
For $y\in\partial\{x\}\cap(\Lambda^+(\ol{B}\,\ol{A}\sigma))^\rr{c}$ we have 
$|S_{A^{j}\sigma}(y)+h|-|S_{A^{j-1}\sigma}(y)+h|=2$, while 
for $y\in\partial\{x\}\cap\Lambda^+(\ol{B}\,\ol{A}\sigma)$ we have the trivial
bound $|S_{A^{j}\sigma}(y)+h|-|S_{A^{j-1}\sigma}(y)+h|\ge-2$.
Since 
$|\partial\{x\}\cap(\Lambda^+(\ol{B}\,\ol{A}\sigma))^\rr{c}|\ge3$ and
$|\partial\{x\}\cap\Lambda^+(\ol{B}\,\ol{A}\sigma)|\le1$, 
by using (\ref{hl}) we get
\begin{equation}
\label{svolta01}
\begin{array}{rl}
E(A^{j-1}\sigma)-E(A^{j}\sigma)=2-2h&+
{\displaystyle
\!\!\!\!\sum_{y\in\partial\{x\}\cap(\Lambda^+(\ol{B}\,\ol{A}\sigma))^\rr{c}}
\!\!\!\!(|S_{A^{j}\sigma}(y)+h|-|S_{A^{j-1}\sigma}(y)+h|)}\\
&
{\displaystyle
+
\!\!\!\!\sum_{y\in\partial\{x\}\cap\Lambda^+(\ol{B}\,\ol{A}\sigma)}
\!\!\!\!(|S_{A^{j}\sigma}(y)+h|-|S_{A^{j-1}\sigma}(y)+h|)
\ge6-2h }\\
\end{array}
\end{equation}
Recall, finally, that by definition the map $A$ decreases the energy;
then, by (\ref{svolta01}), we have
\begin{displaymath}
E(\sigma)\geq E(A^{j-1}\sigma)\geq 
E(A^{j}\sigma)-2h+6\geq E(\ol{A}\sigma)-2h+6
\end{displaymath}
and the bound (\ref{svolta}) follows.

Let now $\sigma\in\cc{S}$ such that $L(\sigma)=2$;
the set $\Lambda^+(\sigma)\setminus\Lambda^+(\ol{B}\,\ol{A}\sigma)$
has two elements $x,y$.
Since $\ol{B}\,\ol{A}\sigma=\ol{B}\,\ol{A}\sigma^y$ and $L(\sigma)=2$, we have 
$L(\sigma^y)=1$;
by using $\ol{A}\sigma^y=\ol{A}\sigma$ and (\ref{svolta}) in
the already proven case we have that
\begin{equation}
\label{svolta02}
E(\sigma)-E(\ol{A}\sigma)=
E(\sigma)-E(\sigma^y)+E(\sigma^y)-E(\ol{A}\sigma)\ge
E(\sigma)-E(\sigma^y)+6-2h
\end{equation}
In order to bound $E(\sigma)-E(\sigma^y)$, we first note that by
(\ref{hl}) we get
\begin{equation}
\label{svolta03}
E(\sigma)-E(\sigma^y)
=
-2h-\sum_{z\in\overline{\{y\}}}
(|S_\sigma(z)+h|-|S_{\sigma^y}(z)+h|)
\end{equation}

We distinguish, now, two cases. We first suppose that
$x\not\in\overline{\{y\}}$, i.e., the two sites $x$ and $y$ are not nearest 
neighbors.
It is easy to prove that $-(|S_\sigma(y)+h|-|S_{\sigma^y}(y)+h|)=+2$. Moreover,
note that the contribution to the sum (\ref{svolta03}) of all the sites in
$\partial\{y\}\cap(\Lambda^+(\ol{B}\,\ol{A}\sigma))^\rr{c}$ is equal to $+2$
excepted for at most one site whose contribution is equal to $-2h$.
Note, also, that 
$|\partial\{y\}\cap(\Lambda^+(\ol{B}\,\ol{A}\sigma))^\rr{c}|\ge3$;
hence, we have that
$E(\sigma)-E(\sigma^y)\ge-2h+(2-2h)+2+2-2h-2$,
where the contribution of the site
$\partial\{y\}\cap\Lambda^+(\ol{B}\,\ol{A}\sigma)$, which possibly exists,
has been bounded trivially by $-2$. The bound (\ref{svolta})
follows immediately.

\begin{figure}[ht]
\vskip 0.5 cm
\begin{center}
\includegraphics[height=2.5cm]{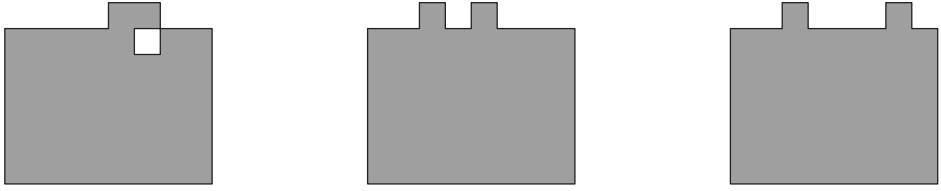}
\caption{The three cases studied in the proof of the Lemma~\ref{t:discesa};
on the left the not trivial one.}
\label{f:trecasi}
\end{center}
\end{figure}

\begin{figure}[ht]
\label{insieme}
\vskip 0.5 cm
\begin{center}
\includegraphics[height=7cm]{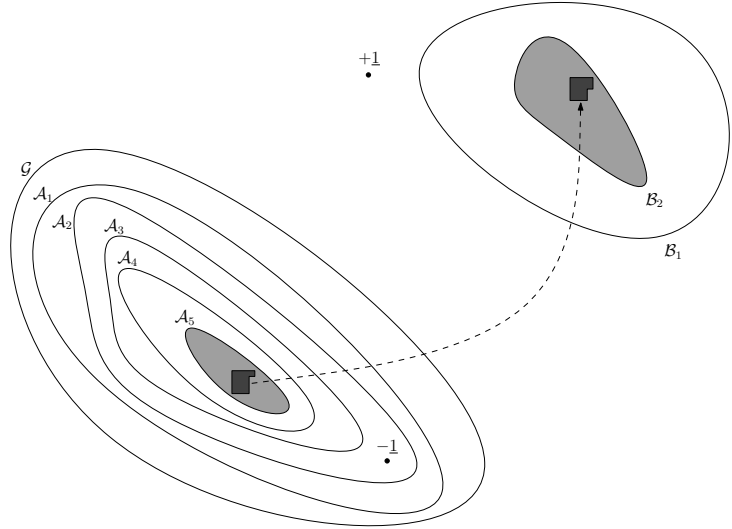}
\caption{Restricted sets on which we evaluate $E(\eta,\zeta)$ in the 
proof of item~\ref{i:insiemeG-2} of Proposition~\ref{t:insiemeG}.}
\label{f:strategia}
\end{center}
\end{figure}

Suppose, now, that $x\in\overline{\{y\}}$, i.e., 
the two sites $x$ and $y$ are adjacent.
The only not trivial case, see Figure~\ref{f:trecasi}, is the one
in which both the sites $x$ and $y$ are at distance one from the set
$\Lambda^+(\ol{B}\,\ol{A}\sigma)$. 
Since the plus spins associated to $x$ and $y$
are flipped by the iterative application of the map $A$ to $\sigma$,
the spin associated to at least one of the two sites in
$\partial\{x,y\}\cap\Lambda^+(\ol{B}\,\ol{A}\sigma)$ is equal to $-1$,
see Figure~\ref{f:trecasi}. Without loss of generality
we let $\partial\{y\}\cap\Lambda^+(\ol{B}\,\ol{A}\sigma)=\{y'\}$ and
$\sigma(y')=-1$.
It is easy to prove that $-(|S_\sigma(y)+h|-|S_{\sigma^y}(y)+h|)=+2$,
$-(|S_\sigma(x)+h|-|S_{\sigma^y}(x)+h|)\ge-2h$,
$-(|S_\sigma(y')+h|-|S_{\sigma^y}(y')+h|)\ge-2$, and
$-(|S_\sigma(z)+h|-|S_{\sigma^y}(z)+h|)=2$ for each
$z\in\partial\{y\}\setminus\{x,y'\}$. Hence, by using (\ref{svolta03}) we get
\begin{equation}
\label{svolta10}
E(\sigma)-E(\sigma^y)\ge-2h+2-2h-2+2+2=4-4h
\end{equation}
The bound (\ref{svolta})
follows by (\ref{svolta10}) and (\ref{svolta02}).

Let, finally, $\sigma\in\cc{S}$ such that $L(\sigma)\ge3$. Let
$i$ a suitable integer such that $L(A^i\sigma)=2$.
The bound (\ref{svolta})
follows easily by using the Lemma~\ref{t:boots} and (\ref{svolta})
applied to $A^i\sigma$.
\qed

\subsec{Energy estimates for rectangular droplets}{s:rectdrop}
\par\noindent
We first state and prove the following  Lemma on some simple geometrical 
properties of rectangles on the lattice. 

\begin{lemma}
\label{lem1}
Let $Q_{l_i,m_i}$, for $i=1,\dots,n$, be pairwise disjoint rectangles
with sides $l_i,m_i\in\bb{N}\setminus\{0\}$, such that $\ell_i\le m_i$ for
$i=1,\dots,n$, and semi--perimeter $p:=\sum_i^n(\ell_i+m_i)$.
\begin{enumerate}
\item
\label{i:lem1-1}
We have
\begin{equation}
\label{mammt}
\frac{1}{4}p^2\geq \sum_{i=1}^n l_i\,m_i
\end{equation}
\item
\label{i:lem1-3}
If there exists a positive integer $k$ such that $\ell_i\le k-1$ and 
$m_i\le k$ for all $i=1,\dots,n$, we have 
\begin{equation}
\label{subcrit}
\sum_{i=1}^n\ell_im_i\le\frac{1}{2}kp-\frac{1}{2}\sum_{i=1}^nm_i
\end{equation}
\item
\label{i:lem1-2}
If $n\geq2$ and $l_i\ge2$ then
\begin{equation}
\label{differenza}
\frac{1}{4}p^2\geq \sum_{i=1}^n l_i\,m_i+p
\end{equation}
\end{enumerate}
\end{lemma}

\smallskip
\par\noindent
\textit{Proof of Lemma~\ref{lem1}.\/}
Item~\ref{i:lem1-1}: we have
\begin{equation*}
\frac{1}{4}p^2=\frac{1}{4}\big(\sum_{i=1}^n(l_i+m_i)\big)^2\geq
 \frac{1}{4}\sum_{i=1}^n(l_i+m_i)^2=\frac{1}{4}\sum_{i=1}^n(l_i-m_i)^2+\sum_{i=1}^nl_i\,m_i
\geq\sum_{i=1}^nl_i\,m_i
\end{equation*}
Item~\ref{i:lem1-3}: we have
\begin{displaymath}
\sum_{i=1}^n\ell_im_i
=2\sum_{i=1}^n\frac{1}{2}l_im_i
\le
\frac{1}{2}\sum_{i=1}^n(k-1)m_i+\frac{1}{2}\sum_{i=1}^n\ell_ik
\le
\frac{1}{2}k\sum_{i=1}^n(\ell_i+m_i)-\frac{1}{2}\sum_{i=1}^nm_i
\end{displaymath}
which implies (\ref{subcrit}).
Item~\ref{i:lem1-2}: note that
\begin{displaymath}
\begin{array}{rcl}
{\displaystyle
 \Big(\frac{1}{2}\sum_{i=1}^n(l_i+m_i)\Big)^2-\sum_i^nl_i\,m_i}
&\!\!\!=\!\!\!&
{\displaystyle
\frac{1}{4}
\Big(\Big(\sum_{i=1}^n(l_i+m_i)\Big)^2-4\sum_i^nl_i\,m_i\Big)}\\
&\!\!\!=\!\!\!&
{\displaystyle
\frac{1}{4}
\Big(\sum_i^n(l_i+m_i)^2-4\sum_{i=1}^n l_i\,m_i
     +\sum_{i\neq j}(l_i+m_i)(l_j+m_j)\Big)}\\
&\!\!\!=\!\!\!&
{\displaystyle
\frac{1}{4}
\Big(\sum_{i=1}^n(l_i-m_i)^2+\sum_{i\neq j}(l_i+m_i)(l_j+m_j)\Big)}\\
&\!\!\!\ge\!\!\!&
{\displaystyle
\frac{4}{4}\sum_{j=1}^n(l_j+m_j)=p}
\end{array}
\end{displaymath}
where in the second step we used $n\ge2$
and in the last step $l_i\wedge m_i\ge2$.
The bound (\ref{differenza}) follows.
\qed

We introduce the notion of semi--perimeter of a multi--rectangular
droplet. Let $n\ge1$ and
$\ell_1,m_1,\dots,\ell_n,m_n$ integers such that
$2\le\ell_1,m_1,\dots,\ell_n,m_n\le L-2$,
$\sigma\in\cc{S}$ a $n$--rectangular droplet with sides
$\ell_1,m_1,\dots,\ell_n,m_n$, we let
\begin{equation}
\label{semiperimetro}
p(\sigma):=\sum_{i=1}^n(\ell_i+m_i)
\end{equation}
be the \textit{semi--perimeter} of the multi--rectangular droplet $\sigma$.

\begin{lemma}
\label{t:retto-ener}
Let $\ell,m$ two integers such that $2\le\ell\le m\le L-2$ and
$\sigma\in\cc{S}$ a rectangular droplet with sides $\ell$ and $m$.
If $\ell\le\lambda-1$, we have
\begin{equation}
\label{retsub}
E(\sigma)-E(\muno)>8\ell>0
\end{equation}
If $\ell\le\lambda-1$ and $m\ge\lambda+1$,
we have
\begin{equation}
\label{enesub}
E(\sigma)-E(\cc{R})\ge4h(1-\delta_h)>0
\end{equation}
where we recall $\cc{R}$ has been defined above Proposition~\ref{t:insiemeG}
and $\delta_h$ below (\ref{lcritica}).

Moreover, for $n\ge1$ integer, for any $n$--rectangular droplet $\eta\in\cc{S}$
with sides $2\le\ell_i\le m_i$ such that
$\ell_i\le\lambda-1$ and $m_i\le\lambda$ for $i=1,\dots,n$,
we have that
\begin{equation}
\label{enesemi}
E(\eta)-E(\muno)
>(4-2h)p(\eta)+\frac{1}{2}\sum_{i=1}^nm_i
\end{equation}
\end{lemma}

\smallskip
\par\noindent
\textit{Proof of Lemma~\ref{t:retto-ener}.\/}
Suppose $\ell\le\lambda-1$:
by using (\ref{energiaret}) we have
$E(\sigma)-E(\muno)=-4h\ell m+8(\ell+m)=(8-4h \ell)+8\ell$;
since $\ell\le\lambda-1$, the lemma follows.
Suppose $\ell\le\lambda-1$ and $m\ge\lambda+1$, by using
(\ref{energiaret}) we have
$E(\sigma)-E(\cc{R})=4h(m-\lambda)[(\lambda-\ell)-\delta_h]$,
which implies (\ref{enesub}).

We finally prove (\ref{enesemi}).
Recall that by hypothesis $\ell_i\le\lambda-1$ and
$m_i\le\lambda$ for any $i=1,\dots,n$; 
by definition of multi--rectangular droplets and by using (\ref{subcrit}) 
with $k=\lambda$, we have 
\begin{equation}
\label{subcrit-c}
|\Lambda^+(\eta)|
\le
\frac{\lambda}{2}\,p(\eta)-\frac{1}{2}\sum_{i=1}^nm_i
\end{equation}
Now, by using (\ref{energiaret}), (\ref{semiperimetro}), (\ref{subcrit}),
and the fact that the
support of a multi--rectangular droplet is made of pairwise not interacting
rectangles, we have that
\begin{displaymath}
E(\eta)-E(\muno)=-4h|\Lambda^+(\eta)|
+8p(\eta)\ge p(\eta)(8-2h\lambda)+\frac{1}{2}\sum_{i=1}^nm_i
\end{displaymath}
which implies (\ref{enesemi}) since
$\lambda<(2/h)+1$.
\qed

\subsec{Relations between configurations in $\cc{G}$ and in 
$\cc{G}^\rr{c}$}{s:rectdrop1}
\par\noindent
Consider $\sigma\in\cc{G}$ and $\eta\in\cc{G}^\rr{c}$, in 
Lemma~\ref{t:basin} we state a property relating the pluses in 
$\eta$ to those in $\ol{B}\,\ol{A}\sigma$ and we bound from below the 
transition rate $\Delta(\sigma,\eta)$ (see(\ref{defdelta})).

\begin{lemma}
\label{t:basin}
Let $\sigma\in\cc{G}$ and $\eta\notin\cc{G}$,
\begin{enumerate}
\item
\label{i:basin-1}
we have
\begin{equation}\label{basin1}
|\Lambda^+(\eta)\setminus\Lambda^+(\ol{B}\,\ol{A}(\sigma))|\geq 2
\end{equation}
\item
\label{i:basin-2}
we have
\begin{equation}\label{basin2}
\Delta(\sigma,\eta)\geq\left\{
\begin{array}{l}
12-4h \,\,\,\,\,\textnormal{for} \,\,\,L(\sigma)=0\\
4-4h \,\,\,\,\,\textnormal{for} \,\,\,L(\sigma)=1
\end{array}
\right.
\end{equation}
\end{enumerate}
\end{lemma}

\smallskip
\par\noindent
\textit{Proof of Lemma~\ref{t:basin}.\/}
Item~\ref{i:basin-1}:
the item follows from the definition of the subcritical set $\cc{G}$.
Indeed, if $|\Lambda^+(\eta)\setminus\Lambda^+(\ol{B}\,\ol{A}(\sigma))|\leq1$,
we have that under the map $A$
the positive spin outside $\Lambda^+(\ol{B}\,\ol{A}\sigma)$ is flipped, so that
$\Lambda^+(\ol{B}\,\ol{A}\eta)\subseteq \Lambda^+(\ol{B}\,\ol{A}\sigma)$. 
Hence $\eta\in\cc{G}$, that is a contradiction.

Item~\ref{i:basin-2}: from (\ref{defdelta}) we get
\begin{equation}
\label{delta3}
\Delta(\sigma,\eta)=
2\sum_{\newatop{z\in\Lambda:}
               {\eta(z)\,S_\sigma(z)<0}}
       |S_\sigma(z)+h|\geq
2\sum_{\newatop{z\in\Lambda\setminus \Lambda^+(\ol{B}\,\ol{A}\sigma):}
       {\eta(z)\,S_\sigma(z)<0}}
       |S_\sigma(z)+h|
\end{equation}
If $L(\sigma)=0$, by (\ref{basin1}),(\ref{delta3}), the theorem
follows. Indeed, in the r.h.s of (\ref{delta3}) there are at least
two terms corresponding to sites $x$ and $y$ such that
$\eta(x)=\eta(y)=1$, and $S_\sigma(x)\leq -3$,\,$S_\sigma(y)\leq-3$.
If $L(\sigma)=1$, from (\ref{basin1}) there exist two sites
\begin{equation}
\{x,y\}\subseteq \Lambda^+(\eta)\setminus \Lambda^+(\ol{B}\,\ol{A}\sigma)
\end{equation}

Note that, since $L(\sigma)=1$,
one has $S_\sigma(x)\leq -1$ and  $S_\sigma(y)\leq -1$. 
From (\ref{delta3}) we have the bound
\begin{equation}
\Delta(\sigma,\eta)\geq 2(1-h)+2(1-h)
\end{equation}
and the theorem follows, see also Figure~\ref{f:tabella}.
\qed

\subsec{Proof of the Proposition~\ref{t:insiemeG}}{s:proof}
\par\noindent
Let $\sigma\in\cc{S}$ and suppose
$\ol{B}\,\ol{A}\sigma\neq\muno$,
there exist $n(\sigma)\in\bb{N}\setminus\{0\}$,
$\ell_i(\sigma),m_i(\sigma)$ integers larger than $2$,
and $x_i(\sigma)\in\Lambda$ for $i=1,\dots,n(\sigma)$ such that
\begin{displaymath}
\Lambda^+(\ol{B}\,\ol{A}\sigma)=\bigcup_{i=1}^{n(\sigma)}
 Q_{\ell_i(\sigma),m_i(\sigma)}(x_i(\sigma))
\end{displaymath}
If $\ol{B}\,\ol{A}\sigma=\muno$ we shall understand
$n(\sigma)=1$, $\ell_1(\sigma)=m_1(\sigma)=0$, and $p(\sigma)=0$,
see also (\ref{semiperimetro}).
Let $\sigma\in\cc{S}$,
we order the droplets in $\Lambda^+(\ol{B}\,\ol{A}\sigma)$ so that
$\ell_i(\sigma)\wedge m_i(\sigma)\ge\lambda$ for $i=1,\dots,k(\sigma)$ and
$\ell_i(\sigma)\wedge m_i(\sigma)\le\lambda-1$ for
$i=k(\sigma)+1,\dots,n(\sigma)$;
note that for $\sigma\in\cc{G}$ we have $k(\sigma)=0$, while for
$\sigma\in\cc{G}^\rr{c}$ we have $k(\sigma)\ge1$.
For the sake of simplicity, for $\sigma\in\cc{G}^\rr{c}$ in the sequel
we shall let
$r_i(\sigma):=\ell_i(\sigma)-\lambda$ and
$q_i(\sigma):=m_i(\sigma)-\lambda$ for $i=1,\dots,k(\sigma)$.

Before starting the proof of the Proposition~\ref{t:insiemeG} we sketch
the main idea.
We shall define the subsets of the configuration space
$\cc{A}_5\subset\cc{A}_4\subset\cc{A}_3\subset\cc{A}_2\subset
 \cc{A}_1\subset\cc{G}$,
$\cc{B}_2\subset\cc{B}_1\subset\cc{G}^\rr{c}$, and reduce the proof
to the computation of $E(\eta,\zeta)$ for $\eta\in\cc{A}_5$ and
$\zeta\in\cc{B}_2$ (see Figure~\ref{f:strategia}).
We recall (\ref{semiperimetro}), (\ref{branch}),
(\ref{ciccio}), and let
\begin{equation}
\label{foliazione}
\begin{array}{ll}
\cc{A}_1:=\{\sigma\in\cc{G}:\,\ell_i(\sigma)\vee m_i(\sigma)\le\lambda
            \textrm{ for } i=1,\dots,n(\sigma)\}
&
\!\!\cc{A}_4:=\{\sigma\in\cc{A}_3:\,n(\sigma)=1\}
\\
\cc{A}_2:=\{\sigma\in\cc{A}_1:\,p(\sigma)\le2\lambda+4,
                              \,L(\sigma)\le4\lambda+42\}
&
\!\!\cc{A}_5:=\{\sigma\in\cc{A}_4:\,p(\sigma)=2\lambda-1\}
\\
\cc{A}_3:=\{\sigma\in\cc{A}_2:\,p(\sigma)\ge2\lambda-50\}
&
\\
\end{array}
\end{equation}
and
\begin{equation}
\label{foliazione2}
\begin{array}{l}
\cc{B}_1:=\{\sigma\in\cc{G}^\rr{c}:\,
\ell_i(\sigma),m_i(\sigma)\le L-2\textrm{ for }i=1,\dots,n(\sigma)\}
\\
{\displaystyle
\cc{B}_2:=\{\sigma\in\cc{B}_1:\,
4hN_B(\sigma)-4h\sum_{i=1}^{k(\sigma)}(r_i(\sigma)+q_i(\sigma)
         +r_i(\sigma)q_i(\sigma))\le10-2h\}
}
\end{array}
\end{equation}
In order to bound $E(\eta,\zeta)$ for $\eta\in\cc{G}$ and 
$\zeta\in\cc{G}^\rr{c}$, we shall use the identity
\begin{equation}
\label{fabriano00}
\begin{array}{rcl}
E(\eta,\zeta)-E(\muno)
&\!=\!&
[E(\eta)-E(\ol{A}\eta)]+[E(\ol{A}\eta)
        -E(\ol{B}\,\ol{A}\eta)]\vphantom{\Big\}}\\
&&+[E(\ol{B}\,\ol{A}\eta)-E(\muno)]+\Delta(\eta,\zeta)\\
\end{array}
\end{equation}
which is a straightforward consequence of the definition (\ref{defdelta}).
Depending on the choice of $\eta$, the different terms in the r.h.s.\ of 
the identity (\ref{fabriano00}) will be properly bounded in order to get 
the theorem. 

\smallskip
\par\noindent
\textit{Proof of Proposition~\ref{t:insiemeG}.\/}
Item~\ref{i:insiemeG-1}.\ The proof is an immediate application of the
definition of the set $\cc{G}$ (see Section~\ref{s:variazionale}).

Items~\ref{i:insiemeG-2}.\ 
\textit{Step 1.\/}
Let $\eta\in\cc{G}\setminus\cc{A}_1$ and $\zeta\in\cc{G}^\rr{c}$.
There exists $i\in\{1,\dots,n(\eta)\}$ such that
$l_i(\eta)\vee m_i(\eta)\ge\lambda+1$; hence, by using 
(\ref{fabriano00}), (\ref{boots}), $N_B(\eta)\ge0$,
(\ref{retsub}), and (\ref{enesub}), we get 
\begin{equation}
\label{fabriano01}
E(\eta,\zeta)-E(\muno)
\ge
[E(\eta)-E(\ol{A}\eta)]+
[E(\cc{R})-E(\muno)]+\Delta(\eta,\zeta)
\end{equation}
where (\ref{retsub}) and (\ref{enesub}) have been applied to each 
non--interacting droplet in $\ol{B}\,\ol{A}\sigma$ to deduce that
$E(\ol{B}\,\ol{A}\sigma)-E(\cc{R})\ge0$.
Now, if $L(\eta)=0$, by using (\ref{fabriano01}), (\ref{filam}), 
$N_A(\eta)\ge0$, (\ref{basin2}), and (\ref{enerre}), we get 
\begin{displaymath}
E(\eta,\zeta)-E(\muno)>
[E(\cc{R})-E(\muno)]+12-4h
>\Gamma
\end{displaymath}
On the other hand, if $L(\eta)\ge1$, by using (\ref{fabriano01}), 
(\ref{svolta}), (\ref{basin2}), and (\ref{enerre}), we get 
\begin{displaymath}
E(\eta,\zeta)-E(\muno)>
6-2h+[E(\cc{R})-E(\muno)]+4-4h
>\Gamma
\end{displaymath}

\smallskip
\par\noindent
\textit{Step 2.\/}
Let $\eta\in\cc{A}_1\setminus\cc{A}_2$ and $\zeta\in\cc{G}^\rr{c}$.
By using (\ref{fabriano00}), (\ref{boots}), and $N_B(\eta)\ge0$, we get 
\begin{equation}
\label{fabriano02}
E(\eta,\zeta)-E(\muno)
\ge
[E(\eta)-E(\ol{A}\eta)]+
[E(\ol{B}\,\ol{A}\eta)-E(\muno)]+\Delta(\eta,\zeta)
\end{equation}
Now, suppose $p(\eta)\ge2\lambda+5$,
by using (\ref{fabriano02}), (\ref{enesemi}), $\Delta(\eta,\zeta)\ge0$, 
the definition (\ref{lcritica}), and (\ref{stimagamma}), we get 
\begin{displaymath}
E(\eta,\zeta)-E(\muno)>(4-2h)(2\lambda+5)>8\lambda+12-14h>\Gamma
\end{displaymath}
provided $h>0$ is chosen smaller than $1/6$.
Suppose, finally, $L(\eta)\ge4\lambda+43$.
If $\ol{B}\,\ol{A}\eta\neq\muno$, by using (\ref{retsub}) we get
$E(\ol{B}\,\ol{A}\eta)-E(\muno)\ge0$; note that this bound holds trivially also 
in the case $\ol{B}\,\ol{A}\eta=\muno$. Hence, 
by using this bound, (\ref{fabriano02}), (\ref{svolta}), and 
(\ref{stimagamma}), we get 
\begin{displaymath}
E(\eta,\zeta)-E(\muno)>10-6h+(2-10h)(4\lambda+43)>\Gamma
\end{displaymath}

\smallskip
\par\noindent
\textit{Step 3.\/}
Let $\eta\in\cc{A}_2$ and $\zeta\in\cc{G}^\rr{c}\setminus\cc{B}_1$.
There exists $i\in\{1,\dots,k(\zeta)\}$ such that
$\ell_i(\zeta)\vee m_i(\zeta)> L-2$.
Since $\eta\in\cc{A}_2$ we have that $p(\eta)\le2\lambda+4$ and
$L(\eta)\le4\lambda+42$, then by using (\ref{subcrit}) with $k=\lambda$ we have
$|\Lambda^+(\eta)|
\le|\Lambda^+(\ol{B}\,\ol{A}\eta)|+L(\eta)
\le\lambda p(\eta)/2+L(\eta)\le\lambda^2+6\lambda+42$.
Given the magnetic field $h>0$, the number of plus spins in $\eta$ is
bounded by a finite number; then we can choose $L=L(h)$ so large that
there exist an horizontal and a vertical stripe winding around the
torus with arbitrarily large width and such that $\eta(x)$ is
equal to $-1$ for each $x$ in such two stripes. Since in $\ol{B}\,\ol{A}\zeta$,
there exists a rectangular droplet of pluses with one of the two side lengths
larger or equal to $L-2$; we choose $L$ so large that
$\Delta(\eta,\zeta)>\Gamma$. By using, finally, (\ref{defdelta}) we get
$E(\eta,\zeta)-E(\muno)>\Gamma$, once we remark that
$E(\eta)-E(\muno)\ge E(\ol{B}\,\ol{A}\eta)-E(\muno)\ge0$.

\smallskip
\par\noindent
\textit{Step 4.\/}
Let $\eta\in\cc{A}_2$ and $\zeta\in\cc{B}_1\setminus\cc{B}_2$.
By using Lemma~\ref{t:boots} and $N_A(\zeta)\ge0$, we have the bound
\begin{equation}
\label{boot3}
E(\zeta)-E(\muno)\geq E(\ol{B}\,\ol{A}\zeta)-E(\muno)+4hN_B(\zeta)
\end{equation}
By (\ref{energiaret}) and (\ref{prot-h}) it follows
\begin{eqnarray}
\label{cin}
E(\ol{B}\,\ol{A}\zeta)-E(\muno)
&=&
-4h\sum_{i=1}^{n(\zeta)}(\lambda+r_i(\zeta))(\lambda+q_i(\zeta))
+8\sum_{i=1}^{n(\zeta)}(2\lambda+r_i(\zeta)+q_i(\zeta))
\nonumber\\
&=&
n(\zeta)(\Gamma-10+6h)
 -\sum_{i=1}^{n(\zeta)}(r_i(\zeta)+q_i(\zeta))(4h\lambda-8)
\nonumber\\
&\phantom{a}&
-4n(\zeta)(h\lambda-2)-4h\sum_{i=1}^{n(\zeta)}r_i(\zeta)q_i(\zeta)
\nonumber\\
&>&
(\Gamma-10+2h)-4h\sum_{i=1}^{n(\zeta)}(r_i(\zeta)+q_i(\zeta)
 +r_i(\zeta)q_i(\zeta))
\end{eqnarray}
where in the last inequality we used (\ref{lcritica}) and the fact
that $\Gamma>10-6h$. Hence, by (\ref{boot3}) and (\ref{cin}), 
we have
\begin{displaymath}
E(\zeta)-E(\muno)\geq
 \Gamma-(10-2h)
 +4hN_B(\zeta)-4h\sum_{i=1}^{n(\zeta)}
  (r_i(\zeta)+q_i(\zeta)+r_i(\zeta)q_i(\zeta))
\end{displaymath}
Since $\zeta\in\cc{B}_2\setminus\cc{B}_1$, we get
$E(\zeta)-E(\muno)>\Gamma$. Finally, by the inequality in 
(\ref{sim-ene}), we get $E(\eta,\zeta)-E(\muno)>\Gamma$.

\smallskip
\par\noindent
\textit{Step 5.\/}
Let $\eta\in\cc{A}_2\setminus\cc{A}_3$ and $\zeta\in\cc{B}_2$.
We note now that $E(\ol{B}\,\ol{A}\eta)-E(\muno)\ge0$, which is trivial
if $\ol{B}\,\ol{A}\eta=\muno$, otherwise it follows immediately
from (\ref{retsub}).
By using this bound, (\ref{fabriano00}), Lemma~\ref{t:boots}, $N_A(\eta)\ge0$,
and $N_B(\eta)\ge0$, we get 
\begin{equation}
\label{fabriano04}
E(\eta,\zeta)-E(\muno)
\ge
\Delta(\eta,\zeta)
\end{equation}

We find, now, a lower bound to $\Delta(\eta,\zeta)$ by multiplying the
the minimum \emph{quantum} $2(1-h)$, see Figure~\ref{f:tabella},
times the number of flips against the drift in the transition from
$\eta$ to $\zeta$. More precisely,
\begin{eqnarray}
\label{delta}
\Delta(\eta,\zeta)
&\ge&
2(1-h)\,|\{x\in\Lambda:\,\eta(x)S_\eta(x)>0,
                       \,\eta(x)\zeta(x)<0\}|
\nonumber\\
&\ge&
2(1-h)(|\Lambda^+(\zeta)|-|\bar{\Lambda}(\eta)|)
\end{eqnarray}
with
\begin{equation}
\label{dank}
\bar{\Lambda}(\eta):=\Lambda^+(\ol{B}\,\ol{A}\eta)\cup
 \overline{\Lambda^+(\eta)\setminus\Lambda^+(\ol{B}\,\ol{A}\eta)}
\end{equation}
where we recall the closure of a subset of the lattice has been defined
in Section~\ref{s:lat}.

Recalling that the application of the map $A$ does not add pluses,
the number of plus spins in the configuration $\zeta$ can be bounded
from below by adding the number of pluses in $\ol{B}\,\ol{A}\zeta$ to the
branch $L(\zeta)$ of $\zeta$ and subtracting the number of pluses $N_B(\zeta)$
added by the bootstrap map $B$. Namely, we have
\begin{displaymath}
|\Lambda^+(\zeta)|
\ge
n(\zeta)\lambda^2+\lambda\sum_{i=1}^{n(\zeta)}(r_i(\zeta)+q_i(\zeta))
+\sum_{i=1}^{n(\zeta)} r_i(\zeta)q_i(\zeta)-N_B(\zeta)+L(\zeta)
\end{displaymath}
Now, by using that $\zeta\in\cc{B}_2$, (\ref{lcritica}), and $L(\zeta)\ge0$,
we get
\begin{eqnarray}
\label{occa}
|\Lambda^+(\zeta)|
&\ge&
\lambda^2+\sum_{i=1}^{n(\zeta)}(\lambda(r_i(\zeta)+q_i(\zeta))
-r_i(\zeta)-q_i(\zeta))
-\frac{10-2h}{4h}+L(\zeta)
\nonumber\\
&\ge&
\lambda^2-\frac{5}{4}\lambda
+\sum_{i=1}^{n(\zeta)}(\lambda-1)(r_i(\zeta)+q_i(\zeta))
\ge
\lambda^2-\frac{5}{4}\lambda
\end{eqnarray}
where we also used $\lambda-1\ge0$.

We next bound from above $|\bar\Lambda(\eta)|$. We first note that
by using (\ref{dank}) and (\ref{branch}) we get
\begin{equation}
\label{zzo}
|\bar{\Lambda}(\eta)|\le|\Lambda^+(\ol{B}\,\ol{A}\eta)|+5L(\eta)
\end{equation}
Now, suppose that $\Lambda^+(\ol{B}\,\ol{A}\eta)\neq\muno$; by using
(\ref{subcrit}) with $k=\lambda$ and exploiting $\eta\in\cc{A}_2$, we conclude
\begin{equation}
\label{pic}
|\bar{\Lambda}(\eta)|\le
\frac{1}{2}\lambda p(\eta)+20\lambda+210
\end{equation}
Suppose, on the other hand, that $\Lambda^+(\ol{B}\,\ol{A}\eta)=\muno$.
By using (\ref{zzo}), we get
$|\bar{\Lambda}(\eta)|\le5L(\eta)\le20\lambda+210$; hence
the bound (\ref{pic}) holds since in this case $p(\eta)=0$.

We finally bound $\Delta(\eta,\zeta)$ by using the
preliminary inequalities (\ref{delta}), (\ref{occa}), and (\ref{pic}); we
have
\begin{equation}
\label{cione}
\Delta(\eta,\zeta)\ge
2(1-h)\Big[
\lambda^2-\frac{85}{4}\lambda-\frac{1}{2}\lambda p(\eta)-210\Big]
\end{equation}

Recall $\eta\in\cc{A}_2\setminus\cc{A}_3$, then $p(\eta)\le2\lambda-51$;
hence by using (\ref{fabriano04}), (\ref{cione}), and (\ref{stimagamma}),
we get
\begin{displaymath}
E(\eta,\zeta)-E(\muno)>\Gamma+\frac{1}{h}-\frac{53}{2}+O(h)>\Gamma
\end{displaymath}
where in the last inequality we have chosen
$h>0$ small enough.

\smallskip
\par\noindent
\textit{Step 6.\/}
Let $\eta\in\cc{A}_3\setminus\cc{A}_4$ and $\zeta\in\cc{B}_2$.
By using (\ref{fabriano00}), Lemma~\ref{t:boots}, $N_A(\eta)\ge0$,
$N_B(\eta)\ge0$, (\ref{energiaret}), and $\Delta(\eta,\zeta)\ge0$, we get
\begin{equation}
\label{fabriano05}
E(\eta,\zeta)-E(\muno)
\ge
E(\ol{B}\,\ol{A}\eta)-E(\muno)=-4h|\Lambda^+(\ol{B}\,\ol{A}\eta)|+p(\eta)
\end{equation}
Now, since $\eta\in\cc{A}_3\setminus\cc{A}_4$, we can use (\ref{differenza})
to obtain
\begin{equation}
\label{llo}
E(\eta,\zeta)-E(\muno)
\ge-h(p(\eta))^2+(4h+8)p(\eta)
\end{equation}
Finally, by exploiting the properties of the parabola on the right--hand
side of (\ref{llo}) and recalling that for
$\eta\in\cc{A}_3\setminus\cc{A}_4$ the semi--perimeter satisfies
the bounds
$2\lambda-50\le p(\eta)\le2\lambda+4$, it is immediate to prove that
the parabola attains its minimum at $p(\eta)=2\lambda-50$; hence,
by using (\ref{llo}) and (\ref{prot-h}),
we get $E(\eta,\zeta)-E(\muno)>\Gamma$ for $h>0$
small enough.

\smallskip
\par\noindent
\textit{Step 7.\/}
Let $\eta\in\cc{A}_4\setminus\cc{A}_5$ and $\zeta\in\cc{B}_2$.
By using 
(\ref{fabriano00}), (\ref{boots}), and $N_B(\eta)\ge0$,
we get the bound
\begin{equation}
\label{fabriano06}
E(\eta,\zeta)-E(\muno)
\ge
[E(\eta)-E(\ol{A}\eta)]+
[E(\ol{B}\,\ol{A}\eta)-E(\muno)]+\Delta(\eta,\zeta)
\end{equation}
Since $\eta\in\cc{A}_4\setminus\cc{A}_5$, we have that $n(\eta)=1$
and then $2\lambda-50\le p(\eta)\le2\lambda-2$.
We repeat, now, the same argument used at Step~6, but,
since $n(\eta)=1$, we have to use
(\ref{mammt}) instead of (\ref{differenza}); we then get
\begin{equation}
\label{fasc}
E(\eta,\zeta)-E(\muno)
\ge
[E(\eta)-E(\ol{A}\eta)]+[\Gamma-10+O(h)]+\Delta(\eta,\zeta)
\end{equation}

Moreover, since 
$n(\eta)=1$ and $p(\eta)\le2\lambda-2$, by using the same arguments
developed in the proof of (\ref{basin1}), we get
\begin{equation}
\label{le-chnot}
|\Lambda^+(\zeta)\setminus\Lambda^+(\ol{B}\,\ol{A}\eta)|\ge3
\end{equation}

To complete the proof of the Step~7, we distinguish four cases by means of
the parameter $L(\eta)$. 
Consider, first, the case $L(\eta)\ge3$; by using 
(\ref{fasc}), (\ref{svolta}), and $\Delta(\eta,\zeta)\ge0$, it follows
immediately $E(\eta,\zeta)-E(\muno)>\Gamma$.

Consider, now, the case $L(\eta)=2$.
We first note that by using (\ref{fasc}) and (\ref{svolta}) we get 
\begin{equation}
\label{fasc01}
E(\eta,\zeta)-E(\muno)
\ge
10-6h+[\Gamma-10+O(h)]+\Delta(\eta,\zeta)
\ge
\Gamma+\Delta(\eta,\zeta)+O(h)
\end{equation}
The result $E(\eta,\zeta)-E(\muno)>\Gamma$ will then be proven once we shall
have obtained the bound $\Delta(\eta,\zeta)\ge2(1-h)$.

To prove such a bound, we note that 
there exist $x,y\in\Lambda^+(\eta)\setminus\Lambda^+(\ol{B}\,\ol{A}\eta)$;
since $x,y\in\Lambda\setminus\Lambda^+(\ol{B}\,\ol{A}\eta)$, by the
definition of the two maps $A$ and $B$, it follows that they cannot
be both stable w.r.t.\ $\eta$ (see Section~\ref{s:moddep}).
If one of the two sites $x$ and $y$, say $x$, is stable w.r.t.\ $\eta$, it is
immediate to prove that $x\in\partial\Lambda^+(\ol{B}\,\ol{A}\eta)$ and
$\{y\}=\partial\{x\}\setminus\Lambda^+(\ol{B}\,\ol{A}\eta)$.
Since $x$ and $y$ are nearest neighbors, it follows that there exist
no site in $\Lambda^-(\eta)\setminus\Lambda^+(\ol{B}\,\ol{A}\eta)$ which
is unstable w.r.t.\ $\eta$; hence, by using (\ref{le-chnot}) and 
(\ref{defdelta}), it follows that $\Delta(\eta,\zeta)\ge2(1-h)$. 
We consider, now, the case when both $x$ and $y$
are unstable w.r.t.\ $\eta$. 
Suppose that either $\zeta(x)=+1$ or $\zeta(y)=+1$;
from (\ref{defdelta}) we have $\Delta(\eta,\zeta)\geq2(1-h)$. 
On the other hand,
if $\zeta(x)=\zeta(y)=-1$, it is easy to see that, since $L(\eta)=2$, we have
$|\Lambda^-_\rr{u}(\eta)\setminus \Lambda^+(\ol{B}\,\ol{A}\eta)|\leq 1$ 
(recall the definition (\ref{lambdi-us})). Then, by using 
(\ref{le-chnot}), it follows that $\Delta\geq2(1-h)$. 

Consider, now, the case $L(\eta)=1$. 
We first note that, by using (\ref{fasc}) and (\ref{svolta}), we get 
\begin{equation}
\label{fasc02}
E(\eta,\zeta)-E(\muno)
\ge
6-2h+[\Gamma-10+O(h)]+\Delta(\eta,\zeta)
\ge
\Gamma-4+\Delta(\eta,\zeta)+O(h)
\end{equation}
The result $E(\eta,\zeta)-E(\muno)>\Gamma$ will then be proven once we shall
have obtained the bound $\Delta(\eta,\zeta)\ge3\cdot2(1-h)$.

To prove such a bound we let $x$ the site such that
$\{x\}:=\Lambda^+(\eta)\setminus\Lambda^+(\ol{B}\,\ol{A}\eta)$.  
Suppose $\zeta(x)=+1$; since $x$ is unstable w.r.t.\ $\eta$, by 
(\ref{defdelta}) and
(\ref{le-chnot}), we have $\Delta(\eta,\zeta)\geq 2(1-h)+2(1-h)+2(1-h)$. 
On the other hand, suppose $\zeta(x)=-1$; 
since $|\Lambda^-_\rr{u}(\eta)\setminus\Lambda^+(\ol{B}\,\ol{A}\eta)|=0$,
by (\ref{le-chnot}) it follows that
$\Delta(\eta,\zeta)\geq 2(1-h)+2(1-h)+2(3-h)$. 

Consider, finally, the case $L(\eta)=0$. 
Recall (\ref{defdelta}); the condition (\ref{le-chnot}) implies 
that $\Delta(\eta,\zeta)\ge3\cdot2(3-h)$. Hence, by using also (\ref{fasc}),
(\ref{filam}), and $N_A(\eta)\ge0$, we get
\begin{displaymath}
E(\eta,\zeta)-E(\muno)\ge\Gamma-10+O(h)+3\cdot 2(3-h)>\Gamma
\end{displaymath}

\smallskip
\par\noindent
\textit{Step 8.\/}
Let $\eta\in\cc{A}_5$ and $\zeta\in\cc{B}_2$. We remark that, since 
$p(\eta)=2\lambda-1$ and $\ell_1\vee m_1\leq\lambda$, we have 
$\ol{B}\,\ol{A}\eta\in\cc{R}$. Hence, by using (\ref{fabriano00}),
(\ref{boots}), and (\ref{enerre}), we get 
\begin{equation}
\label{fabriano}
E(\eta,\zeta)-E(\muno)
\ge
[E(\eta)-E(\ol{A}\eta)]+4hN_B(\eta)
+[\Gamma-10+6h]+\Delta(\eta,\zeta)
\end{equation}

To complete the proof of the Step~8, we distinguish four cases by means of
the parameter $L(\eta)$. 
Consider, first, the case $L(\eta)\ge3$; 
by using (\ref{fabriano}),
(\ref{svolta}), $N_B(\eta)\ge0$, and $\Delta(\eta,\zeta)\ge0$, we get 
\begin{displaymath}
E(\eta,\zeta)-E(\muno)\ge
12-16h+\Gamma-10+6h>\Gamma
\end{displaymath}

Consider, now, the case $L(\eta)=2$; we let $x,y$ be the two sites in 
$\Lambda^+(\eta)\setminus\Lambda^+(\ol{B}\,\ol{A}\eta)$. We first note that,
by using the inequalities (\ref{fabriano}) and (\ref{svolta}), we get the bound 
\begin{equation}
\label{fabriano08}
E(\eta,\zeta)-E(\muno)
\ge
4hN_B(\eta) +\Gamma+\Delta(\eta,\zeta)
\end{equation}
Suppose, first, $N_B(\eta)\ge1$; by (\ref{fabriano08}) and 
$\Delta(\eta,\zeta)\ge0$, we immediately get $E(\eta,\zeta)-E(\muno)>\Gamma$. 
We are then left with the case 
$N_B(\eta)=0$, i.e., $\Lambda^+(\eta)\supset\Lambda^+(\ol{B}\,\ol{A}\eta)$;
by using (\ref{fabriano08}),
the result $E(\eta,\zeta)-E(\muno)>\Gamma$ will be proven once we shall
have obtained the bound $\Delta(\eta,\zeta)\ge2(1-h)$.

We note that 
$|\Lambda^+_\rr{s}(\eta)\setminus\Lambda^+(\ol{B}\,\ol{A}\eta)|\le1$,
indeed if by the way of contradiction $x$ and $y$ belonged both to 
$\Lambda^+_\rr{s}(\eta)\setminus\Lambda^+(\ol{B}\,\ol{A}\eta)$, then it should 
necessarily be $x,y\in\partial\Lambda^+(\ol{B}\,\ol{A}\eta)$ and $\dis(x,y)=1$, 
namely, there would be a two--site protuberance added to the 
$\lambda\times(\lambda-1)$ rectangle of pluses which is present in $\eta$. 
Hence, we would have $\eta\in\cc{C}\subset\cc{G}^\rr{c}$, which is a
contradiction.

Suppose $|\Lambda^+_\rr{s}(\eta)\setminus\Lambda^+(\ol{B}\,\ol{A}\eta)|=1$ and 
let $x$ be the site in 
$\Lambda^+_\rr{s}(\eta)\setminus\Lambda^+(\ol{B}\,\ol{A}\eta)$;
since $x$ is stable w.r.t.\ $\eta$, we must necessarily have 
$x\in\partial\Lambda^+(\ol{B}\,\ol{A}\eta)$ and 
$y\in\partial\{x\}\setminus\overline{\Lambda^+(\ol{B}\,\ol{A}\eta)}$.
Note, also, that this implies
$|\Lambda^-_\rr{u}(\eta)\setminus\Lambda^+(\ol{B}\,\ol{A}\eta)|=0$.
Thus, for $\zeta(y)=+1$, in the sum in (\ref{defdelta}) there is at least 
the term corresponding to $y$; then we have $\Delta(\eta,\zeta)\ge2(1-h)$. 
On the other hand, if it were $\zeta(y)=-1$, recalling (\ref{basin1}) we 
would have 
that in the sum in (\ref{defdelta}) there is at least a term corresponding 
to the flip of the spin associated with a site in $\Lambda^-(\eta)$ which is 
stable w.r.t.\ $\eta$, hence we would have $\Delta(\eta,\zeta)\ge2(1-h)$. 
Suppose, finally, 
$|\Lambda^+_\rr{s}(\eta)\setminus\Lambda^+(\ol{B}\,\ol{A}\eta)|=0$;
it is immediate to prove that 
$|\Lambda^-_\rr{u}(\eta)\setminus\Lambda^+(\ol{B}\,\ol{A}\eta)|\le1$.
Then we get $\Delta(\eta,\zeta)\ge2(1-h)$, since 
from (\ref{basin1}) it follows that 
in the sum in (\ref{defdelta}) there is at least a term corresponding 
to the persistence of the spin associated with a site in 
$\Lambda^+_\rr{u}(\eta)$ or 
to the flip of the spin associated with a site in $\Lambda^-(\eta)$ which is 
stable w.r.t.\ $\eta$. 

Consider, now, the case $L(\eta)=1$. We let $x$ be the site in 
$\Lambda^+(\eta)\setminus\Lambda^+(\ol{B}\,\ol{A}\eta)$, note that $x$ in 
unstable w.r.t\ $\eta$ and $w$ is stable w.r.t.\ $\eta$ for any 
$w\in\Lambda^-(\eta)\setminus\Lambda^+(\ol{B}\,\ol{A}\eta)$. We remark that
by using (\ref{fabriano}), (\ref{svolta}), (\ref{boots}), and $N_B(\eta)\ge0$,
we get the bound 
\begin{equation}
\label{fabriano10}
E(\eta,\zeta)-E(\muno)
\ge
\Gamma-4+4h+\Delta(\eta,\zeta)
\end{equation}
and distinguish different cases depending on the number of plus spins in 
the configuration $\zeta$ which are associated to sites outside the 
support of the configuration $\ol{B}\,\ol{A}\eta$, that is on 
$|\Lambda^+(\zeta)\setminus\Lambda^+(\ol{B}\,\ol{A}\eta)|\ge2$, see 
(\ref{basin1}).

Suppose, first, 
$|\Lambda^+(\zeta)\setminus\Lambda^+(\ol{B}\,\ol{A}\eta)|\ge3$;
since $x\in\Lambda^+_\rr{u}(\eta)$ and $w$ is stable w.r.t.\ $\eta$ for any 
$w\in\Lambda^-(\eta)\setminus\Lambda^+(\ol{B}\,\ol{A}\eta)$, we have 
$\Delta(\eta,\zeta)\ge3\cdot2(1-h)$, for 
in the sum in (\ref{defdelta}) there are at least three terms.

We are left with the case 
$|\Lambda^+(\zeta)\setminus\Lambda^+(\ol{B}\,\ol{A}\eta)|=2$;
we let $\{y,z\}:=\Lambda^+(\zeta)\setminus\Lambda^+(\ol{B}\,\ol{A}\eta)$ and 
notice that it must be necessarily 
$y,z\in\partial\Lambda^+(\ol{B}\,\ol{A}\eta)$ and $\dis(y,z)=1$,
otherwise it would be $\zeta\in\cc{G}$. 
Suppose, first, $\zeta(x)=-1$; since $x\neq y$, $x\neq z$, and $y$ and $z$ are 
nearest neighbors, it follows that at most one of the two sites $y$ and $z$ is 
nearest neighbor of $x$. Then, since in the sum (\ref{defdelta}) there are at 
least two terms and one of them is greater or equal to $2(3-h)$, we have 
$\Delta(\eta,\zeta)\ge2(1-h)+2(3-h)$. 
By the previous inequality and (\ref{fabriano10}) we get
$E(\eta,\zeta)-E(\muno)>\Gamma$.
Suppose, finally, $\zeta(x)=+1$; without loss of generality we let
$y=x$. Since 
$z\in\partial\{x\}\cap\partial\Lambda^+(\ol{B}\,\ol{A}\eta)$, 
by (\ref{defdelta}) we have
$\Delta(\eta,\zeta)\ge2(1-h)+2(1-h)$,
with one of the two terms corresponding to the 
persistence of the plus spin associated to $x$ in $\eta$ and the other 
corresponding to the flip of the minus spin associated to $z$ in $\eta$.
By the previous inequality and (\ref{fabriano10}) we get
$E(\eta,\zeta)-E(\muno)\ge\Gamma$.

Consider, finally, the case $L(\eta)=0$. 
By using (\ref{fabriano}), (\ref{filam}), $N_A(\eta)\ge0$, 
$N_B(\eta)\ge0$, and (\ref{basin2}), we get 
\begin{displaymath}
E(\eta,\zeta)-E(\muno)\ge
\Gamma-10+6h+12-4h>\Gamma
\end{displaymath}

Item~\ref{i:insiemeG-3}.\
Suppose $\zeta\in\cc{C}$ and $\eta\in\pi(\zeta)$, by using 
(\ref{hl}) and (\ref{defdelta}) it follows $E(\eta,\zeta)-E(\muno)=\Gamma$.

Conversely, suppose 
$\eta\in\cc{G}$ and $\zeta\in\cc{G}^\rr{c}$ such that 
$E(\eta,\zeta)-E(\muno)=\Gamma$. By using the results in the proof 
of the Item~\ref{i:insiemeG-2} above, see in particular the Step~8,
we have that it must be necessarily 
$\eta\in\cc{A}_5$, $\zeta\in\cc{B}_2$, $L(\eta)=1$,
$|\Lambda^+(\zeta)\setminus\Lambda^+(\ol{B}\,\ol{A}\eta)|=2$, and 
$\zeta(x)=+1$, with $x$ such that 
$\{x\}=\Lambda^+(\eta)\setminus\Lambda^+(\ol{B}\,\ol{A}\eta)$,
indeed for any different choice of $\eta$ and $\zeta$ it has 
be proven $E(\eta,\zeta)-E(\muno)>\Gamma$.
For configurations $\eta$ and $\zeta$ as above we have also proven that 
$\ol{B}\,\ol{A}\eta\in\cc{R}$, that $\Delta(\eta,\zeta)\ge2\cdot2(1-h)$, and 
that there exists $z\in\partial\{x\}\cap\partial\Lambda^+(\ol{B}\,\ol{A}\eta)$ 
such that $\zeta(z)=+1$.

Now, by using $\ol{B}\,\ol{A}\eta\in\cc{R}$, 
$\Delta(\eta,\zeta)\ge2\cdot2(1-h)$, 
(\ref{fabriano00}), (\ref{svolta}), (\ref{boots}), and 
(\ref{enerre}), we get 
\begin{displaymath}
E(\eta,\zeta)-E(\muno)
\ge
6-2h+4hN_B(\eta)+\Gamma-10+6h+2\cdot2(1-h)
=
\Gamma+4hN_B(\eta)
\end{displaymath}
If it were $N_B(\eta)\ge1$ it would follow 
$E(\eta,\zeta)-E(\muno)>\Gamma$, then it must necessarily be 
$N_B(\eta)=0$.

By the above characterization of $\eta$ we have that $\eta\in\cc{P}$; then, 
by using (\ref{fabriano00}) and the definition of the map $A$ we get the 
following expression for the communication energy $E(\eta,\zeta)$:
\begin{displaymath}
E(\eta,\zeta)-E(\muno)
=
6-2h+\Gamma-10+6h+\Delta(\eta,\zeta)
=
\Gamma-4+4h+\Delta(\eta,\zeta)
\end{displaymath}
Since $\zeta(x)=\zeta(z)=+1$, we have that $\Delta(\eta,\zeta)=2\cdot2(1-h)$ if 
and only if $\zeta(w)=+1$ for all $w\in\Lambda^+(\ol{B}\,\ol{A}\eta)$. 
We then have that $\zeta\in\cc{C}$ and $\eta\in\pi(\zeta)$. 
\qed

\appendice{Review of results in \cite{[MNOS]}}{s:mnos}
\par\noindent
The proof of Theorem~\ref{t:meta} is based on 
general results in \cite[Theorem~4.1, 4.9, and 5.4]{[MNOS]}
concerning the hitting time on the 
set of global minima of the energy for the chain 
started at a metastable state. 
We restate those results in our framework which is slightly different 
from the one considered in that paper (see the discussion at the beginning of
Section~\ref{s:scappo}). 

Recall (\ref{pizza2}). 
Let $\cc{S}^\rr{s}$ be the set of
global minima of the energy (\ref{hl}).
For any $\sigma\in\cc{S}$, let
$\cc{I}_\sigma:=\{\eta\in\cc{S}:\,E(\eta)<E(\sigma)\}$ be the set of states 
with energy below $E(\sigma)$ and
$V_\sigma:=\Phi(\sigma,\cc{I}_\sigma)-E(\sigma)$ be the \textit{stability level
of} $\sigma$.
Set $V_\sigma:=\infty$ if $\cc{I}_\sigma=\emptyset$.
We define the set of \textit{metastable states}
$\cc{S}^\rr{m}:=\{\eta\in\cc{S}:\,V_\eta=
              \max_{\sigma\in\cc{S}\setminus\cc{S}^\rr{s}}V_\sigma\}$.
We say that 
$\cc{W}(\eta,\zeta)\subset\cc{S}$ is a \textit{gate}
for the transition from $\eta\in\cc{S}$ to 
$\zeta\in\cc{S}$ if and only if the two
following conditions are satisfied:
(1) for any $\sigma\in\cc{W}(\eta,\zeta)$ 
there exist a path $\omega\in\Theta(\eta,\zeta)$, such that 
$\Phi_\omega=\Phi(\eta,\zeta)$, and 
$i\in\{2,\dots,|\omega|\}$ such that $\omega_i=\sigma$ and
$E(\omega_{i-1},\omega_i)=\Phi(\eta,\zeta)$;
(2) $\omega\cap\cc{W}(\eta,\zeta)\neq\emptyset$
for any path $\omega=\Theta(\eta,\zeta)$ such that
$\Phi_\omega=\Phi(\eta,\zeta)$.
A function $f:\beta\in\bb{R}\to f(\beta)\in\bb{R}$ is called 
\textit{super--exponentially small} (SES) in the limit $\beta\to\infty$ 
if and only if $\lim_{\beta\to\infty}(1/\beta)\log f(\beta)=-\infty$. 
Given $\sigma\in\cc{S}$ and $A\subset\cc{S}$, finally,
recall the definition of hitting time $\tau^\sigma_A$
given in (\ref{hitting}) and the notation $\bb{E}_\sigma$
introduced just before it.

\begin{theorem}[restatement of Theorem~4.1 in \cite{[MNOS]}]
\label{t:mnos4.1}
Let $\sigma\in\cc{S}^\rr{m}$;
for any $\delta>0$, there exist $\beta_0>0$ and $K>0$ such that,
for any $\beta>\beta_0$,
\begin{equation}
\label{mnos4.1a}
\bb{P}_{\sigma}
 (\tau_{\cc{S}^\rr{s}}^{\sigma}
   <e^{\beta V_\sigma-\beta\delta})
<e^{-K\beta}    
\end{equation}
and
\begin{equation}
\label{mnos4.1b}
\bb{P}_{\sigma}
  (\tau_{\cc{S}^\rr{s}}^{\sigma} 
    >e^{\beta V_\sigma+\beta\delta}) =
\textnormal{SES}  
\end{equation}
\end{theorem}

\begin{theorem}[restatement of Theorem~4.9 in \cite{[MNOS]}]
\label{t:mnos4.9}
Let
$\sigma\in\cc{S}^\rr{m}$, then
\begin{equation}
\label{mnos4.9}
\lim_{\beta\to\infty}
\frac{1}{\beta}\,\log\bb{E}_\sigma[\tau^\sigma_{\cc{S}^\rr{s}}]=V_\sigma
\end{equation}
\end{theorem}

\begin{theorem}[restatement of Theorem~5.4 in \cite{[MNOS]}]
\label{t:mnos5.4}
Let $\sigma,\eta\in\cc{S}$; consider a gate $\cc{W}$
for the transition from $\sigma$ to $\eta$. Then there 
exists $c>0$ such that 
\begin{equation}
\label{mnos5.4}
\bb{P}_\sigma(\tau^\sigma_{\cc{W}}>\tau^\sigma_\eta)
\le e^{-\beta c}
\end{equation}
for $\beta$ large enough.
\end{theorem}

The proof of Theorems~\ref{t:mnos4.1}--\ref{t:mnos5.4} can be achieved by
arguments much similar to the one developed in \cite{[MNOS]}. 
To this purpose the main ingredients are the revised
definition of cycle and the revised statement of 
\cite[Theorem~2.17]{[MNOS]} (see also \cite[Theorem~6.23]{[OV]}) 
and \cite[Theorem 3.1]{[MNOS]}.

Let $A\subset\cc{S}$, consider $\Phi(A,A^\rr{c})$, we say that 
$A$ is a \textit{cycle} if and only if 
\begin{equation}
\label{defciclo}
\max_{\sigma,\eta\in A}\,\Phi(\sigma,\eta)
<
\Phi(A,A^\rr{c})
\end{equation}
Let $\sigma\in\cc{S}$; we say that the \textit{singleton} 
$\{\sigma\}\subset\cc{S}$ is a \textit{trivial cycle} if and only if 
it is not a cycle.
Given a cycle $A\subset\cc{S}$, we denote by $F(A)$ the set of the minima
of the energy in $A$, i.e.,
\begin{equation}
\label{groundt}
F(A):=\{\sigma\in A:\min_{\eta\in A} E(\eta)=E(\sigma)\}
\end{equation}
We also write $E(F(A))=E(\sigma)$ with $\sigma\in F(A)$. Noted that 
$\Phi(\sigma,A^\rr{c})=\Phi(\sigma',A^\rr{c})$ 
for any $\sigma,\sigma'\in F(A)$, we pick $\sigma\in F(A)$ and set 
$\Phi(A):=\Phi(\sigma,A^\rr{c})$.

\begin{theorem}[restatement of Theorem~2.17 \cite{[MNOS]}]
\label{t:mnos2.17}
Let $A\subset\cc{S}$ be a cycle.
For any $\sigma\in A$,
$\eta\in A^\rr{c}$, $\epsilon,\epsilon'>0$, 
$\delta\in(0,\epsilon)$, 
and $\beta>0$ large enough
\begin{equation}
\label{mnos2.17a}
\bb{P}_{\sigma}(\tau_{A^\rr{c}}^\sigma
               <e^{\beta[\Phi(A)-E(F(A))]+\beta\epsilon};\,
                \tau_{A^\rr{c}}^\sigma=\tau_\eta^\sigma)
\ge e^{-\beta[\Phi(\eta,A))-\Phi( A)]-\beta\epsilon'}
\end{equation}
and
\begin{equation}
\label{mnos2.17b}
\bb{P}_{\sigma}({\tau_{A^\rr{c}}^\sigma} 
                 >e^{\beta[\Phi(A)-E(F(A))]-\beta\epsilon})
\ge 1-e^{-\beta\delta}
\end{equation}  
Moreover, there exists $\kappa>0$ such that for any $\sigma,\sigma'\in A$ 
and $\beta$ large enough
\begin{equation}
\label{mnos2.17c}
\bb{P}_\sigma({\tau_{\sigma'}^\sigma<\tau_{A^\rr{c}}^\sigma})
\ge 1-e^{-\beta\kappa}.
\end{equation}
\end{theorem}

Equation (\ref{mnos2.17a}) is a bound from below to the probability that 
the chain exits a cycle $A$ in a time smaller than the exponential of 
$\beta$ times the height $\Phi(A)-E(F(A))$ of the cycle plus $\epsilon$.
In particular, it is stated that for such a probability the estimate is 
optimal when the exit from the cycle is achieved by touching a configuration 
$\eta$ such that $\Phi(\eta,A)$ is equal to $\Phi(A)$. 
In this case, for any $\epsilon'$, provided $\beta$ is chosen large enough, 
such a probability is larger than $\exp\{-\beta\epsilon'\}$. 
Equation (\ref{mnos2.17b}) is a bound from below to the probability that 
the chain exits a cycle $A$ in a time larger than the exponential of 
$\beta$ times the height $\Phi(A)-E(F(A))$ of the cycle minus $\epsilon$;
such a probability is larger than $1-\exp\{-\beta\delta\}$ with 
$\delta\in(0,\epsilon)$.
Equation (\ref{mnos2.17c}) is a bound from below to the probability that 
the chain started at $\sigma\in A$ visits another configuration $\sigma'$ 
belonging to the cycle $A$ before exiting it; 
such a probability is larger than $1-\exp\{-\beta\kappa\}$ with 
$\kappa>0$ not depending on $\sigma$ and $\sigma'$. 

Before stating the revised version of \cite[Theorem~3.1]{[MNOS]},
we introduce the concept of metastable state at level $V\in\bb{R}$. 
We call \textit{metastable set at level} $V\in\bb{R}$
the set of all states with stability level strictly larger than $V$, i.e.,
$\cc{S}_V:=\{\sigma\in\cc{S}:\,V_\sigma>V\}$. 
Any $\sigma\in\cc{S}_V$ is such that for any path $\omega$ starting 
from $\sigma$ and ending in a configuration with energy lower than 
$E(\sigma)$, the quantity $\Phi_\omega-E(\sigma)$, that is  the energy
level reached along the path and measured with respect to $\sigma$,
is lower than $V$.
 
\begin{theorem}[restatement of Theorem~3.1 in \cite{[MNOS]}]
\label{t:mnos3.1}
For any $\epsilon>0$ and $\beta>0$ large enough
\begin{equation}
\label{mnos3.1}
\sup_{\sigma\in\cc{S}} 
\bb{P}_{\sigma}({\tau_{\cc{S}_V}^\sigma
                >e^{\beta V+\beta\epsilon}}) 
=\textnormal{SES}
\end{equation}
\end{theorem}

Equation (\ref{mnos3.1}) states that the probability that 
the chain started at $\sigma\in\cc{S}$ visits a configuration with 
metastability level $V\in\bb{R}$ in a time larger than the 
exponential of $\beta$ times $V$ plus $\epsilon$ is super--exponentially small
in $\beta$. 

The proof of Theorem~\ref{t:mnos3.1} can be achieved by
repeating the same arguments developed in \cite{[MNOS]} and based on 
\cite[Theorem~2.17]{[MNOS]}.
The proof of Theorem~\ref{t:mnos2.17} can be achieved by
repeating the same arguments quoted in \cite{[MNOS]} and developed 
in the proof of \cite[Theorem~6.23]{[OV]} (see also
\cite[Proposition~3.7]{[OS]}).
In particular, in the proof of (\ref{mnos2.17b}), a revised version
of the so called \textit{reversibility property}
\cite[Lemma~3.1]{[OS]} is needed.

\begin{lemma}
\label{t:OSprop2.7}
Let $A\subset\cc{S}$ be a cycle. For any $\sigma\in F(A)$ and
$\epsilon>0$, there exist $\beta_0>0$ and $c>0$, such that for any 
$\beta\ge\beta_0$,
\begin{equation}
\label{OSprop2.7}
\bb{P}_\sigma\big(\tau^\sigma_{A^\rr{c}}\le 
                  e^{\beta[\Phi(A)-E(F(A))]-\beta\epsilon}\big)
\le e^{-c\,\beta}
\end{equation}
\end{lemma}

To prove Lemma~\ref{t:OSprop2.7} 
we develop an argument much similar to the one
proposed by Olivieri and Vares to prove \cite[Lemma~6.22]{[OV]}.
The difference between the two cases is in the fact that the Hamiltonian 
of the model (\ref{markov}) of the present paper depends on the inverse 
temperature $\beta$, while \cite[Lemma~6.22]{[OV]} refers to a model 
(see \cite[Condition~R in Chapter~6]{[OV]})
with Hamiltonian not depending on the temperature.
Note, also, that the reversibility statement (\ref{OSprop2.7}) is 
given in terms of energy--like quantities not depending on $\beta$; on the 
other hand in the proof of the lemma the key property is the reversibility 
of the dynamics w.r.t.\ the Hamiltonian (\ref{ham}). 

\smallskip
\par\noindent
\textit{Proof of Lemma~\ref{t:OSprop2.7}.\/}
First of all we make explicit how hamitonian--like quantities differ
from the corresponding energy--like quantities multiplied times $\beta$. 
More precisely, given $\sigma,\eta\in\cc{S}$, we have
\begin{equation} 
\label{enee1}
H(\sigma)=\beta\,E(\sigma)
          -\sum_{x\in\Lambda}
	   \log\big[1+e^{-2\beta|S_{\sigma}(x)+h|}\big]
	  +|\Lambda|\,\log2
\end{equation}
and
\begin{equation} 
\label{enee2}
H(\sigma,\eta)=\beta\, E(\sigma,\eta)+|\Lambda|\,\log2
\end{equation}
The equality (\ref{enee2}) follows easily from (\ref{enee1}) and 
(\ref{accaede}). 
We prove, now, (\ref{enee1}). 
Using (\ref{ham}), (\ref{hl}), and recalling 
that $\cosh x=\cosh(-x)$ for any $x\in\bb{R}$, we have that 
\begin{displaymath}
\begin{array}{rcl}
H(\sigma)-\beta\,E(\sigma)
&\!\!=\!\!&
{\displaystyle
-\sum_{x\in\Lambda}\log\cosh(\beta|S_{\sigma}(x)+h|)
+\beta\sum_{x\in\Lambda}|S_{\sigma}(x)+h| 
}\vphantom{\bigg\{_\bigg\}}\\
&\!=\!&
{\displaystyle
-\sum_{x\in\Lambda}
   [\log\cosh(\beta|S_{\sigma}(x)+h|)-\log\exp(\beta|S_{\sigma}(x)+h|)] 
}\\
&\!\!=\!\!&
{\displaystyle
-\sum_{x\in\Lambda}
  \log\frac{e^{\beta|S_{\sigma}(x)+h|}+e^{-\beta|S_{\sigma}(x)+h|}}
  {2e^{\beta|S_{\sigma}(x)+h|}}
=-\sum_{x\in\Lambda}\log\frac{1+e^{-2\beta|S_{\sigma}(x)+h|}}{2} 
}
\end{array}
\end{displaymath}
which yields (\ref{enee1}).

Consider, now, an integer $T\ge2$; recalling that the chain is reversible 
with respect to the Gibbs measure $\mu$ defined above (\ref{ham}),
we have that 
\begin{equation}
\label{rever1}
\begin{array}{rcl}
 \bb{P}_\sigma(\tau_{A^\rr{c}}\le T)
&\!\!=\!\!&
{\displaystyle 
  \sum_{\xi\in A^\rr{c}}\big[p(\sigma,\xi)+\sum_{n=1}^{T-1}
   \sum_{\xi_1,\dots,\xi_n\in A} 
   p(\sigma,\xi_1)\cdots p(\xi_{n-1},\xi_n)\,p(\xi_n,\xi)\big]
}\\
&\!\!=\!\!&
{\displaystyle 
 \sum_{\xi\in A^\rr{c}}\big[p(\sigma,\xi)+\sum_{n=1}^{T-1}
  \sum_{\xi_1,\dots,\xi_n\in A} 
   e^{-[H(\xi_n)-H(\sigma)]}p(\xi_n,\xi_{n-1})\cdots
   p(\xi_1,\sigma)p(\xi_n,\xi)\big] 
}\\
\end{array}
\end{equation}
where the detailed balance (\ref{dett}) has been used to invert the 
order of the configurations in the terms 
$p(\sigma,\xi_1),\dots,p(\xi_{n-1},\xi_n)$.
By using the definition (\ref{comm}) of transition Hamiltonian, we have that
$\exp\{-[H(\xi_n)-H(\sigma)]\}\,p(\xi_n,\xi)
 =\exp\{-[H(\xi_n,\xi)-H(\sigma)]\}$.
Noting, also, that 
\begin{displaymath}
\sum_{\xi_1,\dots,\xi_{n-1}\in A} 
  p(\xi_n,\xi_{n-1})\cdots p(\xi_1,\sigma)
=\bb{P}_\sigma(\sigma_n=\xi_n)
\end{displaymath}
we have
\begin{displaymath}
\begin{array}{rcl}
\bb{P}_\sigma(\tau_{A^\rr{c}}\le T)
&\!\!\!=\!\!\!&
{\displaystyle
 \sum_{\xi\in A^\rr{c}}
 \big[e^{-[H(\sigma,\xi)-H(\sigma)]}
 +\sum_{n=1}^{T-1}
  \sum_{\xi_n\in A} 
  e^{-[H(\xi_n,\xi)-H(\sigma)]}\,\bb{P}_\sigma(\sigma_n=\xi_n)\big]
}\\
&\!\!\!\le\!\!\!&
{\displaystyle
 e^{H(\sigma)}
 \sum_{\xi\in A^\rr{c}}
 \big[e^{-H(\sigma,\xi)}
 +\sum_{n=1}^{T-1}
  \sum_{\xi_n\in A} 
   e^{-H(\xi_n,\xi)}\big]
}\\
&\!\!\!=\!\!\!&
{\displaystyle
 e^{H(\sigma)}
 \sum_{\xi\in A^\rr{c}}
 \big[e^{-H(\sigma,\xi)}
 +(T-1)
  \sum_{\zeta\in A} 
   e^{-H(\zeta,\xi)}\big]
\le
 e^{H(\sigma)}\,T\!\!\!
 \sum_{\zeta\in A,\xi\in A^\rr{c}} 
   e^{-H(\zeta,\xi)}
}\\
\end{array}
\end{displaymath}
which yields
\begin{equation}
\label{rever2}
\bb{P}_\sigma(\tau_{A^\rr{c}}\le T)
\le
  e^{H(\sigma)}
  \,T\, |A|\,|A^\rr{c}|\,
  \max_{\zeta\in A,\xi\in A^\rr{c}} 
  e^{-H(\zeta,\xi)}
=
  e^{H(\sigma)}
  \,T\, |A|\,|A^\rr{c}|\,
  e^{
  -\min_{\zeta\in A,\xi\in A^\rr{c}} H(\zeta,\xi)}
\end{equation}
By using, finally, (\ref{enee1}), (\ref{enee2}), and (\ref{rever2}), we get 
\begin{equation}
\label{rever8}
\bb{P}_\sigma(\tau_{A^\rr{c}}\le T)
\le
  e^{\beta E(\sigma)}
  \,T\, |A|\,|A^\rr{c}|\,
  e^{
  -\min_{\zeta\in A,\xi\in A^\rr{c}} E(\zeta,\xi)}
  \exp\Big\{-\sum_{x\in\Lambda}
              \log\big[1+e^{-2\beta|S_{\sigma}(x)+h|}\big]\Big\}
\end{equation}
Since $A$ is a cycle, we have that 
$\min_{\zeta\in A,\xi\in A^\rr{c}} E(\zeta,\xi)=\Phi(A)$. Hence, 
remarked that 
\begin{displaymath}
0<
  \exp\Big\{-\sum_{x\in\Lambda}
              \log\big[1+e^{-2\beta|S_{\sigma}(x)+h|}\big]\Big\}
\le(1+\exp\{-2\beta(5+h)\})^{-|\Lambda|}
\le\Big(\frac{3}{2}\Big)^{|\Lambda|}
\end{displaymath}
for $\beta$ large enough, 
the bound (\ref{mnos2.17b}) 
follows once we take $T=e^{\beta[\Phi(A)-E(F(A))]-\beta\epsilon}$. 
\qed



\end{document}